\documentclass[final,3p,times,fleqn]{elsarticle}

\usepackage{algorithm}
\usepackage{algpseudocode}

\usepackage{multicol}
\usepackage{multirow}
                                                             
\usepackage{amssymb}
\usepackage{amsbsy}
\usepackage{amsmath}
\usepackage{color}
\usepackage{pdflscape}
\usepackage{afterpage}
\usepackage{capt-of}
\usepackage{csquotes}
 \usepackage{tikz}
 \usetikzlibrary{positioning}
 \usepackage{graphicx}
 \usepackage[outdir=./]{epstopdf}

\usepackage[normal,tight,center]{subfigure}
\usepackage{tcolorbox}
 \usepackage{tikz}
 \usetikzlibrary{arrows}
\usepackage{xspace}
\usepackage{graphicx}
\usepackage{media9}
\usetikzlibrary{shapes}
\usepackage{array}
\usepackage{calc}
\usepackage[thinlines]{easytable}
\usepackage{comment}
\usepackage{tabu}
\usepackage{hyperref}
\makeatletter
\newcommand{\mathleft}{\@fleqntrue\@mathmargin2em}
\newcommand{\mathcenter}{\@fleqnfalse}
\makeatother

\newcommand*\mycirc[1]{%
  \begin{tikzpicture}
    \node[draw,circle,inner sep=4pt] {#1};
  \end{tikzpicture}}

\definecolor{dartmouthgreen}{rgb}{0.05, 0.5, 0.06}
\definecolor{mygreen}{rgb}{0.13, 0.65, 0.49}

\tikzstyle{startstop} = [rectangle, rounded corners, minimum width=1cm, minimum height=1cm,text centered, draw=black, fill=blue!30]
\tikzstyle{io} = [trapezium, trapezium left angle=70, trapezium right angle=110, minimum width=3cm, minimum height=1cm, text centered, draw=black, fill=blue!30]
\tikzstyle{process} = [rectangle, rounded corners, minimum width=1cm, minimum height=1cm,text centered, draw=black, fill=blue!30]
\tikzstyle{decision} = [diamond, minimum width=3cm, minimum height=1cm, text centered, draw=black, fill=blue!30]
\tikzstyle{arrow} = [thick,->,>=stealth]

\journal{}

\begin{document}

\begin{frontmatter}



\title{Scalability of the asynchronous discontinuous Galerkin method for compressible flow simulations}

\author[label2]{Shubham K. Goswami\ }
\ead{shubham.goswami@rub.de}
\author[label1]{Dapse Vidyesh}
\ead{vidyeshdapse@alum.iisc.ac.in}
\author[label1]{Konduri Aditya\corref{cor1}}
\ead{konduriadi@iisc.ac.in}
\cortext[cor1]{Corresponding author.}
\address[label1]{Department of Computational and Data Sciences, Indian Institute of Science, Bengaluru, India}
\address[label2]{Faculty of Mathematics, Ruhr University Bochum, Germany}

\begin{abstract}
	The scalability of time-dependent partial differential equation (PDE) solvers based on the discontinuous Galerkin (DG) method is expectedly limited by data communication and synchronization requirements across processing elements (PEs) at extreme scales. To address these challenges, asynchronous computing approaches that relax communication and synchronization at a mathematical level have been proposed.
	In particular, the asynchronous discontinuous Galerkin (ADG) method with asynchrony-tolerant (AT) fluxes has recently been shown to recover high-order accuracy under relaxed communication, supported by detailed analyses of its accuracy and stability.
	However, the scalability of this approach in modern large-scale parallel DG solvers has not yet been systematically investigated.
	In this paper, we address this gap by implementing the ADG method coupled with AT fluxes in the open-source finite element library \texttt{deal.II}. We employ a communication-avoiding algorithm (CAA) that significantly reduces the frequency of inter-process communication while accommodating controlled delays in ghost value exchanges. We first demonstrate that applying standard numerical fluxes in this asynchronous setting degrades the solution to first-order accuracy, irrespective of the polynomial degree. By incorporating AT fluxes that utilize data from multiple previous time levels, we successfully recover the formal high-order accuracy of the DG discretization. The accuracy of the proposed method is rigorously verified using benchmark problems for the compressible Euler equations. Furthermore, we evaluate the performance of the method through extensive strong-scaling studies for both two- and three-dimensional test cases. Our results indicate that the asynchronous DG solver substantially suppresses synchronization overheads, yielding speedups of up to $1.9\times$ in two dimensions and $1.6\times$ in three dimensions compared to a baseline synchronous DG solver. Overall, the proposed approach demonstrates the potential to develop accurate and scalable DG-based PDE solvers, making it a promising approach for large-scale simulations on emerging exascale computing systems.
\end{abstract}

\begin{keyword}
Asynchronous schemes \sep Partial differential equations \sep Massive computations \sep Discontinuous Galerkin method \sep Compressible Euler equations



\end{keyword}

\end{frontmatter}



\section{Introduction}

The discontinuous Galerkin (DG) method has emerged as a powerful framework for the numerical solution of partial differential equations (PDEs), particularly for hyperbolic systems characterized by shocks, sharp gradients, or discontinuities. By combining key features of finite volume and finite element methods, DG achieves high-order accuracy while preserving strict locality of computations through the use of discontinuous basis functions \cite{hesthaven2007dg}. This locality enables element-wise operations and eliminates the need for global linear solves when explicit time integration is employed, resulting in high arithmetic intensity and favorable performance characteristics on modern high-performance computing (HPC) architectures.
%
Despite these advantages, the scalability of DG solvers on distributed-memory systems is increasingly constrained by data movement and synchronization costs. In parallel DG implementations, numerical flux evaluations across processing-element (PE) boundaries require frequent communication of solution data. As simulations scale to large process counts, the surface-to-volume ratio of each subdomain grows, causing communication and synchronization overheads to dominate the runtime. This challenge is further amplified for high-order discretizations, high-dimensional problems, and fine-grained domain decompositions, posing a major obstacle to the efficient use of emerging exascale systems \cite{exadg2020, exadune2020, exa-flexi2023, Munch2021}.

Substantial research effort has therefore been directed toward improving the parallel efficiency of DG-based solvers. Algorithmic strategies such as hybridized discontinuous Galerkin (HDG) methods reduce the number of globally coupled degrees of freedom by introducing additional interface unknowns \cite{cockburn2009unified, roca2013scalable}. Parallel-in-time approaches, including Parareal methods, further increase concurrency by decomposing the temporal dimension alongside space \cite{LIONS2001661, burrage1997parallel, lirias1119269}. In parallel, hardware-driven advances, most notably GPU acceleration, have enabled DG solvers to exploit asynchronous data transfers and performance-portable programming models \cite{xia2015openacc, kirby2020gpu}. 
Matrix-free DG formulations have emerged as a particularly effective approach, delivering high arithmetic intensity and reduced memory traffic, and have been shown to outperform hybridized variants in many large-scale scenarios \cite{kronbichler2019fast, Fehn2019matrix-free-dg-nse, kronbichler-mf-2025, Schussnig2025}.
These developments are embodied in modern DG software frameworks such as \textit{ExaDG}, built on top of the \texttt{deal.II} finite element library, which has demonstrated strong scaling to hundreds of thousands of cores for challenging fluid dynamics benchmarks \cite{dealii96, exadg2020}. Comparable scalability advances have been reported in other DG frameworks, including \textit{DUNE} and \textit{FLEXI}, which emphasize computation--communication overlap and explicit time integration for large-scale compressible and incompressible flow simulations \cite{bastian-dune2021, exadune2020, flexi-2021, exa-flexi2023}. Nevertheless, even these state-of-the-art solvers exhibit noticeable deviations from ideal scaling at extreme process counts, where communication and synchronization costs become dominant.

%
An alternative strategy for mitigating communication bottlenecks is to relax data communication and synchronization requirements between PEs at a mathematical level, allowing communication-dependent computations to proceed using delayed data \cite{Amitai1994275, AMITAI199327, donzis2014, Mittal_PRE_2017}. 
In particular, we are interested in the asynchronous computing approach based on finite-difference schemes introduced in \cite{konduri2012async, donzis2014}, and the subsequent development of asynchrony-tolerant (AT) finite-difference schemes \cite{konduri2017at}. These AT schemes employ wider spatio-temporal stencils to recover high-order accuracy despite relaxed communication.
An accurate strategy for coupling such AT schemes with multi-stage Runge-Kutta time integration was later developed, demonstrating improved scalability at extreme scales \cite{goswami2023lserkat-jcp}.
These methods have been validated for a range of linear and nonlinear problems, including large-scale simulations of compressible turbulence and reacting flows \cite{aditya2019arXiv, komal2020dns-at, komal2023reactions-at, Arumugam2025-lserkat}.

Building on these ideas, asynchronous computing was recently extended to the discontinuous Galerkin framework through the introduction of the asynchronous discontinuous Galerkin (ADG) method \cite{goswami2022asyncdg-aviation, goswami2024-cmame-adg}. In ADG, inter-process communication is relaxed for multiple time steps, and delayed data are used in numerical flux evaluations at PE boundaries. While this approach substantially reduces synchronization overheads, it was shown that the use of standard numerical fluxes restricts the accuracy of ADG schemes to first-order, irrespective of the polynomial degree. To overcome this limitation, asynchrony-tolerant (AT) numerical fluxes were proposed, enabling the recovery of high-order accuracy by leveraging locally stored flux histories. These developments establish the theoretical foundation for accurate asynchronous DG methods, which have been validated for combustion simulations \cite{Arumugam2025-combustion}. However, their practical accuracy and performance for realistic multidimensional problems in modern DG solvers on massively parallel systems remain open questions.

The focus of the present study is the implementation, validation, and performance characterization of the ADG method with asynchrony-tolerant fluxes in a modern, large-scale DG solver. The main contributions of this paper are summarized as follows:
\begin{itemize}
    \item Implement the asynchronous discontinuous Galerkin (ADG) method with asynchrony-tolerant (AT) fluxes in the open-source finite element library \texttt{deal.II}, building upon the matrix-free DG solver provided in \textit{step-76} \cite{step76}.
    \item Employ a communication-avoiding algorithm (CAA) that performs inter-process communication only at prescribed intervals and combine it with AT fluxes to enable asynchronous execution with high-order accuracy.
    \item Verify the accuracy of the ADG method with AT fluxes for higher-dimensional compressible Euler equations.
    \item Perform a comprehensive strong-scaling and profiling analysis to quantify the performance benefits of the CAA-based DG solver relative to a baseline synchronous DG solver.
\end{itemize}

The remainder of this paper is organized as follows. Section~\ref{sec:background} reviews the DG formulation and the synchronous solver framework. Section~\ref{sec:asyncDG} introduces the asynchronous DG method and the communication-avoiding algorithms for standard and AT fluxes. Implementation details within \texttt{deal.II} are discussed in Section~\ref{sec:dealii-solver}. Numerical accuracy and performance results are presented in Section~\ref{sec:numexp}, followed by concluding remarks and future directions in Section~\ref{sec:conclusions}.

\section{Standard discontinuous Galerkin (DG) method}
\label{sec:background}

We begin with a brief overview of the discontinuous Galerkin (DG) formulation for the compressible Euler equations, which are a system of hyperbolic partial differential equations (PDEs) describing inviscid, compressible fluid flow. Solutions of these equations typically exhibit wave-like behavior and may develop discontinuities, such as shocks, even from smooth initial conditions, making them a canonical benchmark for high-order numerical methods.

Let us express the $d$-dimensional compressible Euler equations in the following conservative form
\begin{equation}
    \frac{\partial \boldsymbol{w}}{\partial t} + \nabla \cdot \boldsymbol{F}(\boldsymbol{w}) = \boldsymbol{G}(\boldsymbol{w}),
    \label{eq:euler-dealii}
\end{equation}
defined on a spatial domain $\Omega \subset \mathbb{R}^d$ with appropriate initial and boundary conditions, and $t \in [0, t_{\text{final}}]$.
Here, $\boldsymbol{w} \in \mathbb{R}^{d+2}$ denotes the vector of conserved variables, $\boldsymbol{F}(\boldsymbol{w}) \in \mathbb{R}^{(d+2)\times d}$ the convective flux tensor, and $\boldsymbol{G}(\boldsymbol{w}) \in \mathbb{R}^{d+2}$ a source term, given by
\begin{equation}
    \boldsymbol{w} =
    \begin{bmatrix}
        \rho \\
        \rho \boldsymbol{u} \\
        E
    \end{bmatrix}, \quad
    \boldsymbol{F}(\boldsymbol{w}) =
    \begin{bmatrix}
        \rho \boldsymbol{u} \\
        \rho \boldsymbol{u} \otimes \boldsymbol{u} + \mathbb{I} p \\
        (E + p)\boldsymbol{u}
    \end{bmatrix}, \quad
    \boldsymbol{G}(\boldsymbol{w}) =
    \begin{bmatrix}
        0 \\
        \rho \boldsymbol{g} \\
        \rho \boldsymbol{u} \cdot \boldsymbol{g}
    \end{bmatrix}.
    \label{eq:euler-parameters-dealii}
\end{equation}
In these expressions, $\rho$ is the density, $\boldsymbol{u} \in \mathbb{R}^d$ the velocity vector, and $E$ the total energy, defined as
\begin{equation}
    E = \frac{p}{\gamma - 1} + \frac{\rho}{2}\|\boldsymbol{u}\|^2,
\end{equation}
where $p$ denotes the pressure and $\gamma$ is the ratio of specific heats (taken as $\gamma = 1.4$ for air). The identity matrix $\mathbb{I} \in \mathbb{R}^{d \times d}$ and the outer product $\otimes$ follow standard notation, while $\boldsymbol{g}$ represents acceleration due to gravity or, more generally, an external force per unit mass acting on the fluid.

\subsection{DG spatial discretization}

To construct a numerical approximation, the spatial domain is partitioned into $N_E$ non-overlapping elements,
$
\Omega \approx \Omega_h = \bigcup_{e=1}^{N_E} \Omega_e.
$
Within the DG framework, the solution is approximated locally on each element by polynomials, allowing for discontinuities across element interfaces. Let $\{\boldsymbol{\varphi}_j^e\}_{j=1}^{N_{\text{dof}}}$ denote a set of polynomial basis functions of degree $N_p$ on element $\Omega_e$, where $N_{\text{dof}}$ is the number of local degrees of freedom (DoFs) per element. The element-wise DG approximation is then written as
\begin{equation}
    \boldsymbol{w}_h^e(\boldsymbol{x}, t)
    = \sum_{j=1}^{N_{\text{dof}}} \widehat{\boldsymbol{w}}_j^{\,e}(t)\, \boldsymbol{\varphi}_j^e(\boldsymbol{x}),
    \label{eq:DGsolution-dealii}
\end{equation}
with $\widehat{\boldsymbol{w}}_j^{\,e}(t)$ denoting the time-dependent local DoFs. The global DG solution is obtained as the direct sum of the element-wise approximations,
$
\boldsymbol{w}_h(\boldsymbol{x}, t) \approx \bigoplus_{e=1}^{N_E} \boldsymbol{w}_h^e(\boldsymbol{x}, t).
$

Substituting $\boldsymbol{w}_h$ into Eq.~\eqref{eq:euler-dealii} yields the residual
\begin{equation}
    \boldsymbol{\mathcal{R}}_h =
    \frac{\partial \boldsymbol{w}_h}{\partial t}
    + \nabla \cdot \boldsymbol{F}(\boldsymbol{w}_h)
    - \boldsymbol{G}(\boldsymbol{w}_h),
    \label{eq:residual-dealii}
\end{equation}
which must be minimized to obtain a numerical solution with a desired accuracy. Following the Galerkin approach, we choose test functions from the same space as the basis functions and impose orthogonality of the residual with respect to the $L^2$ inner product. This leads to the element-wise formulation
\begin{equation}
    \left(
    \boldsymbol{\varphi}_i^e,
    \frac{\partial \boldsymbol{w}_h^e}{\partial t}
    \right)_{\Omega_e}
    + \left(
    \boldsymbol{\varphi}_i^e,
    \nabla \cdot \boldsymbol{F}(\boldsymbol{w}_h^e)
    \right)_{\Omega_e}
    - \left(
    \boldsymbol{\varphi}_i^e,
    \boldsymbol{G}(\boldsymbol{w}_h^e)
    \right)_{\Omega_e}
    = 0.
    \label{eq:innerproduct-dealii}
\end{equation}

Applying integration by parts to the flux divergence term yields the standard DG weak form
\begin{equation}
    \left(
    \boldsymbol{\varphi}_i^e,
    \frac{\partial \boldsymbol{w}_h^e}{\partial t}
    \right)_{\Omega_e}
    - \left(
    \nabla \boldsymbol{\varphi}_i^e,
    \boldsymbol{F}(\boldsymbol{w}_h^e)
    \right)_{\Omega_e}
    + \left\langle
    \boldsymbol{\varphi}_i^e,
    \boldsymbol{n} \cdot \widehat{\boldsymbol{F}}(\boldsymbol{w}_h)
    \right\rangle_{\partial \Omega_e}
    - \left(
    \boldsymbol{\varphi}_i^e,
    \boldsymbol{G}(\boldsymbol{w}_h^e)
    \right)_{\Omega_e}
    = 0,
    \label{eq:weakform-dealii}
\end{equation}
where $\boldsymbol{n}$ denotes the outward unit normal and $\widehat{\boldsymbol{F}}$ is a numerical flux enforcing inter-element coupling.

This formulation can be written compactly as
\begin{equation}
    \boldsymbol{\mathcal{M}}_e
    \frac{d \boldsymbol{w}_h^e}{dt}
    = \mathcal{L}_h(t, \boldsymbol{w}_h^e),
    \label{eq:weakform-compact-dealii}
\end{equation}
with $\boldsymbol{\mathcal{M}}_e$ denoting the element-local mass matrix and $\mathcal{L}_h$ the spatial DG operator incorporating both volume and surface contributions.

\subsection{Numerical flux}

\begin{figure}[h!]
    \centering
    \includegraphics[width=0.325\linewidth]{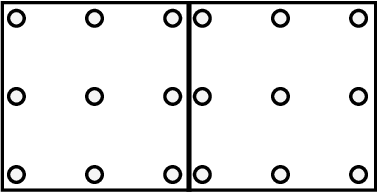}
    \put(-118,-10){$\Omega_{e-1}$}
    \put(-43,-10){$\Omega_{e}$}
    \put(-100,35){$\boldsymbol{w}^-$}
    \put(-63,35){$\boldsymbol{w}^+$}
    \put(-95,80){$\widehat{\boldsymbol{F}}(\boldsymbol{w}^-, \boldsymbol{w}^+) $}
\caption{\small{Illustration of numerical flux computation across a common edge of two quadrilateral elements in a two-dimensional domain. Circles denote the local nodal degrees of freedom of the elements.}}
\label{fig:numericalflux}
\end{figure}
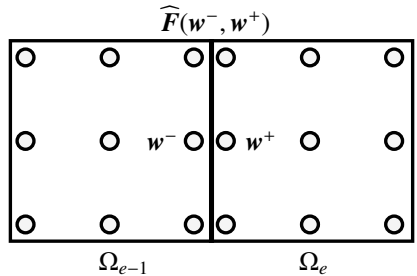

Due to the discontinuous nature of the DG basis functions, the solution obtained over the discretized domain is, in general, discontinuous across element interfaces. As a result, at the common boundary between two neighboring elements, multiple solution values are defined, one from each element sharing the interface. To illustrate this, Fig.~\ref{fig:numericalflux} shows two adjacent elements, $\Omega_{e-1}$ and $\Omega_e$, in a two-dimensional setting, represented here as quadrilateral elements with quadratic basis functions. At the shared face, the solution interface value from the left element is denoted by $\boldsymbol{w}^-$, while the interface value from the right element is denoted by $\boldsymbol{w}^+$. In general, these two values are not equal, leading to ambiguous flux evaluations at the interface.

To ensure a unique, conservative coupling between neighboring elements, a single-valued numerical flux function
$
\widehat{\boldsymbol{F}}(\boldsymbol{w})
=
\widehat{\boldsymbol{F}}(\boldsymbol{w}^-, \boldsymbol{w}^+)
$
is introduced. This flux weakly enforces inter-element continuity, guarantees local conservation, and provides the necessary information exchange between elements in the DG formulation.

In this work, we employ the local Lax-Friedrichs (Rusanov) flux~\cite{dealii96}, defined as
\begin{equation}
    \widehat{\boldsymbol{F}}(\boldsymbol{w}^-,\boldsymbol{w}^+)
    =
    \frac{1}{2}
    \left[
        \boldsymbol{F}(\boldsymbol{w}^-)
        +
        \boldsymbol{F}(\boldsymbol{w}^+)
    \right]
    +
    \frac{\lambda}{2}
    \left(
        \boldsymbol{w}^- - \boldsymbol{w}^+
    \right)
    \otimes \boldsymbol{n}^-,
    \label{eq:lfflux}
\end{equation}
where $\boldsymbol{n}^-$ denotes the outward unit normal associated with the interior state $\boldsymbol{w}^-$. The dissipation parameter $\lambda$ is chosen as
\begin{equation}
    \lambda
    =
    \frac{1}{2}
    \max
    \left(
        \sqrt{\|\boldsymbol{u}^-\|^2 + (c^-)^2},
        \sqrt{\|\boldsymbol{u}^+\|^2 + (c^+)^2}
    \right),
    \label{eq:lfflux-lambda}
\end{equation}
with $c$ denoting the local speed of sound. This choice ensures stability and robustness of the DG discretization in the presence of strong gradients and discontinuities, which are common in compressible flow simulations.

\subsection{Time integration}
\label{sec:time-integration}

The semi-discrete system in Eq.~\eqref{eq:weakform-compact-dealii} constitutes a set of ordinary differential equations in time. To preserve the locality and scalability of the DG formulation, explicit time-integration schemes are typically employed. In this work, we use low-storage explicit Runge-Kutta (LSERK) methods, which are widely adopted in large-scale DG solvers due to their favorable stability and memory characteristics.

A generic $s$-stage two-storage LSERK scheme can be written as
\begin{align}
    \boldsymbol{k}_m^e &=
    \boldsymbol{\mathcal{M}}_e^{-1}
    \mathcal{L}_h(t^n + \partial_m \Delta t, \boldsymbol{r}_m^e), \nonumber \\
    \boldsymbol{r}_{m+1}^e &=
    \boldsymbol{w}_h^{e,n}
    + \Delta t\, a_m\, \boldsymbol{k}_m^e, \nonumber \\
    \boldsymbol{w}_h^{e,n+1} &=
    \boldsymbol{w}_h^{e,n}
    + \Delta t\, b_m\, \boldsymbol{k}_m^e,
    \quad m = 1,\dots,s,
    \label{eq:lserk-dg}
\end{align}
where the coefficients $a_m$, $b_m$, and $\partial_m$ are determined from the corresponding Butcher tableau \cite{KENNEDY2000}.

In a serial implementation, this fully discrete scheme advances the solution by looping over all elements and evaluating the spatial operator locally. Because the scheme is explicit, no global linear solve is required. Instead, element-level computations can be performed independently, apart from the data dependencies introduced by the numerical fluxes. This locality makes DG methods particularly well-suited for parallelization, enabling efficient large-scale simulations through concurrent execution across elements.

\subsection{Parallel implementation}
\label{sec:parallel}

The parallel implementation of the discontinuous Galerkin (DG) method in this work targets distributed-memory systems using the Message Passing Interface (MPI). The computational domain is decomposed into multiple subdomains and distributed across processing elements (PEs), each of which owns a subset of the mesh elements. Based on data dependencies, local elements are classified as \emph{interior elements}, whose DG operator evaluations rely solely on locally available data, and \emph{PE-boundary elements}, for which numerical flux evaluations require solution values from neighboring PEs.
The implementation builds upon the distributed-memory parallel infrastructure provided by the open-source finite element library \texttt{deal.II}. This framework offers explicit control over ghost-value exchange and enables a clear separation between communication and computation, which is essential for exposing communication-computation overlap in large-scale DG solvers. Detailed solver-specific aspects of the \texttt{deal.II} implementation are discussed in Sec.~\ref{sec:dealii-solver}.

Communication between PEs is required only for evaluating numerical fluxes on PE-boundary elements. Solution values associated with neighboring elements are exchanged via MPI and stored in ghost (buffer) elements. The parallel communication model follows a standard ghost-exchange pattern, where ghost-value updates are explicitly initiated and completed outside the element-local computational kernels. This design decouples communication from computation and enables partial overlap of data exchange with local work.
In practice, each PE initiates nonblocking point-to-point communication to exchange ghost values with its neighboring PEs and then proceeds with the evaluation of DG contributions for interior elements that do not depend on remote data. Once these computations are completed, a synchronization step ensures that all required ghost values have been received. The PE then evaluates the contributions for its boundary elements using the updated ghost data. This element-centric execution model preserves locality, minimizes idle time, and forms the basis for overlapping communication with computation.

\begin{figure}[h!]
    \includegraphics[width=0.40\linewidth]{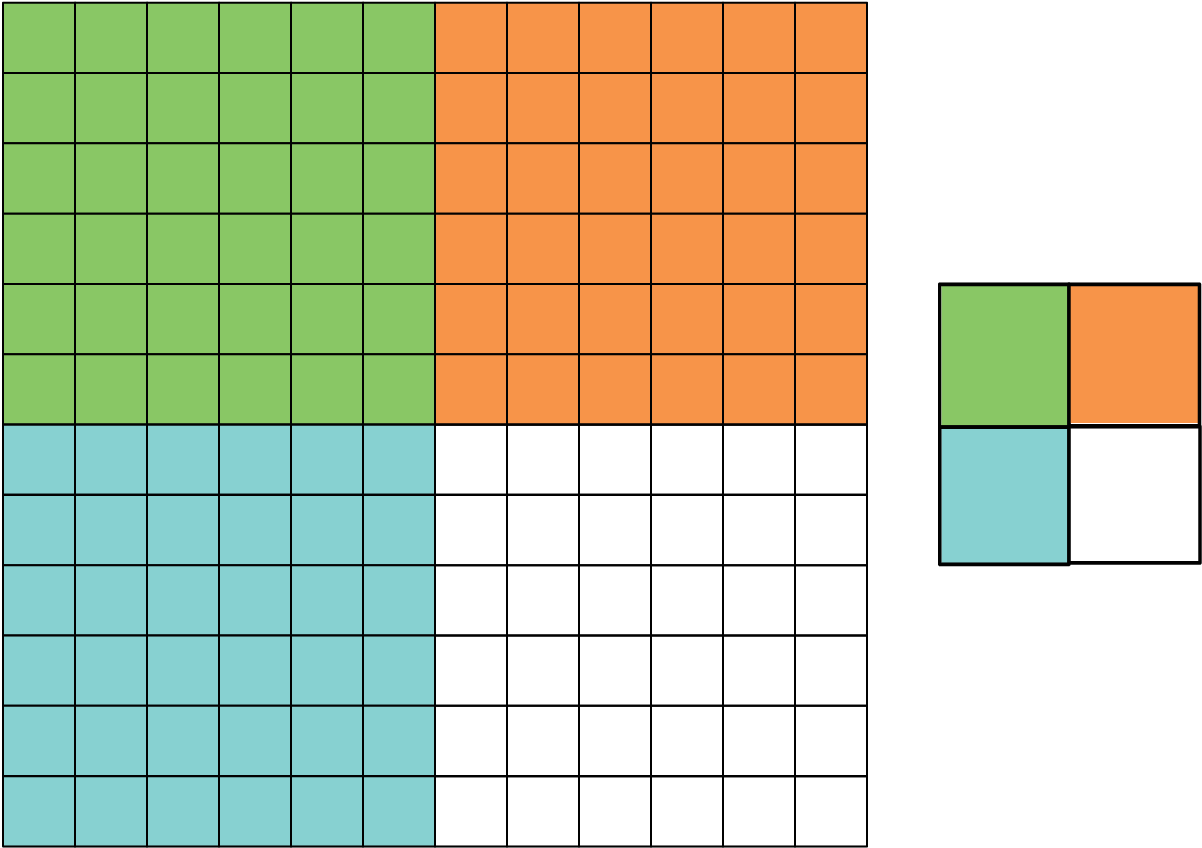}
    \begin{picture}(0,0)
        \put(-105,-8){\small{(a)}}
        \put(-42,75){\small{PE-0}}
        \put(-21,75){\small{PE-1}}
        \put(-42,52){\small{PE-2}}
        \put(-21,52){\small{PE-3}}
    \end{picture}
    \hspace{0.2cm}
    \includegraphics[width=0.28\linewidth]{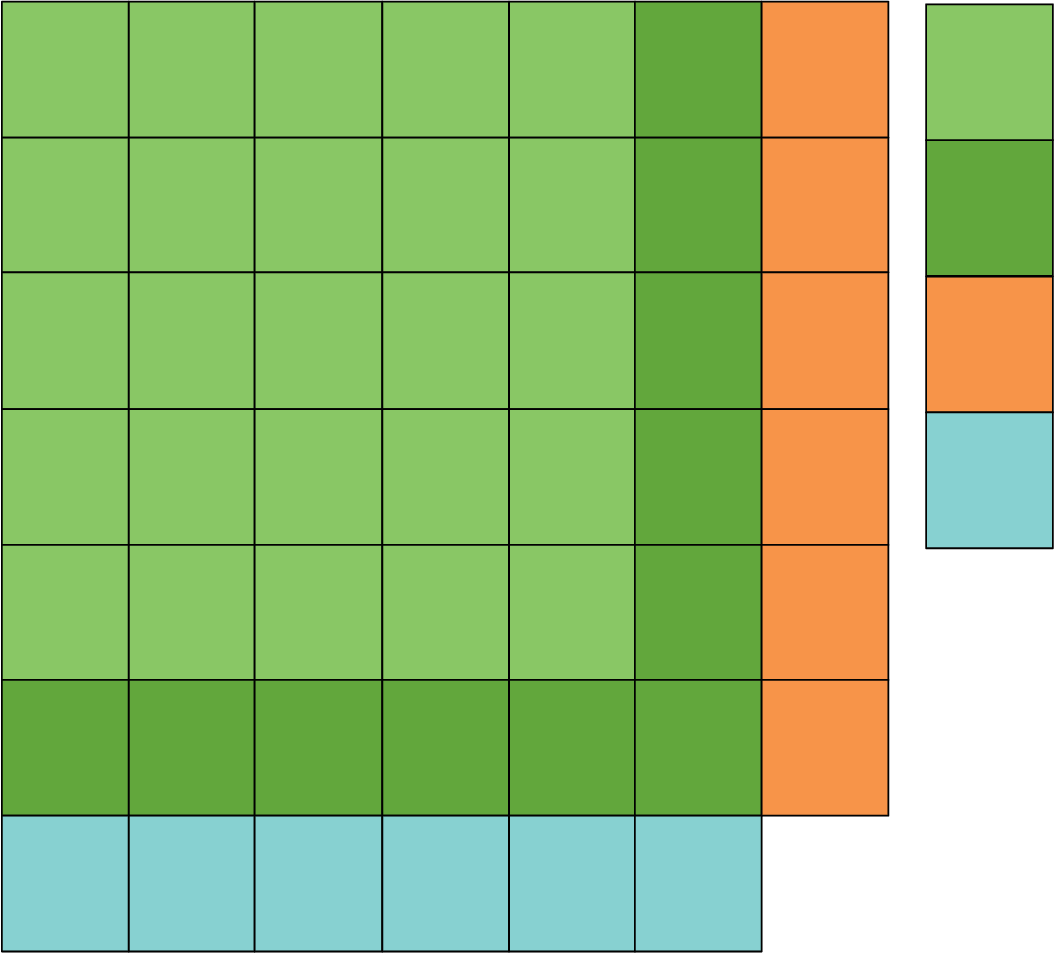}
    \begin{picture}(0,0)
        \put(-60,-8){\small{(b)}}
        \put(0,107){\small{Interior elements of PE-0}}
        \put(0,90){\small{PE boundary elements of PE-0}}
        \put(0,72){\small{Buffer elements with entries from PE-1}}
        \put(0,55){\small{Buffer elements with entries from PE-2}}
    \end{picture}
\caption{\small{(a) Domain decomposition of $144$ elements among four PEs: PE-0 (green), PE-1 (orange), PE-2 (blue), and PE-3 (white); (b) classification of interior and PE-boundary elements in the subdomain of PE-0.}}
\label{fig:Domain-decomposition}
\end{figure}
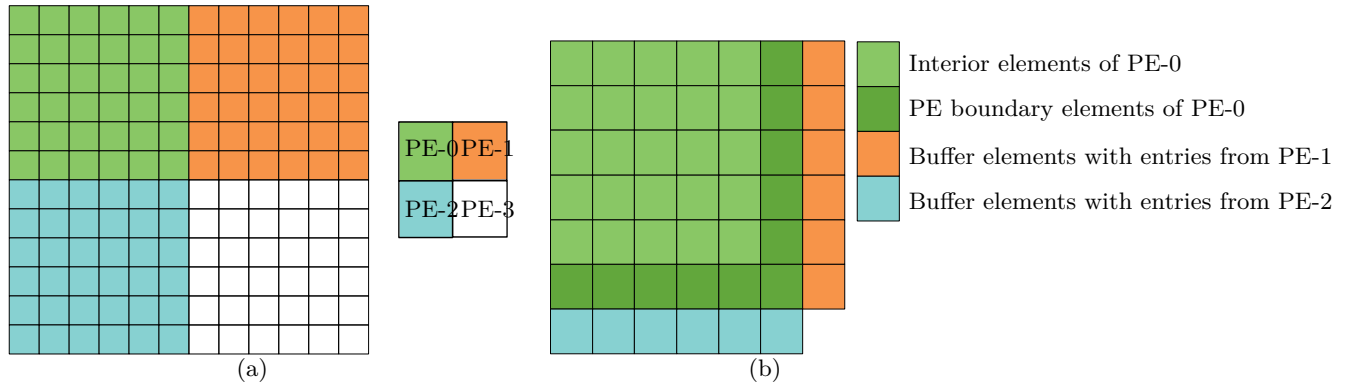

Figure~\ref{fig:Domain-decomposition} illustrates this domain decomposition for a two-dimensional example. Part (a) of the figure exhibits the domain decomposition of $144$ elements among 4 PEs, namely, `PE-0', `PE-1', `PE-2', and `PE-3', where the elements of these PEs are represented by green, orange, blue and white colors, respectively. Each of the PEs has 36 local elements. Part (b) of the figure depicts the subdomain of PE-0, demonstrating the interior elements in light green and the boundary elements of PE-0 that rely on values from neighboring PEs for flux computation in dark green. The required values from different PEs are communicated and stored in buffer elements. The buffer elements that store the values from PE-1 are shown in orange, and for storing the communicated values from elements of PE-2, the buffer elements are shown in blue.

This parallel execution model serves as the foundation for both the synchronous and asynchronous algorithms considered in this work. In the following subsection, we describe the standard synchronous algorithm, in which communication and synchronization are enforced at every Runge-Kutta stage.

\subsubsection{Synchronous algorithm}

The synchronous algorithm (SA) for the discontinuous Galerkin (DG) method combined with a low-storage explicit Runge–Kutta (RK) time integration scheme is summarized in Algorithm~\ref{algo:sa}. This execution model represents the baseline parallel strategy employed in a DG-based parallel PDE solver.
Following domain decomposition across processing elements (PEs), the simulation proceeds through a time-stepping loop that dominates the overall computational cost. Each time step consists of multiple RK stages. At the beginning of every RK stage, non-blocking MPI communication is initiated to exchange ghost values associated with PE-boundary elements. While this communication is in progress, the algorithm evaluates all DG contributions, both volume and face integrals, for interior elements in $\Omega_I$, whose numerical fluxes depend exclusively on locally owned degrees of freedom.
After completing the interior-element computations, a synchronization step is enforced to ensure that all communicated ghost values have been received. Once synchronization is complete, the algorithm proceeds to evaluate the DG operator for PE-boundary elements in $\Omega_B$, where numerical fluxes at inter-process interfaces require the most recent data from neighboring PEs. Using these updated ghost values, the stage solution is assembled and the solution is advanced to the next RK stage.

\begin{algorithm}[h!]
    \centering
    \small
\caption{{Synchronous algorithm (SA). Baseline DG-RK implementation with communication and synchronization at every Runge-Kutta stage. ($\Omega_I$: interior elements not requiring MPI ghost data; $\Omega_B$: PE-boundary elements)}}
\label{algo:sa}
\begin{algorithmic}[1]
\State \textbf{Initialization}
\For{\texttt{each time step $n$}}  
    \For{\texttt{each RK stage $m$}}  
        \State \textbf{Initiate MPI communication} of ghost values for PE-boundary elements
        \Statex \textcolor{gray}{\texttt{/* Compute contributions for interior elements */}}
        \For{\texttt{each element $\Omega_e$ in $\Omega_I$}}
            \State Compute volume contribution using latest local values
            \For{\texttt{each face $f$ in $\partial \Omega_e$}}
                \State Compute numerical flux using latest local values:
                $\boldsymbol{\widehat{F}}^{\,n+\partial_m}
                = \boldsymbol{\widehat{F}}\!\left(
                (\boldsymbol{w}_h^{\,n+\partial_m})^{-},
                (\boldsymbol{w}_h^{\,n+\partial_m})^{+}
                \right)
                $
            \EndFor
            \State Update local stage solution for $\Omega_e$
        \EndFor
        \State \textbf{Finish MPI communication} (synchronize ghost values)
        \Statex \textcolor{gray}{\texttt{/* Compute contributions for PE-boundary elements */}}
        \For{\texttt{each element $\Omega_e$ in $\Omega_B$}}
            \State Compute volume contribution using latest local values
            \For{\texttt{each face $f$ in $\partial \Omega_e$}}
                \State Compute numerical flux using latest values:
                $\boldsymbol{\widehat{F}}^{\,n+\partial_m}
                = \boldsymbol{\widehat{F}}\!\left(
                (\boldsymbol{w}_h^{\,n+\partial_m})^{-},
                (\boldsymbol{w}_h^{\,n+\partial_m})^{+}
                \right)
                $
            \EndFor
            \State Update local stage solution for $\Omega_e$
        \EndFor
        \State \textbf{Advance solution to next RK stage}
    \EndFor
\EndFor
\end{algorithmic}
\end{algorithm}

Although non-blocking communication allows partial overlap between communication and computation, the requirement to synchronize ghost values at every RK stage introduces frequent global synchronization points. As a result, the synchronous algorithm can become increasingly communication-bound at large process counts, particularly for high-order discretizations and large-scale simulations. This limitation motivates the development of asynchronous discontinuous Galerkin method discussed in the next section.

\section{Asynchronous DG method}
\label{sec:asyncDG}

Despite extensive optimization efforts and the use of computation-communication overlap, the communication and synchronization requirements enforced at the end of each Runge-Kutta (RK) stage introduce substantial overhead in large-scale DG solvers. At extreme process counts, this overhead increasingly dominates the overall runtime and limits scalability. To address this issue, a new asynchronous approach for the discontinuous Galerkin (DG) method has been proposed that relaxes communication and synchronization constraints at the mathematical level~\cite{goswami2024-cmame-adg}. This approach allows the solver to continue advancing in time even when data from neighboring PEs are not yet available, by temporarily relying on previously communicated values.

To illustrate the idea, we consider a fully discrete update equation corresponding to the semi-discrete system in Eq.~\eqref{eq:weakform-compact-dealii}, using an explicit Euler scheme for time integration. The update equation for advancing the solution from time level $n$ to $n+1$ with step size $\Delta t$ can be represented as
\begin{equation}
    \boldsymbol{w}_h^{e,n+1}
    =
    \boldsymbol{w}_h^{e,n}
    +
    \Delta t \,
    \boldsymbol{\mathcal{M}}_e^{-1}
    \mathcal{L}_h \left( t^n, \boldsymbol{w}_h^{e,n} \right),
    \label{eq:euler-dg-compact}
\end{equation}
where the right-hand-side operator uses the numerical flux
$\widehat{\boldsymbol{F}}_e^{\,n}
=
\widehat{\boldsymbol{F}}\!\left(
(\boldsymbol{w}_e^-)^n,
(\boldsymbol{w}_e^+)^n
\right)$
evaluated with solution values from the current time level.

In contrast, the asynchronous discontinuous Galerkin (ADG) method permits the use of delayed data at PE boundaries when communication is avoided. The update equation can then be written as
\begin{equation}
    \boldsymbol{w}_h^{e,n+1}
    =
    \boldsymbol{w}_h^{e,n}
    +
    \Delta t \,
    \boldsymbol{\mathcal{M}}_e^{-1}
    \mathcal{L}_h \left(
    t^n,
    \boldsymbol{w}_h^{e,n},
    \boldsymbol{w}_h^{e,n-\tilde{k}}
    \right),
    \label{eq:euler-adg-compact}
\end{equation}
where the numerical flux at PE-boundary elements is evaluated using values from an earlier time level $n-\tilde{k}$,
\begin{equation}
    \widehat{\boldsymbol{F}}_e^{\,\text{std},n}
    =
    \widehat{\boldsymbol{F}}_e\!\left(
    (\boldsymbol{w}_h^{e,n-\tilde{k}})^-,
    (\boldsymbol{w}_h^{e,n-\tilde{k}})^+
    \right)
    =
    \widehat{\boldsymbol{F}}_e^{\,n-\tilde{k}} .
    \label{eq:delayed-flux}
\end{equation}
Here, $\tilde{k}$ denotes the delay with respect to the current time level $n$, and is bounded by a parameter $L$, called the maximum allowable delay, such that, $\tilde{k} \in \{0, 1, \dots, L-1\}$. The standard synchronous DG method is recovered as the special case $\tilde{k}=0$.

\begin{figure}[h!]
\vspace{0.5cm}
\begin{multicols}{2}
    \includegraphics[width=1.03\linewidth]{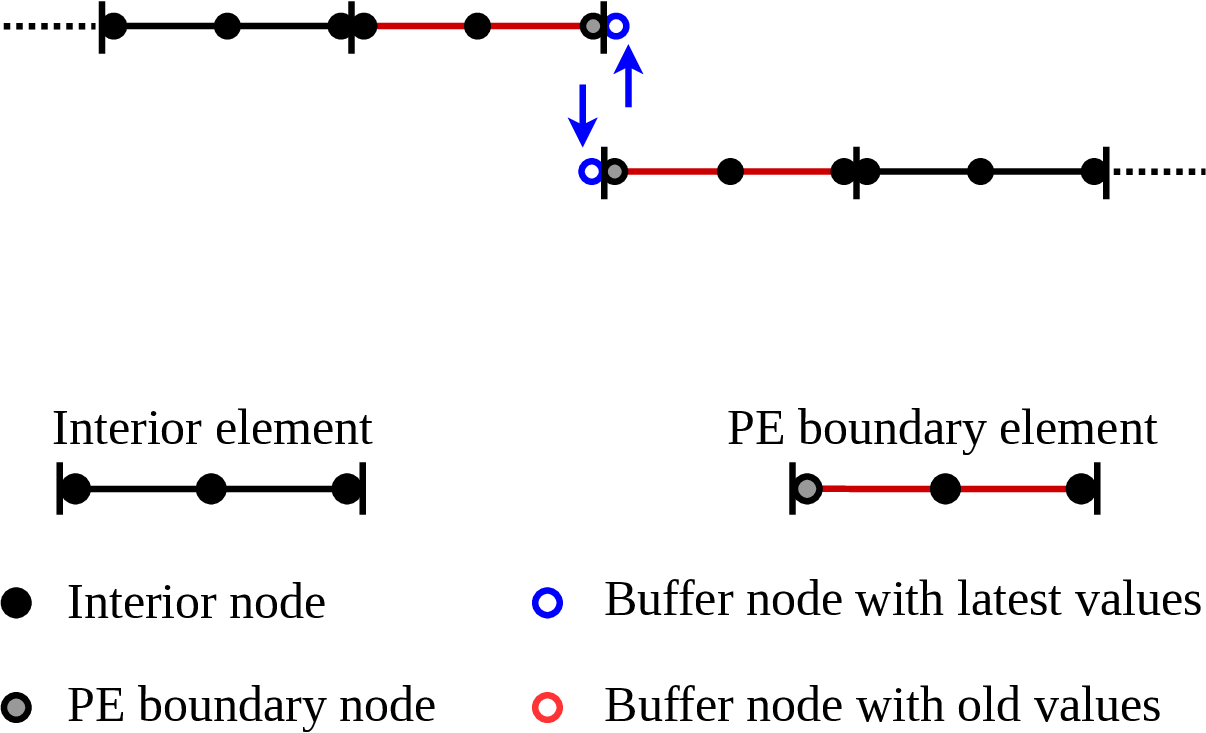}
    \begin{picture}(0,0)
    \put(38,120){\textcolor{blue}{\small{Communication}}}
    \put(131,139){\textcolor{blue}{\small{Synchronization}}}
    \put(90,152){\textcolor{black}{\small{$(w_e^-)^n$}}}
    \put(90,107){\textcolor{blue}{\small{$(w_e^-)^n$}}}
    \put(116,152){\textcolor{blue}{\small{$(w_e^+)^n$}}}
    \put(116,107){\textcolor{black}{\small{$(w_e^+)^n$}}}
    \put(87,134){\small{$\Omega_{e-1}$}}
    \put(39,134){\small{$\Omega_{e-2}$}}
    \put(136,124){\small{$\Omega_{e}$}}
    \put(182,124){\small{$\Omega_{e+1}$}}
    \put(14,130){\small{$x_{e-2}$}}
    \put(62,130){\small{$x_{e-1}$}}
    \put(113,129){\small{$x_{e}$}}
    \put(113,101){\small{$x_{e}$}}
    \put(160,101){\small{$x_{e+1}$}}
    \put(206,101){\small{$x_{e+2}$}}
    \put(50,162){\footnotesize{PE-0}}
    \put(168,86){\footnotesize{PE-1}}
    \put(100,168){\small{(a)}}
    \end{picture}
    \par
    \includegraphics[width=0.85\linewidth]{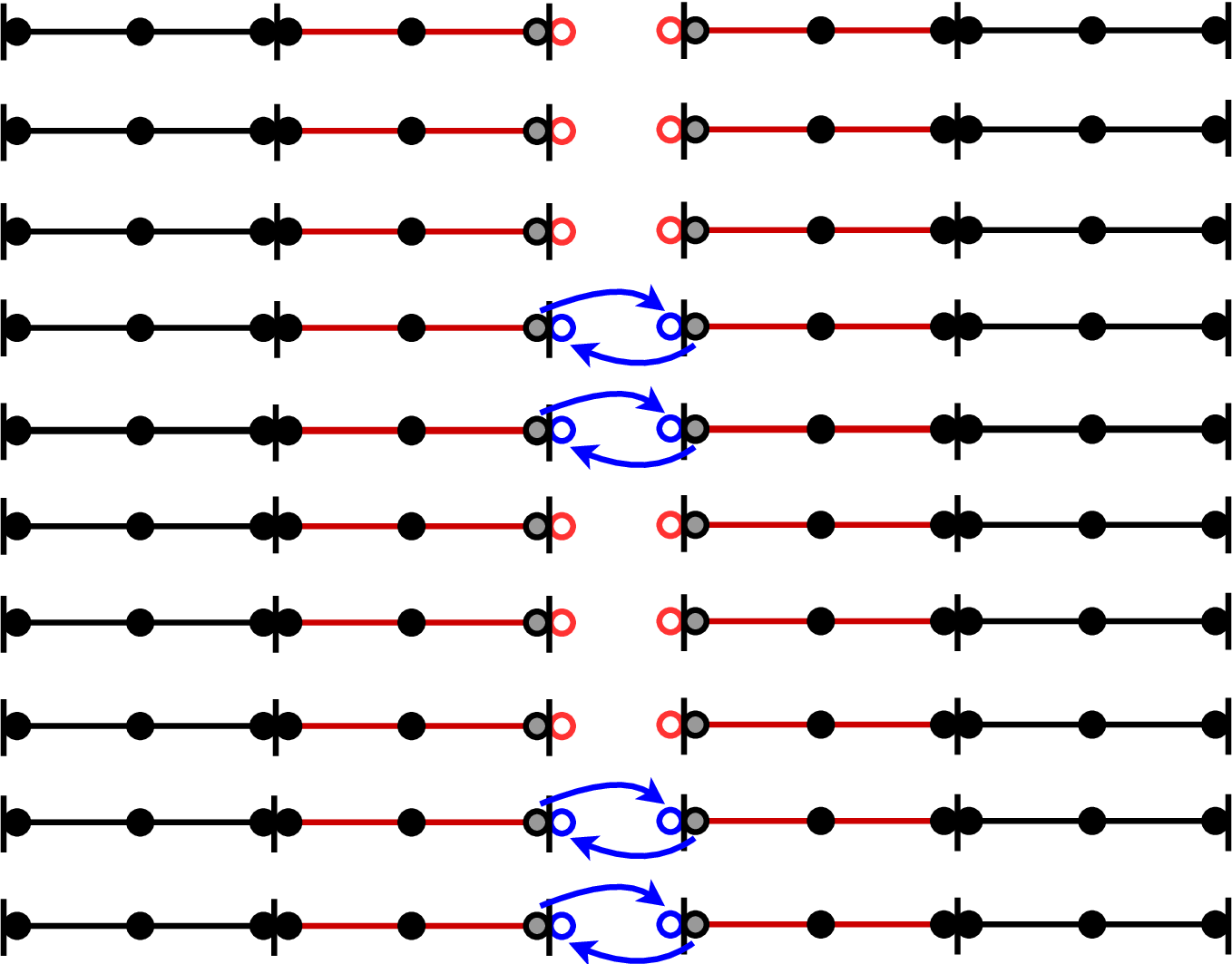}   
    \begin{picture}(0,0)
    \put(-103,56){\textcolor{red}{\small{No}}}
    \put(-128,42){\textcolor{red}{\small{communication}}}
    \put(-103,131){\textcolor{red}{\small{No}}}
    \put(-128,117){\textcolor{red}{\small{communication}}}
    \put(-156,152){\textcolor{black}{\footnotesize{PE-0}}}
    \put(-52,152){\textcolor{black}{\footnotesize{PE-1}}}
    \put(0,4){\textcolor{black}{\small{$t^0$}}}
    \put(0,19){\textcolor{black}{\small{$t^1$}}}
    \put(0,34){\textcolor{black}{\small{$t^2$}}}
    \put(0,50){\textcolor{black}{\small{$t^3$}}}
    \put(0,65){\textcolor{black}{\small{$t^4$}}}
    \put(0,79){\textcolor{black}{\small{$t^5$}}}
    \put(0,95){\textcolor{black}{\small{$t^6$}}}
    \put(0,110){\textcolor{black}{\small{$t^7$}}}
    \put(0,125){\textcolor{black}{\small{$t^8$}}}
    \put(0,141){\textcolor{black}{\small{$t^9$}}}
    \put(-175,11){\textcolor{blue}{\small{Communication}}}
    \put(-83,11){\textcolor{blue}{\small{Synchronization}}}
    \put(-175,87){\textcolor{blue}{\small{Communication}}}
    \put(-83,87){\textcolor{blue}{\small{Synchronization}}}
    \put(-177,-5){\textcolor{black}{\small{$\Omega_{e-2}$}}}
    \put(-135,-5){\textcolor{black}{\small{$\Omega_{e-1}$}}}
    \put(-70,-5){\textcolor{black}{\small{$\Omega_e$}}}
    \put(-31,-5){\textcolor{black}{\small{$\Omega_{e+1}$}}}
    \put(-195,-10){\textcolor{black}{\small{$x_{e-2}$}}}
    \put(-155,-10){\textcolor{black}{\small{$x_{e-1}$}}}
    \put(-110,-10){\textcolor{black}{\small{$x_e$}}}
    \put(-90,-10){\textcolor{black}{\small{$x_{e}$}}}
    \put(-50,-10){\textcolor{black}{\small{$x_{e+1}$}}}
    \put(-10,-10){\textcolor{black}{\small{$x_{e+2}$}}}
    \put(-120,148){\textcolor{black}{\small{$w_e^-$}}}
    \put(-85,148){\textcolor{black}{\small{$w_{e}^+$}}}
    \put(-100,164){\small{(b)}}
    \end{picture}
\end{multicols}
\caption{\small{Illustration of a discretized one-dimensional domain decomposed into two processing elements (PEs): (a) synchronous execution and (b) asynchronous execution based on the communication-avoiding algorithm with maximum allowable delay $L = 4$.}}
\label{fig:domain-sa-vs-caa}
\end{figure}
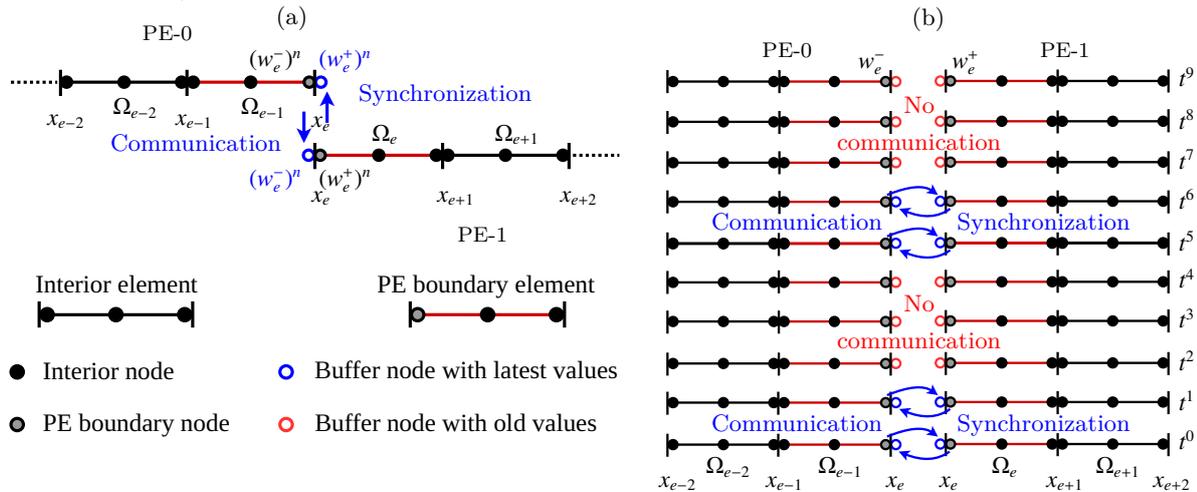

The effect of delayed boundary updates is illustrated in Fig.~\ref{fig:domain-sa-vs-caa} for a one-dimensional domain decomposed between two processing elements, denoted PE-0 and PE-1. The elements $\Omega_{e-1}$ and $\Omega_e$ represent the right and left boundary elements of PE-0 and PE-1, respectively. Figure~\ref{fig:domain-sa-vs-caa}(a) shows the standard synchronous implementation, in which boundary values are exchanged at every time step and stored at buffer nodes (shown in blue). Fig.~\ref{fig:domain-sa-vs-caa}(b) depicts an asynchronous execution based on a communication-avoiding algorithm (CAA)~\cite{komal2020dns-at,goswami2023lserkat-jcp}.
In this illustrative example, PEs communicate and synchronize boundary values for two consecutive time steps, followed by three time steps without communication, corresponding to the maximum allowable delay of $L=4$. This specific pattern is chosen solely for visualization purposes; in practice, different communication intervals may be used.

At time $t^{0}$, boundary values $u^{0}$ are communicated and stored in the buffers, and are subsequently used to compute $u^{1}$ at the PE-boundary elements. Communication is repeated at $t^{1}$ to update the buffers with $u^{1}$, which is then used to compute $u^{2}$ at $t^{2}$. These updates follow the standard DG formulation. Starting from $t^{2}$, communication is suspended and the buffer values are no longer refreshed; these delayed values are indicated in red in the figure. The update at $t^{3}$ therefore uses boundary values $u^{1}$, corresponding to a delay $\tilde{k}=1$. Subsequent updates at $t^{4}$ and $t^{5}$ proceed with delays $\tilde{k}=2$ and $\tilde{k}=3$, respectively.
After this communication-free phase, boundary values are refreshed at $t^{5}$, and the updates at $t^{6}$ and $t^{7}$ again follow the standard DG method. This pattern repeats throughout the simulation. Interior elements always use the standard DG update, as all required neighboring values are locally available. Only PE-boundary elements alternate between synchronous and asynchronous updates, depending on the availability of the latest boundary data in the buffers.

\subsection{Communication-avoiding algorithm with the standard flux}
\label{sec:caa-std}

Contrary to the synchronous DG method, the asynchronous DG (ADG) method allows communication to be relaxed for a limited number of time steps. To realize this idea in practice, we employ the \emph{communication-avoiding algorithm} (CAA), previously introduced in \cite{komal2020dns-at, goswami2023lserkat-jcp}. The CAA performs inter-process communication only periodically, once every $L$ time steps, while allowing the solver to continue advancing in time using previously communicated data during the intervening steps. Algorithm~\ref{algo:caa-std} summarizes the resulting execution pattern.
At the beginning of each time step $n$, the algorithm determines whether communication should be performed by evaluating the condition $n \% L = 0$. Based on this condition, a boolean flag \textit{communication} is set. When communication is enabled, ghost-value exchange is explicitly initiated and completed via MPI calls, ensuring that all ghost buffers contain data corresponding to the most recent time level. When communication is disabled, these calls are skipped and the solver proceeds without synchronization.

As in the synchronous algorithm, computations are organized in a nested loop over Runge-Kutta (RK) stages and elements. Interior elements always use the standard DG update, since all required neighboring values are locally available. For PE-boundary elements, however, numerical flux evaluations may be affected by ghost values that are no longer up-to-date when communication is avoided. To account for this, the CAA introduces a delay parameter $k$, which represents the number of time steps elapsed since the most recent communication.
The delay parameter $k$ is initialized to zero whenever communication is performed and is incremented by one at each subsequent time step for which communication is skipped. For example, if $L = 5$ and communication occurs at time step $n-1$, then the numerical flux computed at time level $n-1$ is reused at time steps $n$ through $n+4$. Consequently, the numerical flux used at any time level $n' \in \{n, \dots, n+4\}$ is given by $\boldsymbol{\widehat{F}}^{\,n'-k-1}$, with $k$ taking the respective values $0,1,\dots,4$.
Within each RK stage, interior elements are updated first and independently of communication. For PE-boundary elements, volume contributions are always evaluated using locally available data. For face nodes that do not lie on PE boundaries, or when communication is enabled, the numerical flux is computed using the latest solution values. When communication is disabled and a face node lies on a PE boundary, the numerical flux is not recomputed; instead, a previously stored flux value is reused. These flux values are stored in a local buffer during the first RK stage of a time step when communication is enabled and are subsequently reused during the following time steps, until the communication flag is reset to \texttt{true}.

In this variant of the CAA, referred to as \emph{CAA with the standard flux}, delayed flux values are used directly in the numerical flux computation at PE boundaries. As a result, only a single set of boundary flux values from the most recent communication step needs to be stored, leading to minimal additional memory overhead proportional to the number of PE-boundary degrees of freedom. While this approach enables communication avoidance and can significantly improve scalability, it is known to reduce the formal accuracy of the scheme due to the use of delayed boundary information \cite{goswami2024-cmame-adg}. In the next subsection, we describe how this limitation is addressed through the use of asynchrony-tolerant numerical fluxes.

\begin{algorithm}[h!]
\centering
\small
\caption{{Communication-avoiding algorithm (CAA) with the standard flux. ($\Omega_I$: interior elements not requiring MPI ghost data; $\Omega_B$: PE-boundary elements)}}
\label{algo:caa-std}
\begin{algorithmic}[1]

\State \textbf{Initialization}
\State Set delay counter $k \leftarrow 0$

\For{\texttt{each time step $n$}}

    \Statex \textcolor{gray}{\texttt{/* Set communication flag and update delay parameter */}}
    \If{$(n \% L) == 0$}
        \State \texttt{communication} $\leftarrow$ \texttt{true}
        \State $k \leftarrow 0$
    \Else
        \State \texttt{communication} $\leftarrow$ \texttt{false}
        \State $k \leftarrow k + 1$
    \EndIf

    \For{\texttt{each RK stage $m$}}

        \If{\texttt{communication} is \texttt{true}}
            \State \textbf{Initiate MPI communication} of ghost values
        \EndIf

        \Statex \textcolor{gray}{\texttt{/* Compute contributions for interior elements */}}
        \For{\texttt{each element $\Omega_e \in \Omega_I$}} 
            \State Compute volume and face contributions using latest local values
            \State Update local RK-stage solution for $\Omega_e$
        \EndFor

        \If{\texttt{communication} is \texttt{true}}
            \State \textbf{Finish MPI communication} (synchronize ghost values)
        \EndIf

        \Statex \textcolor{gray}{\texttt{/* Compute contributions for PE-boundary elements */}}
        \For{\texttt{each element $\Omega_e \in \Omega_B$}}  
            \State Compute volume contribution using latest local values
            \For{\texttt{each face node $f \in \partial \Omega_e$}}
                \If{\texttt{$f$ is not a PE-boundary node} \textbf{or} \texttt{communication} is \texttt{true}}
                    \State Compute standard numerical flux using latest available values
                    \If{$m = 0$ \textbf{and} $f$ lies on a PE boundary}
                        \State Store numerical flux in \texttt{flux\_buffer}
                    \EndIf
                \Else
                    \State Use delayed numerical flux from \texttt{flux\_buffer} at time level $n-k$
                \EndIf
            \EndFor
            \State Update local RK-stage solution for $\Omega_e$
        \EndFor

        \State \textbf{Advance solution to the next RK stage}

    \EndFor
\EndFor

\end{algorithmic}
\end{algorithm}

\subsection{Asynchrony-tolerant (AT) fluxes}
\label{sec:atflux}
In general, a discontinuous Galerkin method using basis polynomials of degree $N_p$ coupled with an optimal choice of numerical flux function exhibits a convergence rate of $\mathcal{O}\left(h^{N_p+1}\right)$ in the $L_2$ norm. However, although the asynchronous discontinuous Galerkin (ADG) method maintains stability and consistency with the same numerical flux, it displays poor accuracy. A thorough error analysis based on a statistical framework has been conducted in the previous work \cite{goswami2024-cmame-adg} to investigate this discrepancy in accuracy. The analysis reveals that the use of delayed values in computing numerical fluxes at PE boundaries restricts the accuracy of the ADG scheme to first-order, regardless of the degree of the basis polynomials employed.

Subsequently, a new flux termed the asynchrony-tolerant (AT) flux has been introduced to recover the accuracy of the ADG schemes. For a time level $n$, the AT flux, denoted as $\widehat{\boldsymbol{F}}_e^{\text{at},n}$, utilizes previously computed numerical fluxes $\widehat{\boldsymbol{F}}_e^{n-\tilde{k}}$, $\widehat{\boldsymbol{F}}_e^{n-\tilde{k}-1}, \dots , \widehat{\boldsymbol{F}}_e^{n-\tilde{k}-N_p}$, achieving an expected accuracy of  $\mathcal{O}\left(h^{N_p+1}\right)$, which can be expressed as
\begin{equation}
   \widehat{\boldsymbol{F}}_{e}^{\,\text{at},n} = \sum_{l=L_1}^{L_2}\tilde{c}^l  \widehat{\boldsymbol{F}}_{e}^{\,n-l} = \widehat{\boldsymbol{F}}_e^{\,n} + \mathcal{O}\left(h^{N_p+1}\right). 
   \label{eq:at-flux}
\end{equation}
Here, the two limits $L_1$ and $L_2$ represent the extent of consecutive time levels required for approximating the numerical flux. The lower limit is $L_1 = \tilde{k}$ based on the latest value $\boldsymbol{w}_h^{e,n-\tilde{k}}$ in the buffer, whereas $L_2$ depends on the desired accuracy of the AT flux. The unknown coefficients $\tilde{c}^l$ can be determined with the help of the following constraints
\begin{align}
    \sum_{l = L_1}^{L_2} \tilde{c}^l \frac{(-l \Delta t)^\zeta }{\zeta !} = \begin{cases}
        1, & \zeta = 0 \\
        0, & 0 < \zeta < \dfrac{N_p+1}{r}
    \end{cases}.
    \label{eq:at-flux-constraints}
\end{align}
Parameter $r$, derived from the stability analysis of the numerical scheme, satisfies the CFL relation $\Delta t \sim \Delta x^r$. For $r = 1$, the upper limit is $L_2 = \tilde{k} + N_p$, and the respective coefficients for the AT flux for the second and third-order accuracy, are reported in Table~\ref{tab:at-flux-coeffs}.

The construction of the AT flux in Eq.~\eqref{eq:at-flux} is inherently multi-step in nature, as it relies on numerical fluxes from multiple previous time levels. When combined with multi-stage explicit Runge-Kutta (RK) schemes, special care is required to consistently account for the intermediate stage times. The coupling of multi-level asynchrony-tolerant schemes with low-storage multi-stage RK methods has been systematically analyzed in \cite{goswami2023lserkat-jcp}, where two different approaches were proposed.
The first is a \emph{naive} approach, in which delayed fluxes are directly incorporated at each RK stage by accounting for the fractional advancement within the time step. While straightforward to implement and robust in practice, this approach limits the overall temporal accuracy to at most third order, regardless of the formal order of the underlying AT scheme. The second approach introduces a more elaborate high-order \emph{accurate} approach that preserves the full accuracy of both the multi-level AT scheme and the RK integrator, at the expense of increased algorithmic complexity.

In the present work, we restrict our implementation to the \emph{naive} approach and focus on demonstrating its effectiveness for second- and third-order accurate ADG schemes. Specifically, for an $s$-stage explicit RK method, the delay parameter used in the AT flux coefficients is updated at every stage according to
$$
\tilde{k}_m = \tilde{k} + \delta_m, \qquad m = 0,1,\dots,s-1,
$$
where $k$ denotes the integer-valued delay associated with the time step, and $\delta_m$ represents the fractional time-step offset corresponding to the $m$-th RK stage. The stage-specific delays $k_m$ are then used to evaluate the AT flux coefficients $\tilde{c}^l$ in Eq.~\eqref{eq:at-flux}.

This formulation allows fractional time-step advancements to be consistently incorporated into the AT flux evaluation without modifying the structure of the RK scheme. Although it restricts the attainable order of accuracy, it is sufficient for the second- and third-order schemes considered in this study and enables a clean integration of AT fluxes into existing low-storage RK-based DG solvers. A detailed discussion of both approaches and their theoretical properties can be found in \cite{goswami2023lserkat-jcp}.

\begin{table}[h!]
\caption{Coefficients of second- and third-order asynchrony-tolerant (AT) fluxes.}
\label{tab:at-flux-coeffs}
\begin{center}
\begin{tabular}{ | c | l | }
\hline
Order & Coefficients $\tilde{c}^l$ with $\tilde{k} = k$ \\
\hline
{} & {} \\
2 & $\tilde{c}^{k} = (k + 1)$, $\tilde{c}^{k+1} = -k $ \\
3 & $\tilde{c}^{k} = \dfrac{(k^2 + 3k + 2)}{2}$, $\tilde{c}^{k+1} = -(k^2 + 2k)$, $ \tilde{c}^{k+2} = \dfrac{(k^2 + k)}{2}$ \\
{} & {} \\
\hline
\end{tabular}
\end{center}
\end{table}

\subsection{communication-avoiding algorithm with AT fluxes}
\label{sec:caa-at}

The communication-avoiding algorithm (CAA) combined with the asynchrony-tolerant (AT) flux realizes an accurate asynchronous discontinuous Galerkin (ADG) scheme and is summarized in Algorithm~\ref{algo:caa-at}. In contrast to the CAA with the standard flux, this variant employs AT fluxes at PE-boundary elements during communication-free phases in order to recover the formal order of accuracy of the DG discretization in the presence of delayed boundary data.
As introduced in Sec.~\ref{sec:atflux}, the AT flux at a given time level requires access to numerical fluxes from $N_p+1$ consecutive past time levels at process boundaries. Consequently, the communication pattern in CAA-AT differs from that of the standard-flux variant. Specifically, communication is enabled for $N_p+1$ consecutive time steps to populate the boundary flux history, followed by $L$ time steps during which communication is avoided. This pattern is implemented at the beginning of each time step using the condition
$
n \bmod (L + N_p + 1) < (N_p + 1),
$
which sets a boolean \textit{communication} flag. When this condition is satisfied, ghost-value exchange is performed and standard numerical fluxes at PE-boundary faces are computed and stored. When the condition is not satisfied, communication is skipped and AT fluxes are constructed from the stored flux history.

As in the synchronous algorithm and the CAA with the standard flux, computations are organized in nested loops over Runge-Kutta (RK) stages and elements. Interior elements always employ the standard DG update, since all required neighboring values are locally available. For PE-boundary elements, the treatment of numerical fluxes depends on the communication state. If communication is enabled, standard numerical fluxes are computed using the latest ghost values. During communication-free time steps, numerical fluxes at PE-boundary faces are evaluated using the AT flux formulation, which combines previously stored flux values.
The effective delay entering the AT flux at RK stage $m$ is given by
$
k_m = k + \partial_m,
$
where $k$ denotes the number of consecutive time steps elapsed since the most recent communication and $\partial_m$ is the fractional RK-stage offset associated with the time-integration scheme. The delay counter $k$ is reset to zero whenever communication is performed and incremented by one at each subsequent time step for which communication is avoided. Numerical fluxes computed during communication phases are stored at the first RK stage and shifted in a rolling buffer to maintain the required history for AT flux evaluation.

In a $d$-dimensional domain, the number of nodes on a PE boundary scales as $(N_p+1)^{d-1}$. Accordingly, the additional memory required by the CAA-AT approach scales with the number of PE-boundary nodes, the number of stored time levels ($N_p+1$), and the number of solution variables. This overhead remains modest relative to the total solution storage and is independent of the number of interior degrees of freedom.
By combining periodic communication with asynchrony-tolerant fluxes, the CAA-AT approach preserves local conservation and recovers the high-order accuracy of the underlying DG discretization while significantly reducing the frequency of communication and synchronization. This makes the method particularly well-suited for large-scale simulations in communication-dominated regimes, as demonstrated by the numerical results presented in Sec.~\ref{sec:numexp}.

\begin{algorithm}[h!]
\centering
\small
\caption{{Communication-avoiding algorithm (CAA) with asynchrony-tolerant (AT) fluxes. ($\Omega_I$: interior elements not requiring MPI ghost data; $\Omega_B$: PE-boundary elements)}}
\label{algo:caa-at}
\begin{algorithmic}[1]

\State \textbf{Initialization}
\State Set delay counter $k \leftarrow 0$

\For{\texttt{each time step $n$}}

    \Statex \textcolor{gray}{\texttt{/* Set communication flag and update delay parameter */}}
    \If{$n \% (L + N_p + 1) < (N_p + 1)$}
        \State \texttt{communication} $\leftarrow$ \texttt{true}
        \State $k \leftarrow 0$
    \Else
        \State \texttt{communication} $\leftarrow$ \texttt{false}
        \State $k \leftarrow k + 1$
    \EndIf

    \For{\texttt{each RK stage $m$}}

        \If{\texttt{communication} is \texttt{true}}
            \State \textbf{Initiate MPI communication} of ghost values
        \EndIf

        \Statex \textcolor{gray}{\texttt{/* Compute contributions for interior elements */}}
        \For{\texttt{each element $\Omega_e \in \Omega_I$}}
            \State Compute volume and face contributions using latest local values
            \State Update local RK-stage solution for $\Omega_e$
        \EndFor

        \If{\texttt{communication} is \texttt{true}}
            \State \textbf{Finish MPI communication} (synchronize ghost values)
        \EndIf

        \Statex \textcolor{gray}{\texttt{/* Compute contributions for PE-boundary elements */}}
        \For{\texttt{each element $\Omega_e \in \Omega_B$}}

            \State Compute volume contribution using latest local values

            \For{\texttt{each face node $f \in \partial \Omega_e$}}

                \If{\texttt{$f$ is not a PE-boundary node} \textbf{or} \texttt{communication} is \texttt{true}}
                    \State Compute standard numerical flux using latest available values
                    \If{$m = 0$ \textbf{and} $f$ lies on a PE boundary}
                        \State Shift previously stored fluxes: \texttt{flux\_buffer}($l+1$) $\leftarrow$ \texttt{flux\_buffer}($l$),
                        for $l = 0,\dots,N_p-1$
                        \State Store numerical flux in \texttt{flux\_buffer}(0)
                    \EndIf
                \Else
                    \State Set stage-dependent delay $k_m \leftarrow k + \partial_m$
                    \State Compute AT flux using coefficients corresponding to $k_m$:
                    $
                    \boldsymbol{\widehat{F}}^{\,n+\partial_m}
                    = \sum_{l=0}^{N_p} \tilde{c}^{\,k_m + l}\,
                      \texttt{flux\_buffer}(l)
                    $
                \EndIf

            \EndFor

            \State Update local RK-stage solution for $\Omega_e$
        \EndFor

        \State \textbf{Advance solution to next RK stage}

    \EndFor
\EndFor

\end{algorithmic}
\end{algorithm}


\section{DG solver based on the \texttt{deal.II} library}
\label{sec:dealii-solver}

Our implementation uses \texttt{deal.II}, an open-source finite-element library written in C++ that provides extensive support for high-order discretizations, matrix-free operator evaluation, and scalable parallel execution on distributed-memory systems \cite{dealii96}. In addition to a wide range of built-in solvers, \texttt{deal.II} offers a flexible infrastructure for developing custom application codes, making it well-suited for research on advanced numerical methods and parallel algorithms.
To concretely realize the parallel DG execution model described in Sec.~\ref{sec:parallel}, we build upon the \textit{step-76} tutorial solver provided with \texttt{deal.II}, which is designed for the solution of the compressible Euler equations using a discontinuous Galerkin (DG) discretization in space and low-storage explicit Runge-Kutta (LSERK) schemes for time integration. The DG formulation for the compressible Euler equations has been reviewed in Sec.~\ref{sec:background}, and the LSERK coefficients used in the solver are listed in Table~\ref{table:lserk-scheme}. The solver includes test cases for both two- and three-dimensional flows and supports multiple numerical fluxes, including the local Lax-Friedrichs and Harten-Lax-van Leer-Contact (HLLC) fluxes.

The evaluation of the DG operator in \textit{step-76} is performed in a fully matrix-free fashion. Both volume and face integrals, as well as the application of the inverse mass matrix, are computed without assembling global matrices, relying instead on sum-factorization techniques to reduce computational complexity and memory traffic. This matrix-free, element-centric execution model enables efficient use of modern CPU architectures and exposes a high degree of parallelism across elements.

Parallelization is realized through domain decomposition across processing elements (PEs) using MPI. To facilitate overlap of communication and computation, \texttt{deal.II} internally partitions the locally owned elements into three groups, commonly referred to as \textit{part-0}, \textit{part-1}, and \textit{part-2}. Elements in \textit{part-1} correspond to PE-boundary elements whose numerical fluxes depend on degrees of freedom from neighboring PEs and therefore require MPI communication. Whereas, elements in \textit{part-0} and \textit{part-2} are interior elements that can be processed independently of inter-process data.
At each Runge-Kutta stage, the solver initiates the exchange of ghost values required for PE-boundary elements by calling \texttt{vector\_update\_ghosts\_start()}, which triggers non-blocking MPI communication using the MPI calls \texttt{MPI\_Isend/Irecv}. While this communication is in progress, the solver proceeds with the evaluation of DG contributions for elements in \textit{part-0}, which consist of interior elements whose computations are completely independent of ghost data. 
Once the computations for \textit{part-0} are complete, communication is finalized using the call \texttt{vector\_update\_ghosts\_finish()}, which performs \texttt{MPI\_Wait/all} operation, ensuring that all required ghost values are available. The solver then evaluates the DG operator for elements in \textit{part-1}, corresponding to PE-boundary elements whose numerical fluxes depend on data from neighboring processes.
After completing the element-local DG operator evaluation, the solver initiates the accumulation of contributions to shared degrees of freedom by calling \texttt{vector\_compress\_start()}. This operation is required primarily for continuous finite-element discretizations, where degrees of freedom are shared across element boundaries. While this communication is ongoing, the solver proceeds with computations for elements in \textit{part-2}, which again do not depend on inter-process data. Finally, the communication is completed via \texttt{vector\_compress\_finish()}, finalizing the update of the distributed solution vectors.

This structured execution pattern enables partial overlap of communication and computation at multiple points within each Runge-Kutta stage. While synchronization is still required at well-defined points, the partitioning into \textit{part-0}, \textit{part-1}, and \textit{part-2} reduces idle time and mitigates—but does not eliminate—the scalability limitations imposed by communication and synchronization in the synchronous DG algorithm.
In this work, the synchronous DG solver provided by \textit{step-76} serves as the baseline against which we compare the proposed asynchronous DG solvers based on the communication-avoiding algorithms described in Secs.~\ref{sec:caa-std} and~\ref{sec:caa-at}. The same matrix-free discretization, numerical fluxes, and time integration schemes are used in both cases, ensuring that any observed performance differences can be attributed solely to changes in the communication and synchronization strategy.

\begin{table}[h!]
\caption{Coefficients of $s$-stage, $o$-order low-storage explicit Runge-Kutta schemes, $\text{LSERK}(s,o)$, used for time integration.}
\label{table:lserk-scheme}
\begin{center}
\setlength{\tabcolsep}{6pt}
\renewcommand{\arraystretch}{1.2}
\begin{tabular}{|c|ccc|}
\hline
\textbf{Scheme} & $\boldsymbol{a_m}$ & $\boldsymbol{b_m}$ & $\boldsymbol{c_m}$ \\
\hline
\multicolumn{4}{|c|}{LSERK(2,2)} \\
\hline
Stage 1 & 1.000000000 & 0.500000000 & 0.000000000 \\
Stage 2 & --          & 0.500000000 & 1.000000000 \\
\hline
\multicolumn{4}{|c|}{LSERK(3,3)} \\
\hline
Stage 1 & 0.755726352 & 0.245170287 & 0.000000000 \\
Stage 2 & 0.386954477 & 0.184896052 & 0.755726352 \\
Stage 3 & --          & 0.569933661 & 0.632125000 \\
\hline
\end{tabular}
\end{center}
\end{table}

\section{Numerical results}
\label{sec:numexp}

In this section, we assess the accuracy and scalability of the asynchronous discontinuous Galerkin (ADG) method with both standard and asynchrony-tolerant (AT) numerical fluxes, and compare its performance against the baseline synchronous DG method. The ADG solver in \texttt{deal.II} is implemented using the communication-avoiding algorithms described in Sections~\ref{sec:caa-std} and~\ref{sec:caa-at}.

\subsection{Experimental setup}

All numerical experiments are conducted using the DG solver \textit{step-76} for the compressible Euler equations provided by the \texttt{deal.II} library. The solver includes two benchmark configurations: the two-dimensional \textit{isentropic vortex} test case and the three-dimensional \textit{flow around a cylinder} test case, which are described below.

\begin{figure}[h!]
    \centering
    \includegraphics[width=0.4\linewidth]{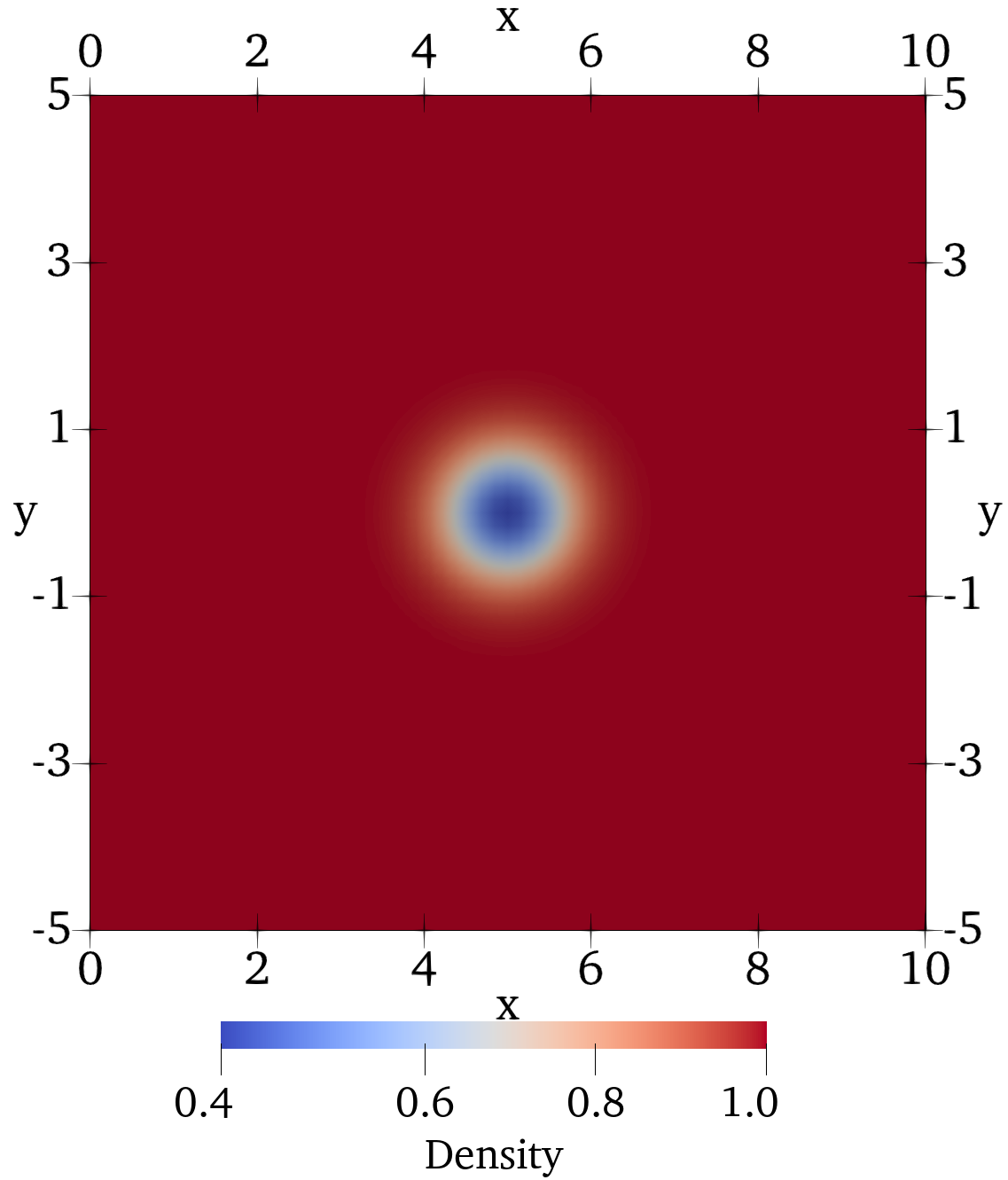}
    \caption{\small{Initial density profile for the two-dimensional domain for the \textit{isentropic vortex} test case.}}
    \label{fig:testcase0-initial-solution}
\end{figure}

\paragraph{\textbf{Two-dimensional \textit{isentropic vortex} test case}}
The first benchmark is a two-dimensional problem defined on a rectangular domain $[0,10] \times [-5,5]$ with Dirichlet boundary conditions. It consists of an isentropic vortex transported through a uniform background flow field (see Fig.~\ref{fig:testcase0-initial-solution}), with the exact solution given by
\begin{align}
    u &= 1 - \beta e^{(1-r^2)} \frac{y-y_0}{2\pi}, \nonumber \\
    v &= \beta e^{(1-r^2)} \frac{x-x_0}{2\pi}, \nonumber \\
    \rho &= \left( 1 - \left( \frac{\gamma - 1}{16\gamma \pi^2} \right) \beta^2 e^{2(1-r^2)} \right)^{\frac{1}{\gamma-1}}, \nonumber \\
    p &= \rho^{\gamma},
\end{align}
where $r = \sqrt{(x - t - x_0)^2 + (y - y_0)^2}$, $x_0 = 5$, $y_0 = 0$, $\beta = 5$, and $\gamma = 1.4$ \cite{hesthaven2007dg}. This analytical solution is used to prescribe both the initial and boundary conditions.

\paragraph{\textbf{Three-dimensional \textit{flow around a cylinder} test case}}
The second benchmark is a three-dimensional channel flow with an embedded circular cylinder. The computational domain is defined as $[0,2.2] \times [0,0.41] \times [0,0.41]$, with a cylinder of diameter $0.1$ placed parallel to the $z$-axis and centered at $(0.2, 0.2, 0)$. A constant inflow condition is prescribed at the inlet using Dirichlet boundary conditions, while a subsonic outflow condition is imposed at the outlet by prescribing the energy and extrapolating mass and momentum from the interior. The top and bottom channel walls, as well as the cylinder surface, satisfy a no-penetration boundary condition.

The initial condition corresponds to a uniform subsonic flow with density $\rho = 1$, velocity $\boldsymbol{u} = (0.4, 0, 0)$, and total energy
$
E = \dfrac{c^2}{\gamma(\gamma - 1)} + \dfrac{1}{2} \rho \| \boldsymbol{u} \|^2,
$
resulting in a Mach number of $Ma = 0.307$. This benchmark follows the classical setup described in \cite{schafer1996}.

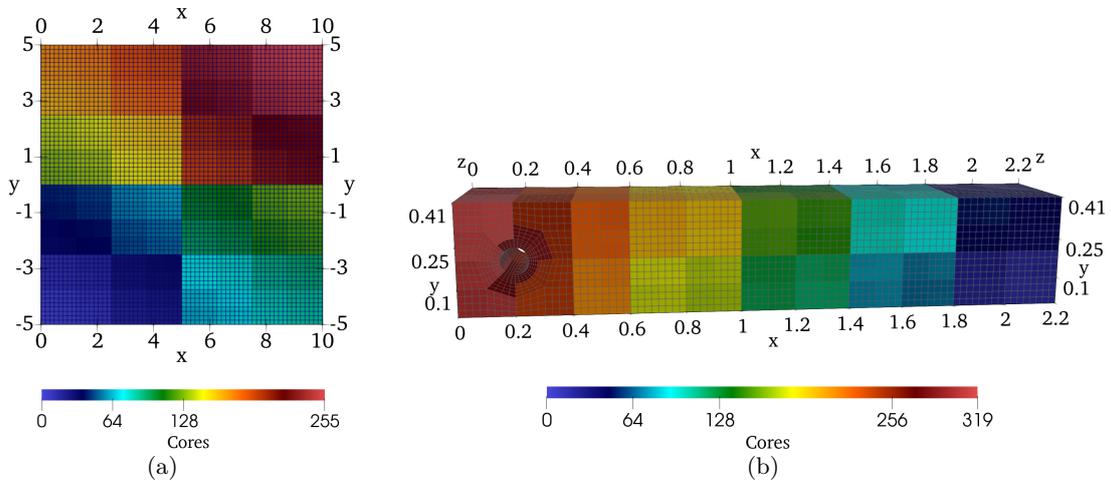
\begin{figure}[h!]
    \centering
    \includegraphics[width=0.28\linewidth]{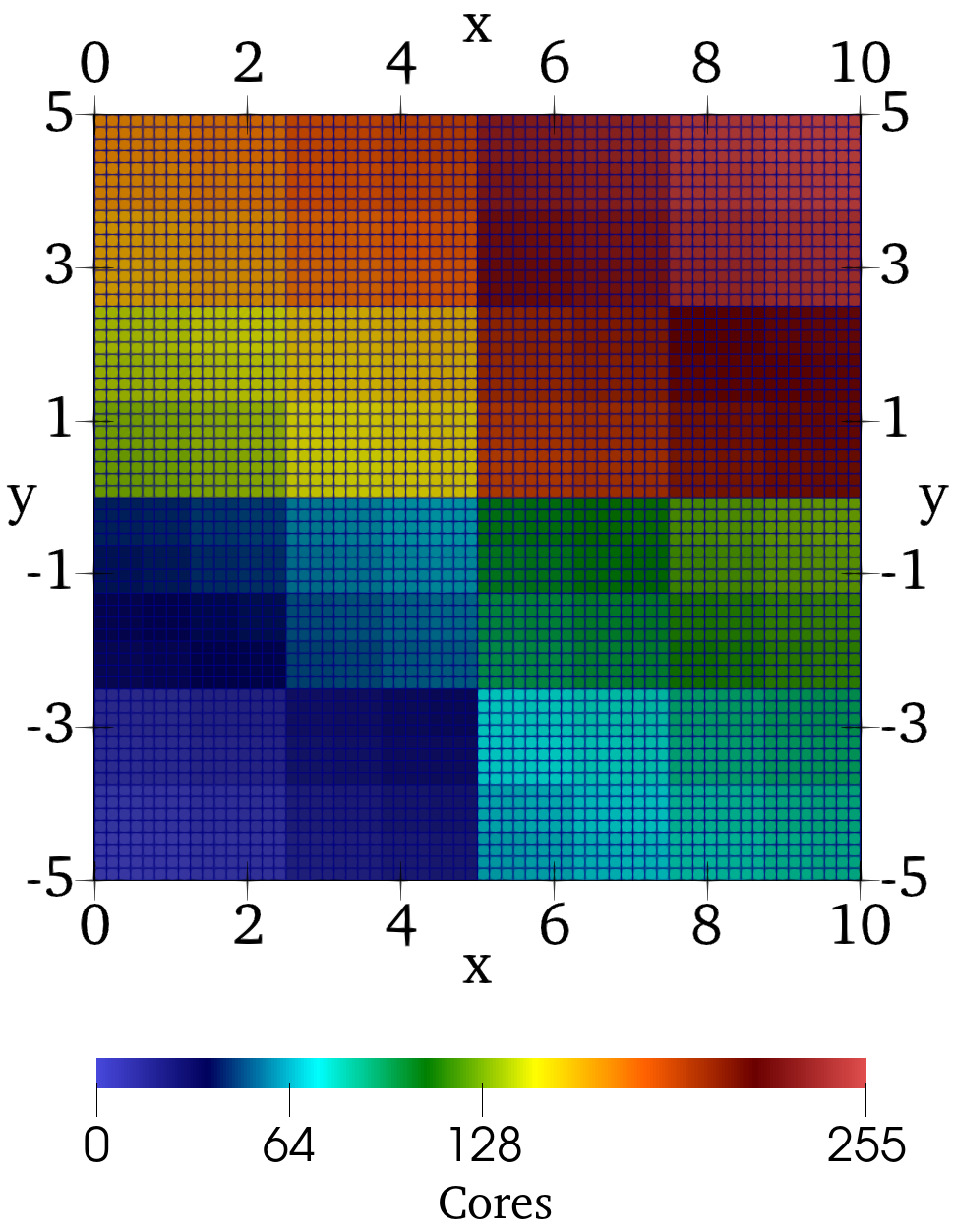}
    \begin{picture}(0,0)
        \put(-82,-8){\small{(a)}}
    \end{picture}
    \hspace{0.3cm}
    \includegraphics[width=0.56\linewidth]{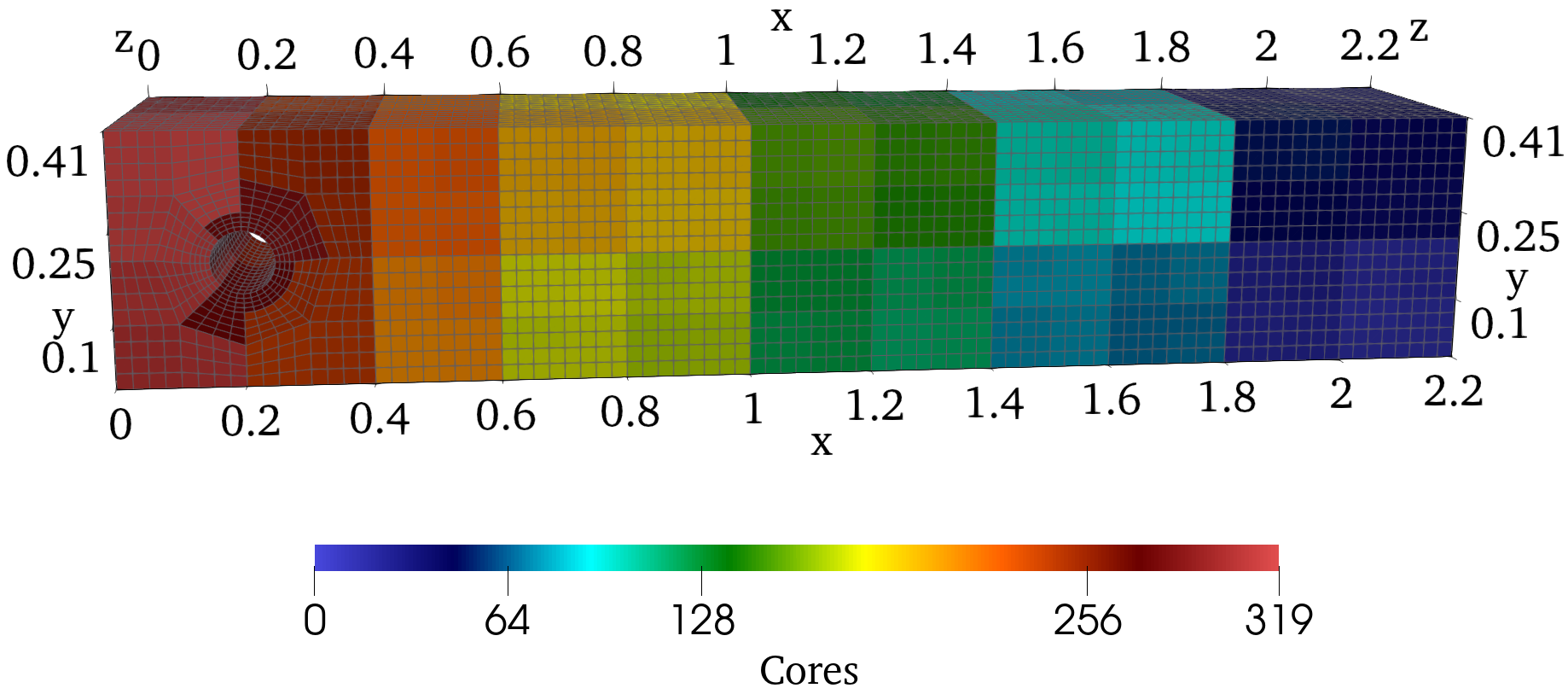}
    \begin{picture}(0,0)
        \put(-138,-8){\small{(b)}}
    \end{picture}
\caption{\small{Domain decompositions for (a) the two-dimensional \textit{isentropic vortex} test case with $4096$ elements distributed across $256$ PEs and (b) the three-dimensional \textit{flow around a cylinder} test case with $25,600$ elements distributed across $320$ PEs.}}
\label{fig:domain-decomposition-param}
\end{figure}

\paragraph{\textbf{Computational platform}}
All simulations are performed on the PARAM Pravega supercomputer at the Indian Institute of Science (IISc), Bengaluru, India. Each compute node is equipped with dual-socket Intel Xeon Cascade Lake 8268 processors (48 cores per node), 4~GB of RAM per core, and is connected via a fat-tree interconnect. In total, the system consists of 428 compute nodes.
The experimental configuration varies key numerical parameters, including the polynomial degree $N_p$, the number of elements per spatial direction, and the coefficients of the low-storage explicit Runge-Kutta (LSERK) scheme. Mesh resolution is controlled by a global refinement parameter $G$, where each increment in $G$ doubles the number of elements along each coordinate direction.
Figure~\ref{fig:domain-decomposition-param} illustrates representative domain decompositions for both test cases. For the two-dimensional \textit{isentropic vortex} test case, part~(a) shows a configuration with $G = 4$, resulting in 4096 elements distributed evenly across 256 MPI processes. For the three-dimensional \textit{flow around a cylinder} test case, part~(b) depicts a setup with $G = 2$, comprising 25,600 elements distributed over 320 MPI processes.

\begin{table}[h!]
\centering
\caption{Problem configurations for the two-dimensional \textit{isentropic vortex} and three-dimensional \textit{flow around a cylinder} test cases with polynomial degree $N_p = 2$.}
\setlength{\tabcolsep}{6pt}
\renewcommand{\arraystretch}{1.2}

\begin{tabular}{|c|c|c|c|c|c|}
\hline
\multicolumn{6}{|c|}{\textbf{2D: \textit{isentropic vortex} test case}} \\
\hline
$G$      & 3      & 4      & 5       & 6       & 7 \\
\hline
$N_E$    & 1,024  & 4,096  & 16,384  & 65,536  & 262,144 \\
DoFs     & 36,864 & 147,456 & 589,824 & 2.4M    & 9.4M \\
\hline
\multicolumn{6}{|c|}{\textbf{3D: \textit{flow around a cylinder} test case}} \\
\hline
$G$      & 1      & 2       & 3        & 4        & -- \\
\hline
$N_E$    & 3,200  & 25,600  & 204,800  & 1.6M & -- \\
DoFs     & 432,000 & 3.5M   & 27.6M    & 221.2M   & -- \\
\hline
\end{tabular}

\label{tab:problem-configurations}
\end{table}

\subsection{Profiling}
\label{sec:profiling}

This subsection presents a detailed profiling analysis of the \textit{step-76} solver in \texttt{deal.II}, with the objective of quantifying the relative costs of computation and communication and identifying the dominant scalability bottlenecks. In particular, we examine the time spent in ghost-value exchange, vector compression, and element-local DG operator evaluations, which together account for the majority of the runtime at large process counts.
The profiling results are obtained using user-defined timers embedded in the solver. Based on the global refinement parameter $G$ and the polynomial degree $N_p$, we consider representative configurations that are sufficiently large to expose communication overheads. For the two-dimensional \textit{isentropic vortex} test case, we analyze the configurations $G = 6, N_p = 2$ (65,536 elements, 2.4 million DoFs) and $G = 7, N_p = 2$ (262,144 elements, 9.4 million DoFs). For the three-dimensional \textit{flow around a cylinder} test case, we use the configurations $G = 3, N_p = 2$ (204,800 elements, 27.6 million DoFs) and $G = 4, N_p = 2$ (1.6 million elements, 221.2 million DoFs).
The two-dimensional cases are run for 5000 time steps, while the three-dimensional cases are run for 1000 time steps. All profiling results are based on multiple independent executions to account for run-to-run variability. Each configuration is executed five times, and for every run the solver reports the minimum, maximum, and average time spent in each measured operation. To obtain robust timing estimates, we report the mean of the average times computed over these five runs for each configuration and process count.

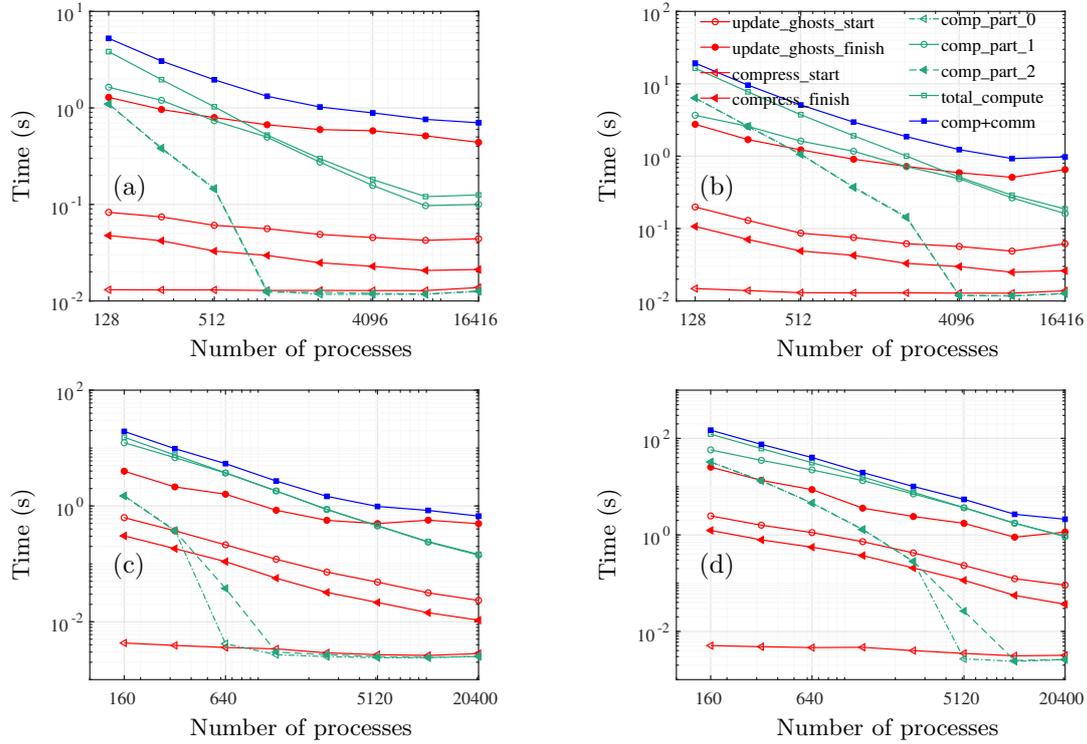
\begin{figure}[h!]
    \centering
    \includegraphics[width=6.6cm]{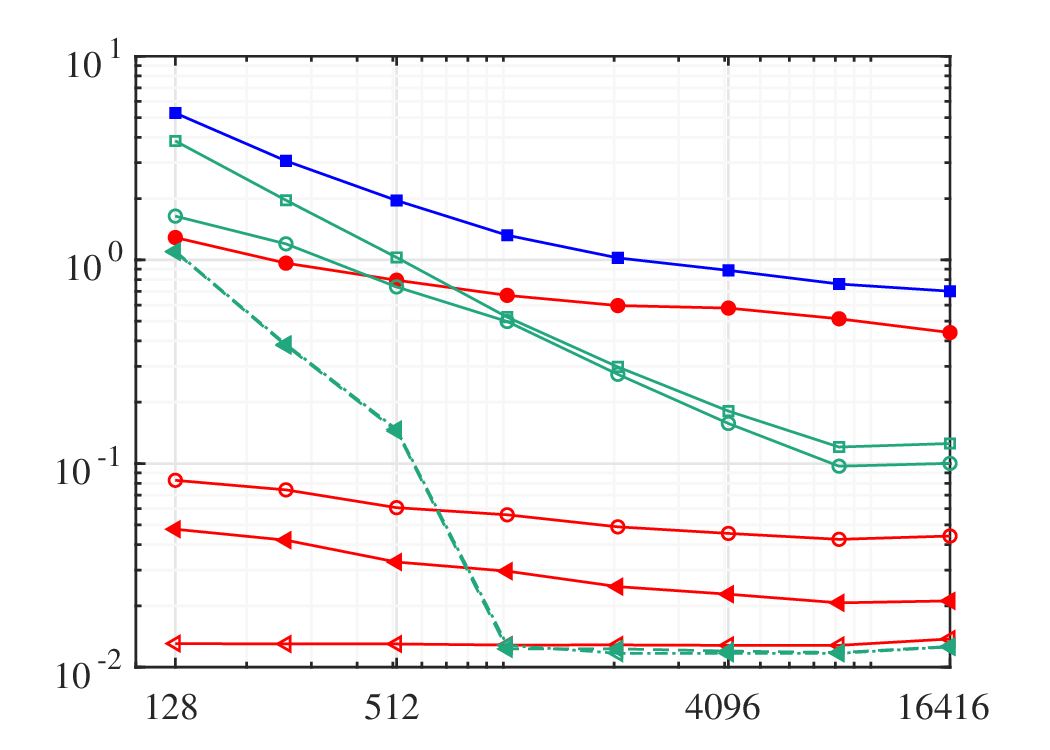}
    \put(-155, 55){(a)}
    \put(-193,52){\small{\rotatebox{90}{Time (s)}}}
    \put(-126,-6){\small{Number of processes}}
    \hspace{1cm}
    \includegraphics[width=6.6cm]{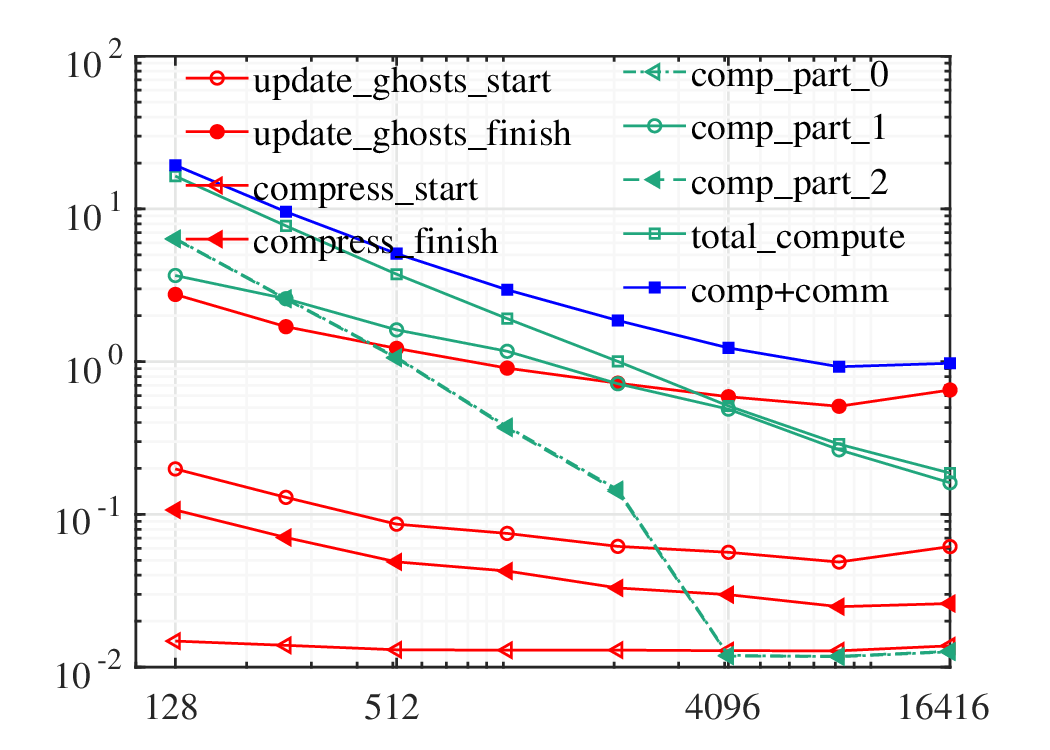}
    \put(-155, 55){(b)}
    \put(-193,52){\small{\rotatebox{90}{Time (s)}}}
    \put(-126,-6){\small{Number of processes}}\\
    \includegraphics[width=6.6cm]{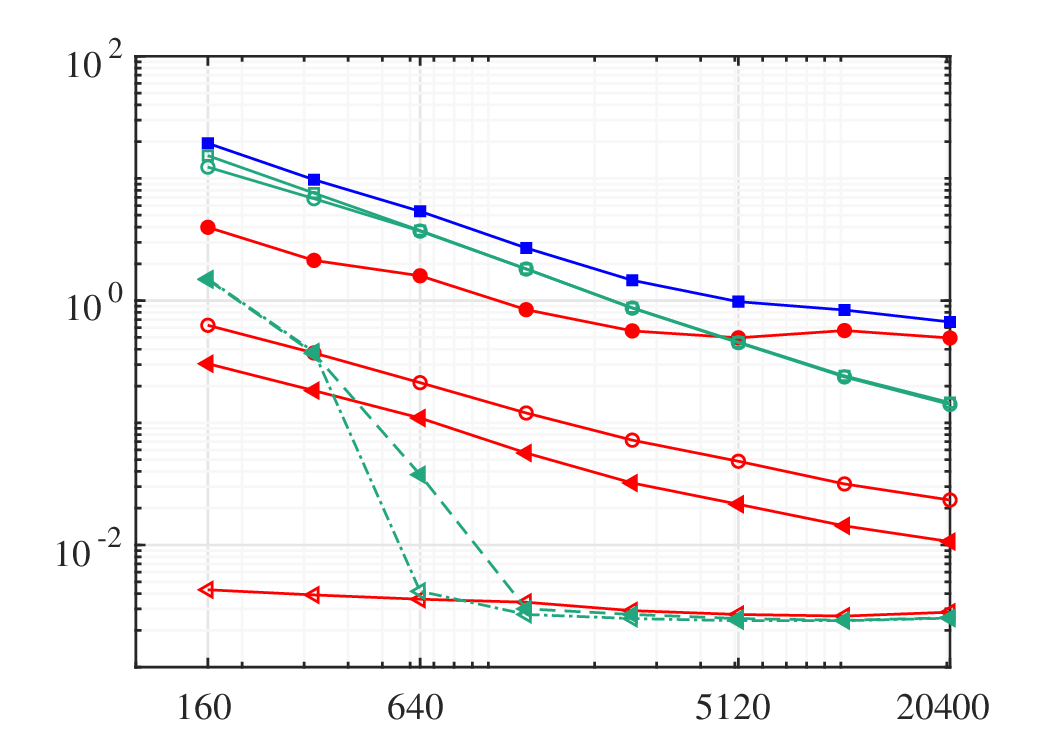}
    \put(-155, 55){(c)}
    \put(-193,52){\small{\rotatebox{90}{Time (s)}}}
    \put(-126,-6){\small{Number of processes}}
    \hspace{1cm}
    \includegraphics[width=6.6cm]{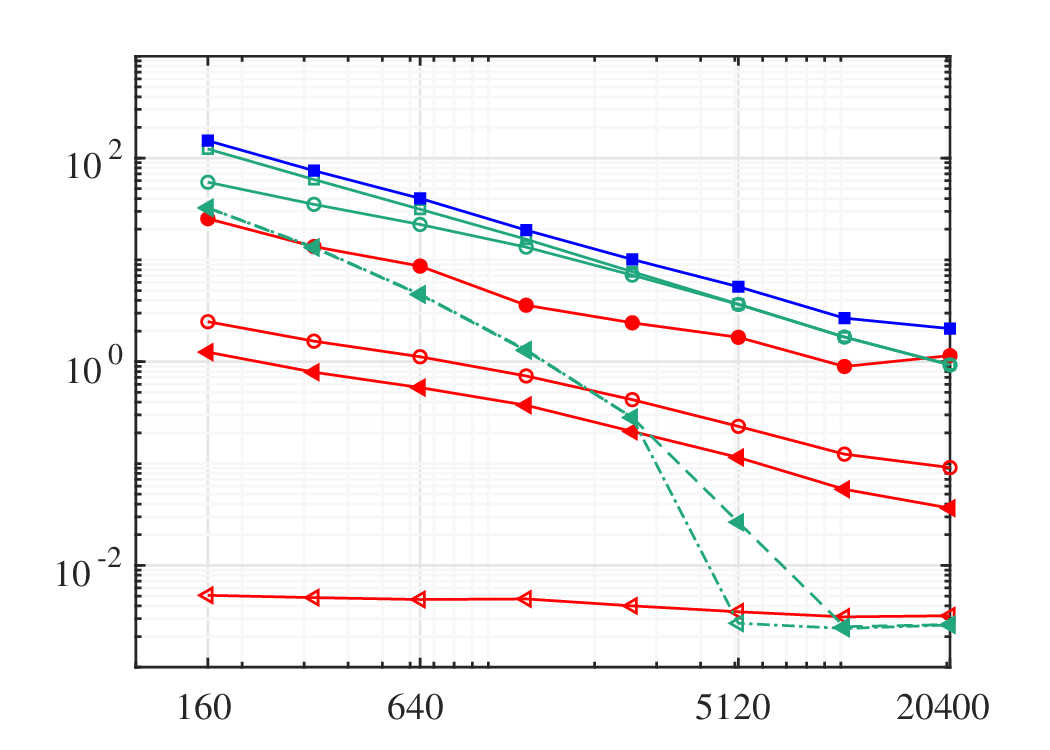}
    \put(-155, 55){(d)}
    \put(-193,52){\small{\rotatebox{90}{Time (s)}}}
    \put(-126,-6){\small{Number of processes}}
\caption{\small{Breakdown of communication and computation times for (a) 2.4M DoFs and (b) 9.4M DoFs for the two-dimensional \textit{isentropic vortex} test case, and (c) 27.6M DoFs and (d) 221.2M DoFs for the three-dimensional \textit{flow around a cylinder} test case.}}
\label{fig:profiling-communication-sa}
\end{figure}

Figure~\ref{fig:profiling-communication-sa} presents a breakdown of computation and communication times as a function of the number of MPI processes. We report the time spent in ghost-value exchange performed by \texttt{vector\_update\_ghosts\_start} and \texttt{vector\_update\_ghosts\_finish}, vector compression (\texttt{vector\_compress\_start} and \texttt{vector\_compress\_finish}), and element-local DG operator evaluations grouped according to the internal partitions \textit{part-0}, \textit{part-1}, and \textit{part-2}. In addition, we show the total computation time, defined as the sum of the three compute partitions, and the combined computation and communication time.
We first consider the two-dimensional results shown in Fig.~\ref{fig:profiling-communication-sa}(a) and (b), corresponding to the configurations $G = 6, N_p = 2$ and $G = 7, N_p = 2$, respectively. The number of MPI processes ranges from 128 to 16,416, corresponding to 4 to 342 compute nodes. For the largest configuration, 48 processes per node are used, whereas all smaller configurations employ 32 processes per node. For all process counts, the total computation time exhibits near-ideal strong scaling for both cases, reflecting the efficiency of the matrix-free DG operator evaluation. For the lower-resolution case ($G = 6, N_p = 2$), the total computation time saturates beyond 8192 processes, as the number of elements per process becomes very small and the number of PE-boundary elements is reduced to as few as four at the highest process count. In this regime, further reductions in computation time are no longer possible, and fixed overheads dominate.
More importantly, communication costs begin to overwhelm computation significantly earlier. For $G = 6, N_p = 2$, the time spent in ghost-value synchronization (\texttt{vector\_update\_ghosts\_finish}) becomes comparable to or larger than the total computation time beyond 512 MPI processes. A similar trend is observed for the higher-resolution case $G = 7, N_p = 2$, where communication overtakes computation beyond approximately 4096 processes. As a result, deviations from ideal strong scaling in the total runtime are observed for both configurations.

The three-dimensional results are shown in Fig.~\ref{fig:profiling-communication-sa}(c) and (d) for the configurations $G = 3, N_p = 2$ and $G = 4, N_p = 2$, respectively. Here, the number of MPI processes ranges from 160 to 20,400, corresponding to 4 to 425 compute nodes. The largest configuration uses 48 processes per node, while the remaining cases use 40 processes per node. Similar trends are observed as in the two-dimensional case: the total computation time initially scales well with increasing process count, but communication costs associated with ghost-value exchanges increasingly dominate at scale. For the $G = 3, N_p = 2$ configuration, communication becomes the primary contributor to runtime beyond approximately 5120 processes, while for the larger $G = 4, N_p = 2$ configuration this transition occurs at around 20,400 processes.

A further insight is obtained by examining the individual compute partitions. The computation times for \textit{part-0} and \textit{part-2}, which correspond to interior elements, decrease sharply once the number of interior elements per process becomes very small. In this regime, the workload shifts almost entirely to PE-boundary elements handled in \textit{part-1}, substantially reducing opportunities for computation–communication overlap. Consequently, the remaining element-local computations become dominated by fixed overheads rather than floating-point work. However, the time spent in vector compression remains consistently small across all configurations. Since vector compression primarily targets continuous finite-element discretizations and contributes negligibly to the overall runtime in the present DG setting, it is omitted from further performance analysis.
Overall, these profiling results clearly demonstrate that, beyond a problem-dependent process count, communication and synchronization associated with ghost-value exchanges overwhelm computation and become the dominant bottleneck, ultimately limiting the scalability of the synchronous DG solver.


\subsection{Accuracy}
\label{sec:accuracy-dealii}

This subsection presents a comparative accuracy study of three implementations: the synchronous algorithm (SA) based on the standard discontinuous Galerkin (DG) method, and two communication-avoiding algorithm (CAA) variants based on the asynchronous discontinuous Galerkin (ADG) method using standard and asynchrony-tolerant (AT) numerical fluxes. The comparison is carried out using the two-dimensional \textit{isentropic vortex} test case, for which an analytical solution is available, enabling a quantitative error assessment.
The computational framework employs DG basis functions of degree $N_p$ in space coupled with a low-storage explicit Runge-Kutta scheme of order $q = N_p + 1$ for time integration. We denote this combination as the DG$(N_p)$-LSERK$q$ scheme. The corresponding asynchronous formulations are referred to as ADG$(N_p)$-LSERK$q$ when the standard numerical flux is used, and ADG$(N_p)$-AT$q$-LSERK$q$ when the AT flux is employed.

\begin{figure}[h!]
    \centering
    \includegraphics[width=0.3\linewidth]{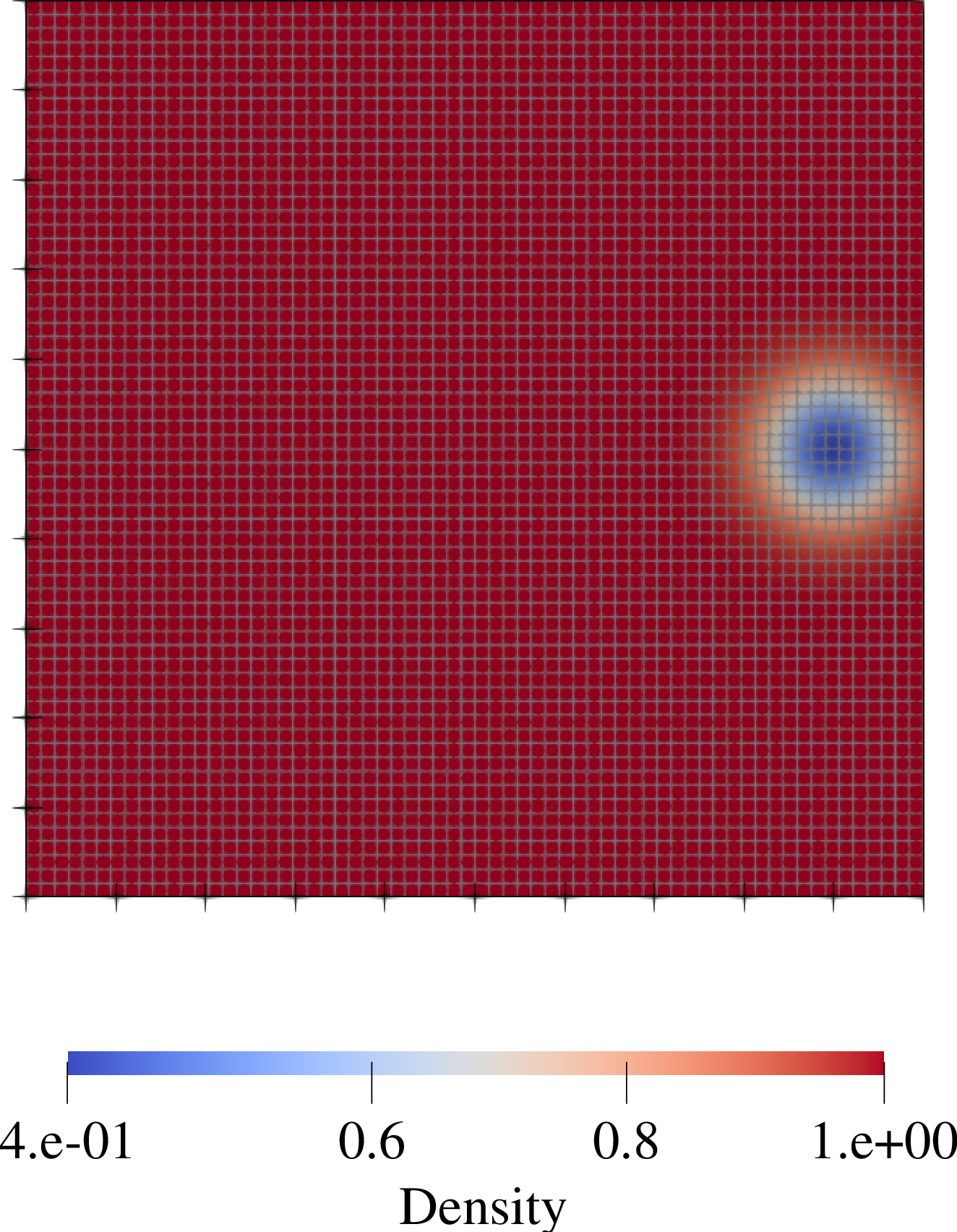}
    \begin{picture}(0,0)
        \put(-110,130){\footnotesize{(a)}}
        \put(-140,40){\footnotesize{0}}
        \put(-115,40){\footnotesize{2}}
        \put(-88,40){\footnotesize{4}}
        \put(-62,40){\footnotesize{6}}
        \put(-36,40){\footnotesize{8}}
        \put(-14,40){\footnotesize{10}}
        \put(-74,35){\footnotesize{$x$}}
        \put(-150,47){\footnotesize{-5}}
        \put(-150,73){\footnotesize{-3}}
        \put(-150,99){\footnotesize{-1}}
        \put(-148,125){\footnotesize{1}}
        \put(-148,151){\footnotesize{3}}
        \put(-148,177){\footnotesize{5}}
        \put(-155,112){\footnotesize{$y$}}
    \end{picture}
    \includegraphics[width=0.3\linewidth]{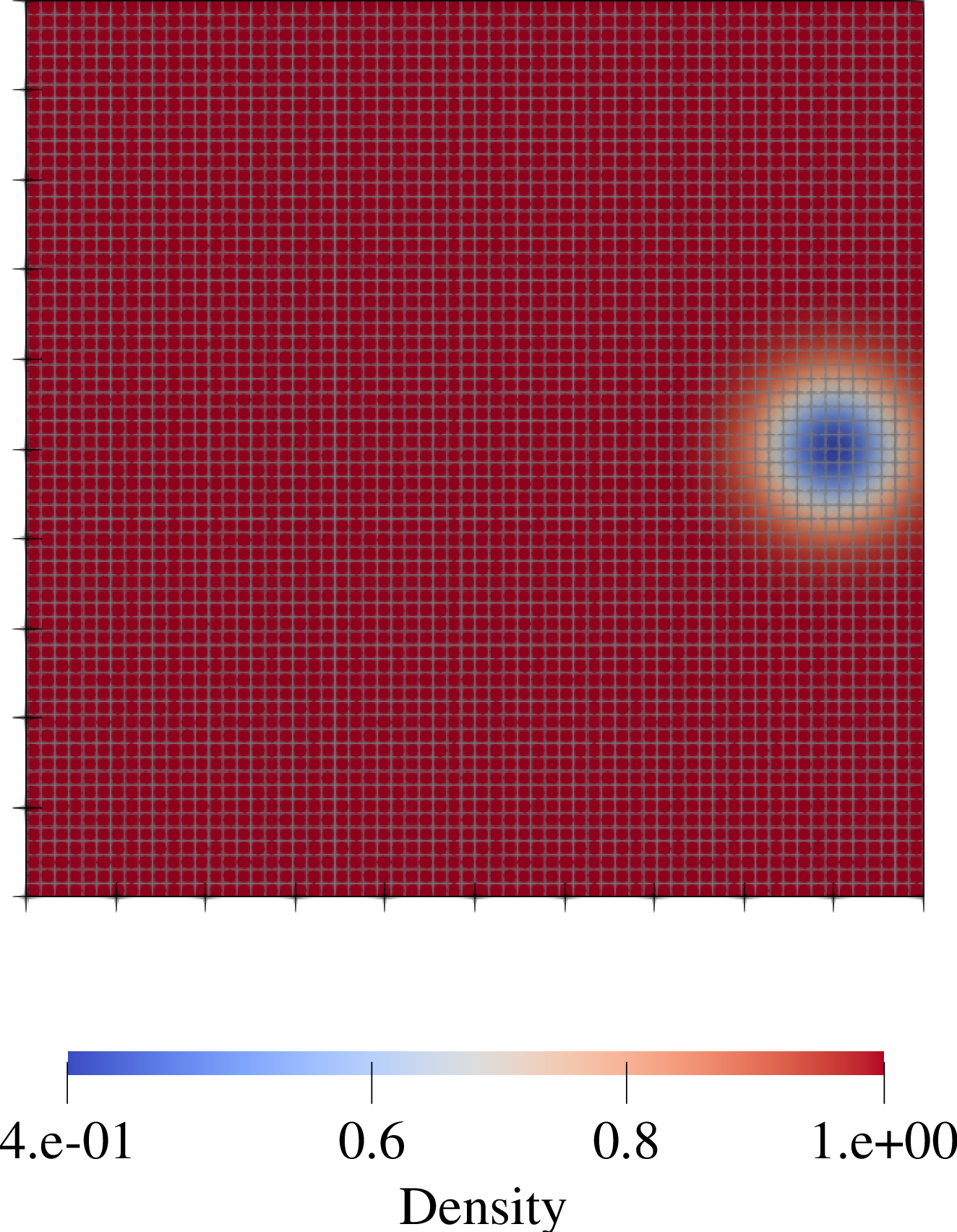}
    \begin{picture}(0,0)
        \put(-110,130){\footnotesize{(b)}}
        \put(-140,40){\footnotesize{0}}
        \put(-115,40){\footnotesize{2}}
        \put(-88,40){\footnotesize{4}}
        \put(-62,40){\footnotesize{6}}
        \put(-36,40){\footnotesize{8}}
        \put(-14,40){\footnotesize{10}}
        \put(-74,35){\footnotesize{$x$}}
    \end{picture}
    \includegraphics[width=0.3\linewidth]{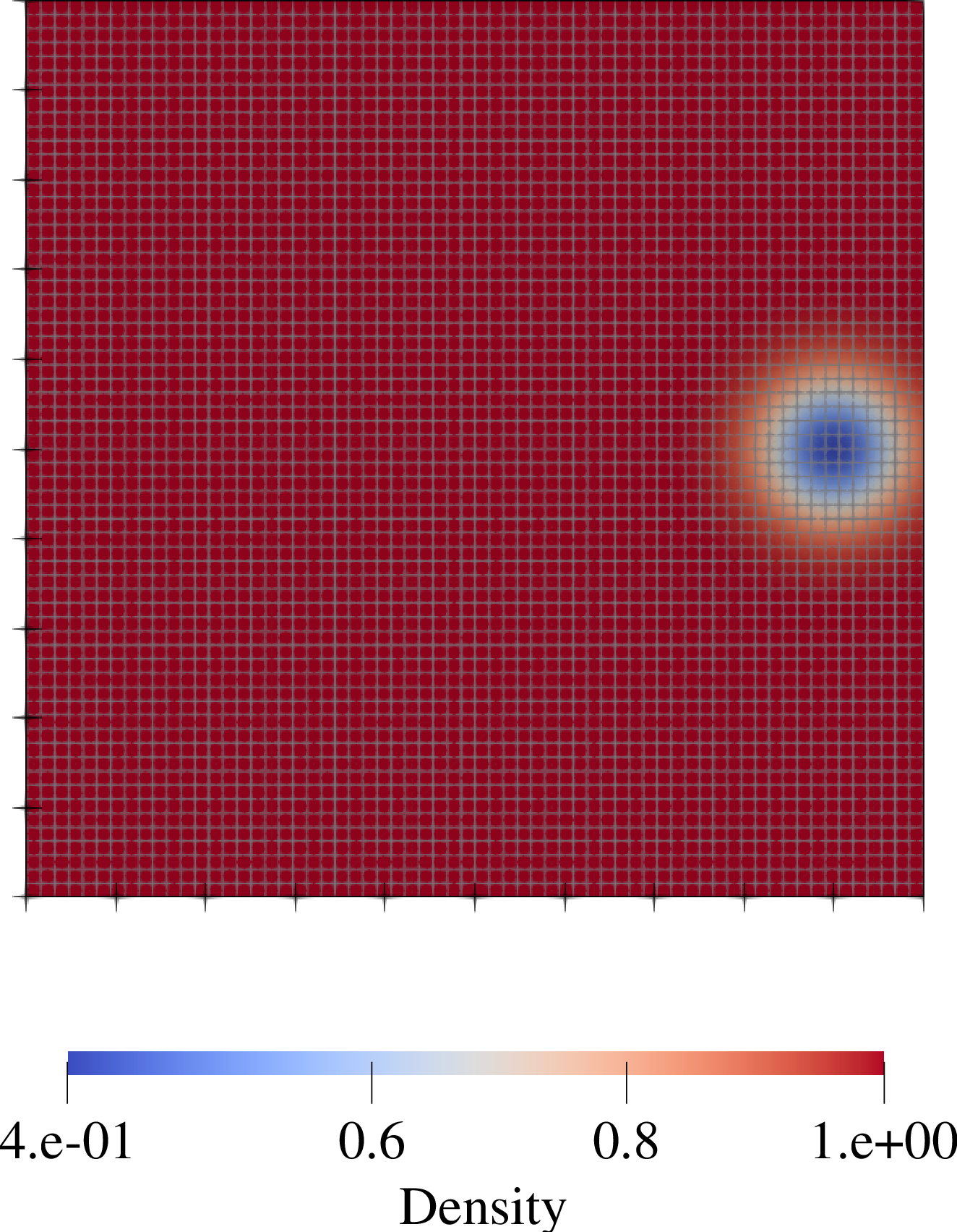}
    \begin{picture}(0,0)
        \put(-110,130){\footnotesize{(c)}}
        \put(-140,40){\footnotesize{0}}
        \put(-115,40){\footnotesize{2}}
        \put(-88,40){\footnotesize{4}}
        \put(-62,40){\footnotesize{6}}
        \put(-36,40){\footnotesize{8}}
        \put(-14,40){\footnotesize{10}}
        \put(-74,35){\footnotesize{$x$}}
    \end{picture}
    \includegraphics[width=0.3\linewidth]{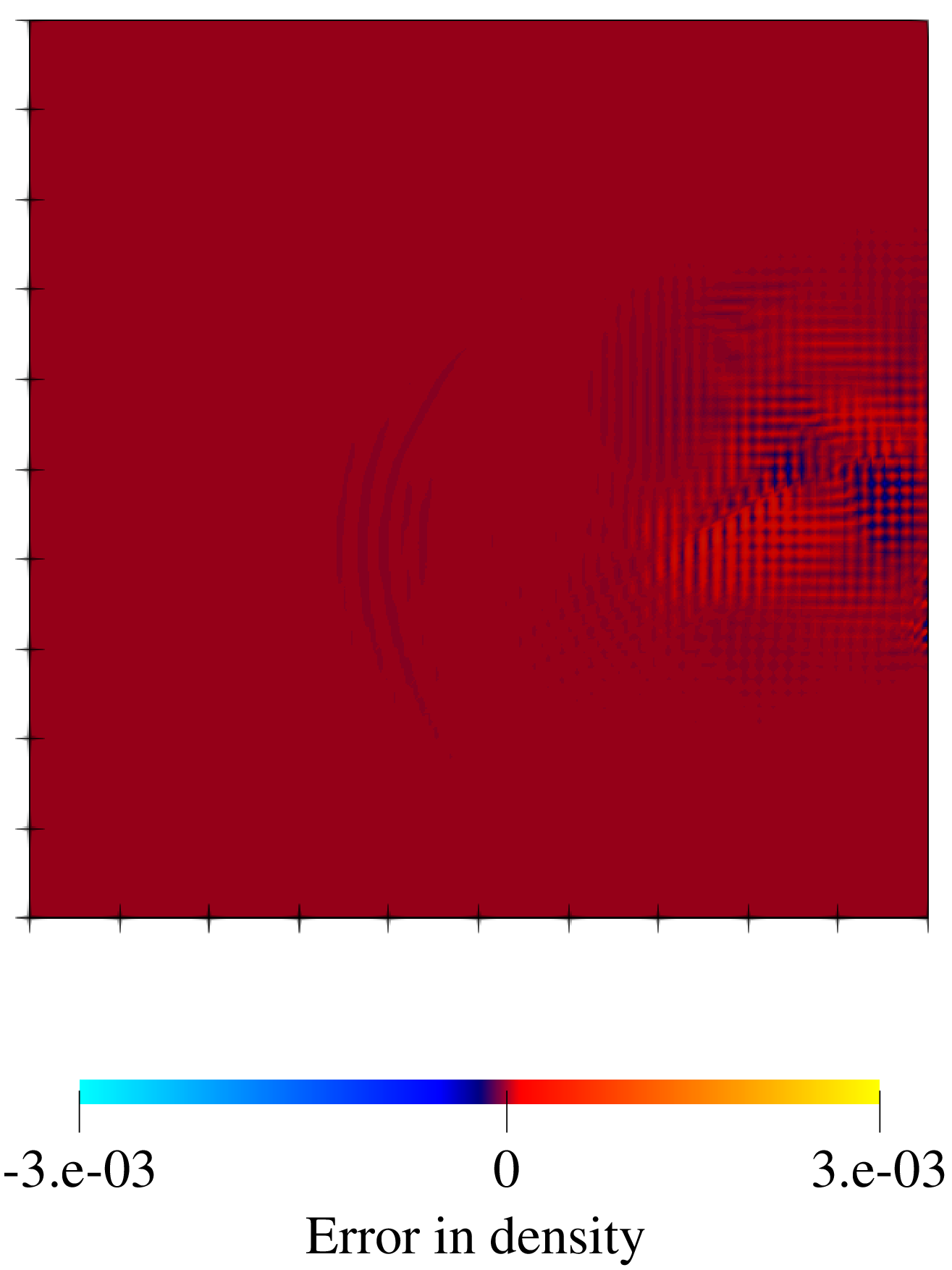}
    \begin{picture}(0,0)
        \put(-110,130){\footnotesize{(d)}}
        \put(-139,43){\footnotesize{0}}
        \put(-114,43){\footnotesize{2}}
        \put(-87,43){\footnotesize{4}}
        \put(-61,43){\footnotesize{6}}
        \put(-34,43){\footnotesize{8}}
        \put(-13,43){\footnotesize{10}}
        \put(-74,38){\footnotesize{$x$}}
        \put(-150,50){\footnotesize{-5}}
        \put(-150,76){\footnotesize{-3}}
        \put(-150,102){\footnotesize{-1}}
        \put(-148,128){\footnotesize{1}}
        \put(-148,155){\footnotesize{3}}
        \put(-148,181){\footnotesize{5}}
        \put(-155,116){\footnotesize{$y$}}
    \end{picture}
    \includegraphics[width=0.3\linewidth]{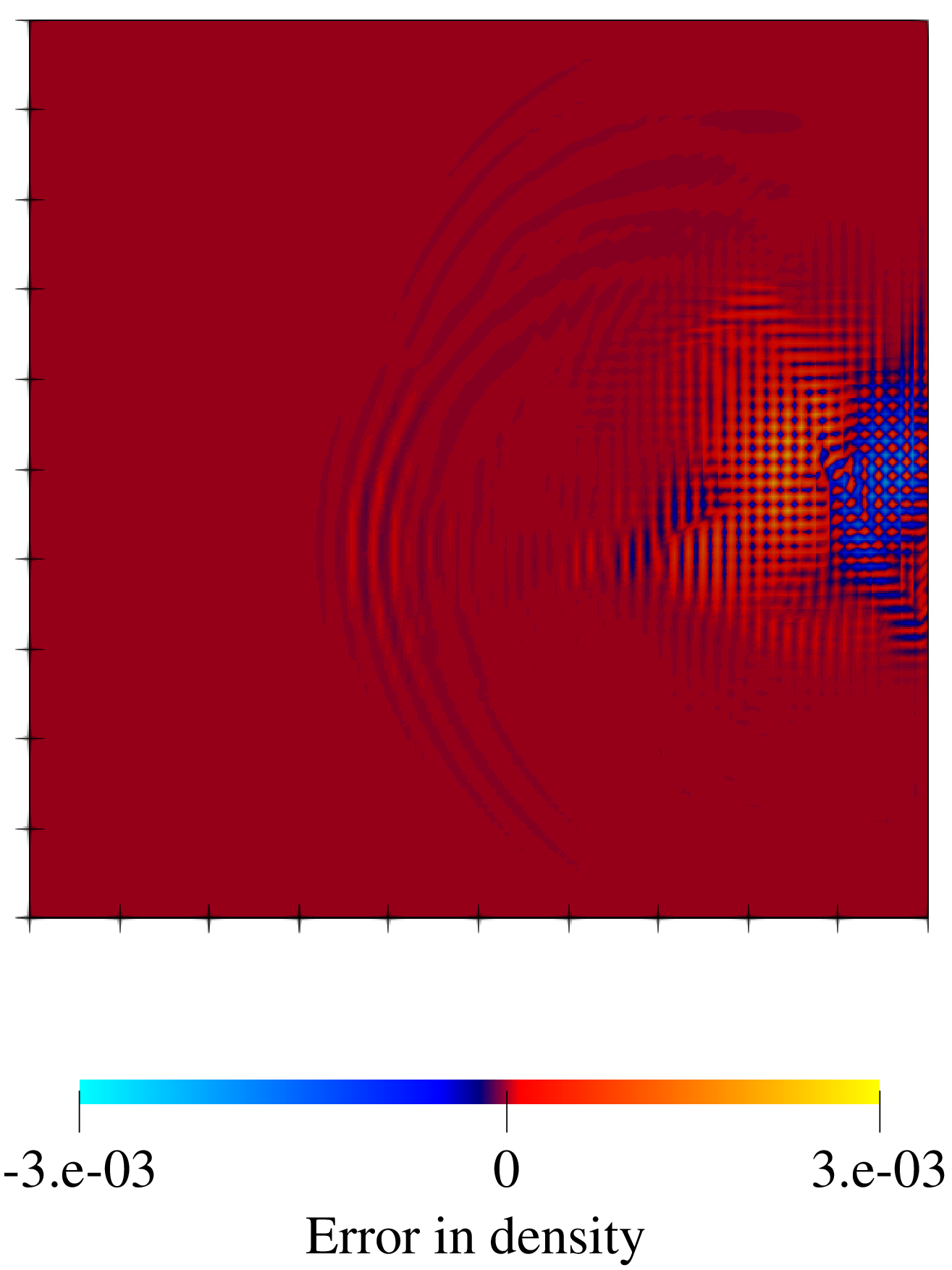}
    \begin{picture}(0,0)
        \put(-110,130){\footnotesize{(e)}}
        \put(-139,43){\footnotesize{0}}
        \put(-114,43){\footnotesize{2}}
        \put(-87,43){\footnotesize{4}}
        \put(-61,43){\footnotesize{6}}
        \put(-34,43){\footnotesize{8}}
        \put(-13,43){\footnotesize{10}}
        \put(-74,38){\footnotesize{$x$}}
    \end{picture}
    \includegraphics[width=0.3\linewidth]{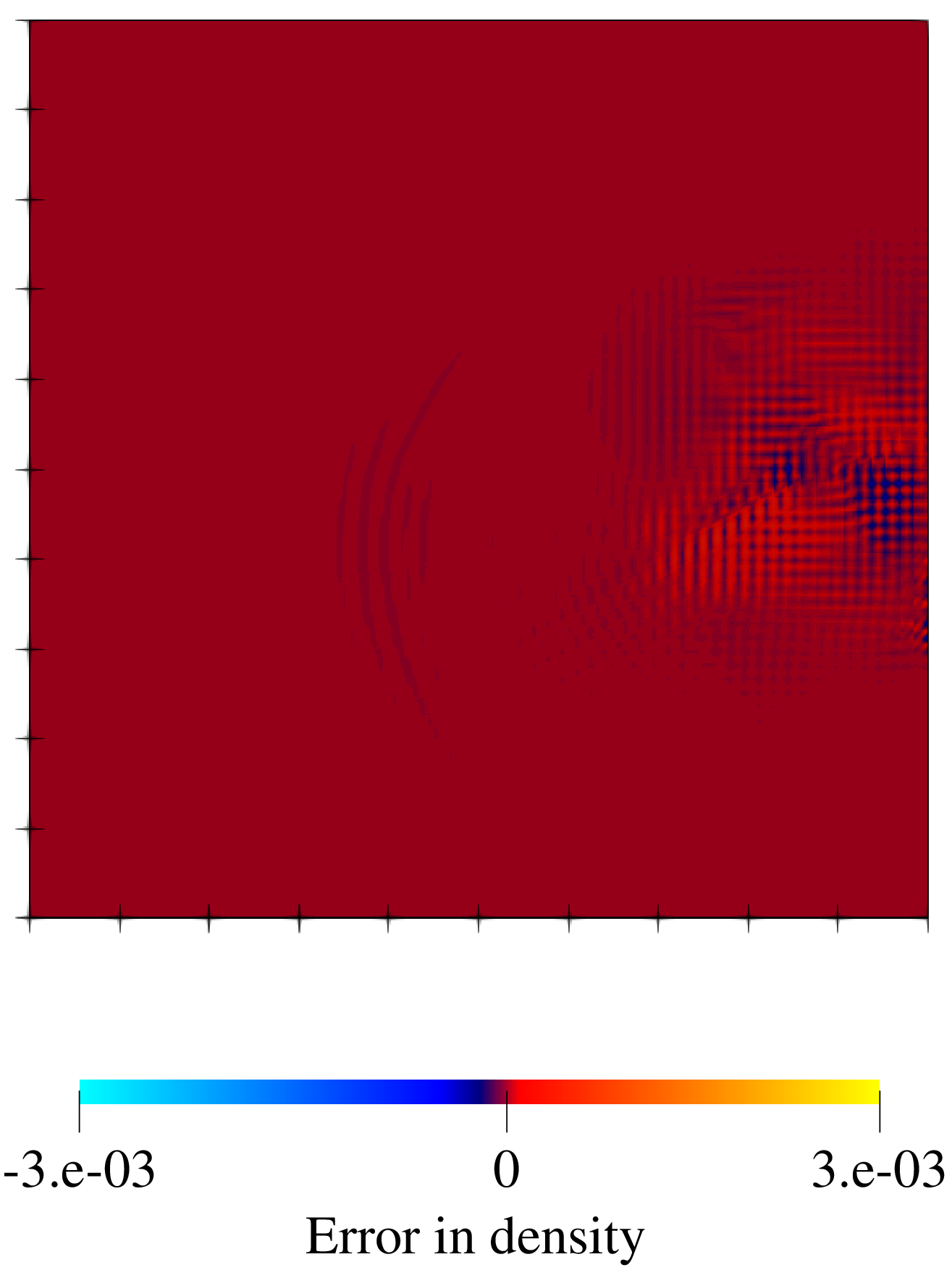}
    \begin{picture}(0,0)
        \put(-110,130){\footnotesize{(f)}}
        \put(-139,43){\footnotesize{0}}
        \put(-114,43){\footnotesize{2}}
        \put(-87,43){\footnotesize{4}}
        \put(-61,43){\footnotesize{6}}
        \put(-34,43){\footnotesize{8}}
        \put(-13,43){\footnotesize{10}}
        \put(-74,38){\footnotesize{$x$}}
    \end{picture}
\caption{\small{Density field (a-c) and corresponding error distribution (d-f) at $t_{\text{final}} = 4.0$ for the two-dimensional \textit{isentropic vortex} test case using synchronous DG(2)-RK3, ADG(2)-RK3, and ADG(2)-AT3-RK3 schemes (left to right). Simulation parameters: $G = 4$ ($N_E = 4096$, 147,456 DoFs), $P = 256$,  $\sigma = 0.05$ for DG(2)-RK3 and ADG(2)-AT3-RK3, $\sigma = 0.03$ for ADG(2)-RK3, and maximum allowable delay $L = 4$.}}
\label{fig:time-evolution-2d}
\end{figure}
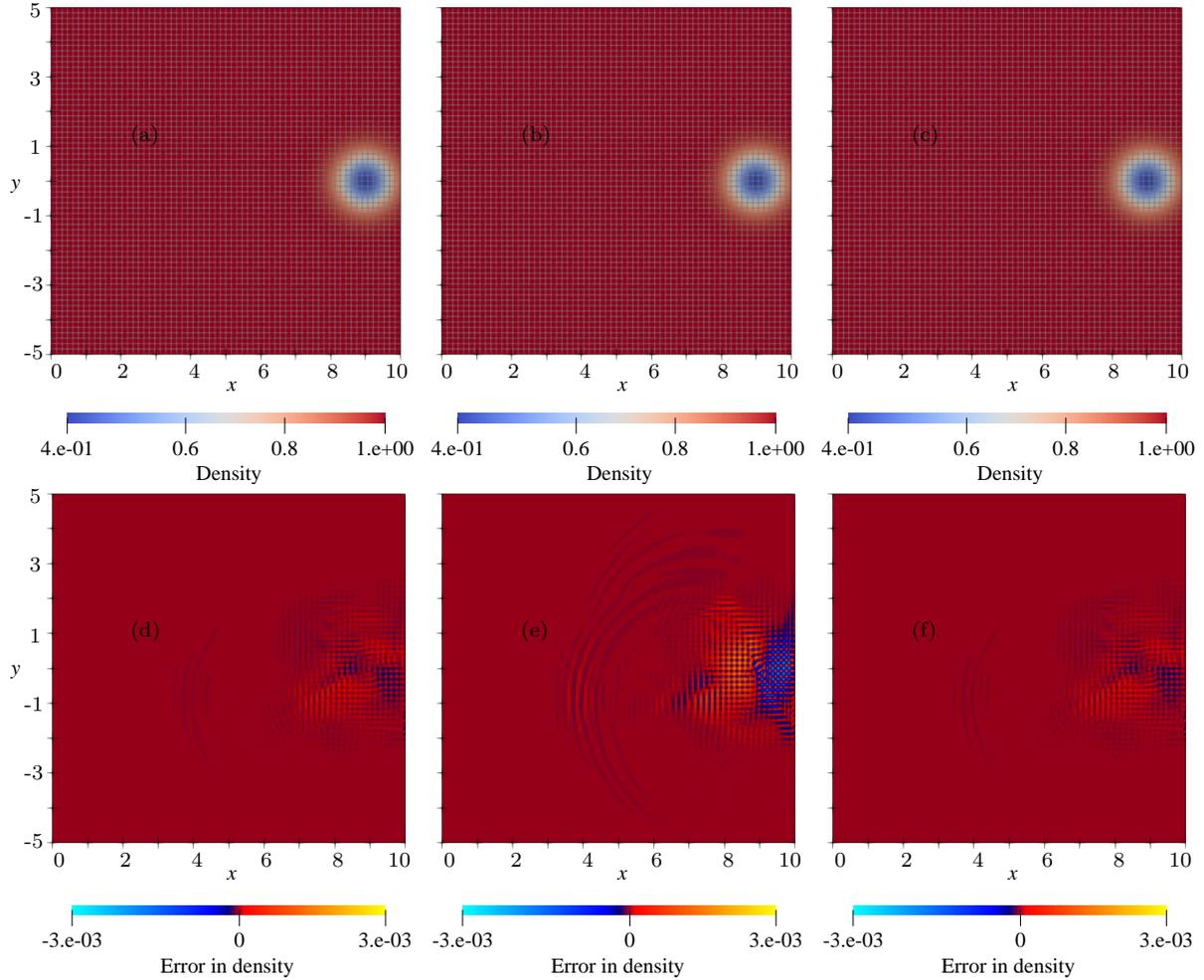

It is important to note that asynchronous DG schemes are subject to more restrictive stability limits compared to their synchronous counterparts, as discussed in detail in \cite{goswami2024-cmame-adg}. In particular, the maximum stable CFL number ($\sigma$) decreases as the allowable communication delay $L$ increases. However, the use of AT fluxes improves the stability properties of ADG schemes. For example, for a CFL value of $\sigma = 0.05$, the ADG(2)-LSERK3 scheme with $L = 4$ becomes unstable, whereas, for the same $L$ and CFL value, the ADG(2)-AT3-LSERK3 scheme remains stable.
For all accuracy studies presented here, we therefore fix the maximum allowable delay to $L = 4$. A conservative CFL number of $\sigma = 0.05$ is used for the DG and ADG-AT schemes, which lies well within the stability limits of the ADG(2)-AT3-LSERK3 formulation. For the ADG scheme with standard fluxes, a smaller CFL value of $\sigma = 0.03$ is employed to ensure stability.

Figure~\ref{fig:time-evolution-2d} shows the density fields obtained using the DG(2)-LSERK3, ADG(2)-LSERK3, and ADG(2)-AT3-LSERK3 schemes at $t_{\text{final}} = 4.0$, together with their corresponding error distributions. The simulations use a mesh refinement level of $G = 4$, resulting in 4096 elements distributed across 256 MPI processes. While the density fields produced by all three schemes appear visually indistinguishable, the error distribution for the ADG(2)-LSERK3 scheme exhibits noticeably larger errors compared to the other two cases. This observation indicates that the use of standard numerical fluxes within the CAA framework leads to a degradation in accuracy. Whereas, incorporating AT fluxes reduces the error magnitude to a level comparable to that of the synchronous DG solver.

\begin{figure}[h!]
    \centering
    \includegraphics[width=0.32\linewidth]{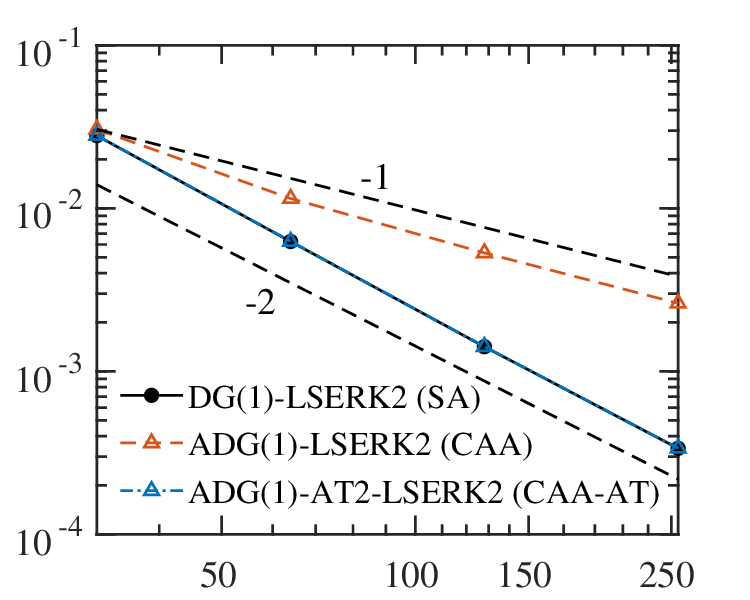}
    \begin{picture}(0,0)
        \put(-50,100){\small{(a)}}
        \put(-160,33){\small{\rotatebox{90}{Error in density}}}
        \put(-109,-8){\small{Number of elements}}
    \end{picture}
    \includegraphics[width=0.32\linewidth]{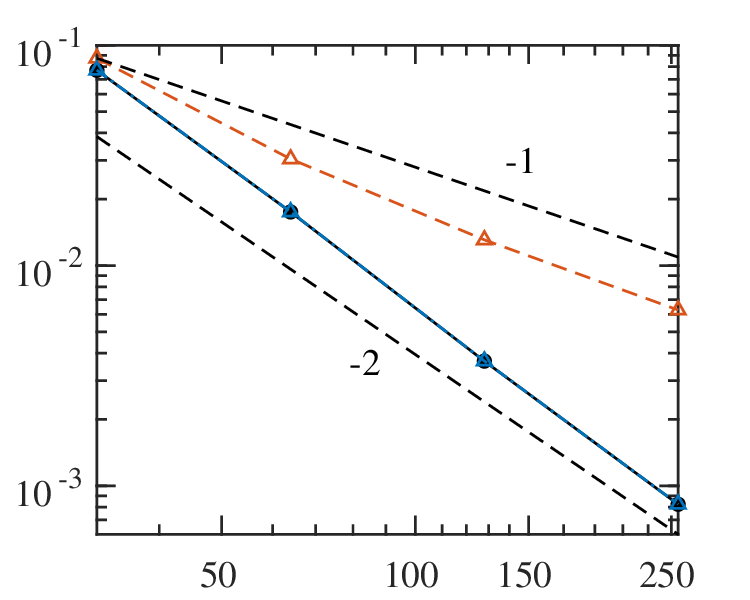}
    \begin{picture}(0,0)
        \put(-50,100){\small{(b)}}
        \put(-160,25){\small{\rotatebox{90}{Error in momentum}}}
        \put(-109,-8){\small{Number of elements}}
    \end{picture}
    \includegraphics[width=0.32\linewidth]{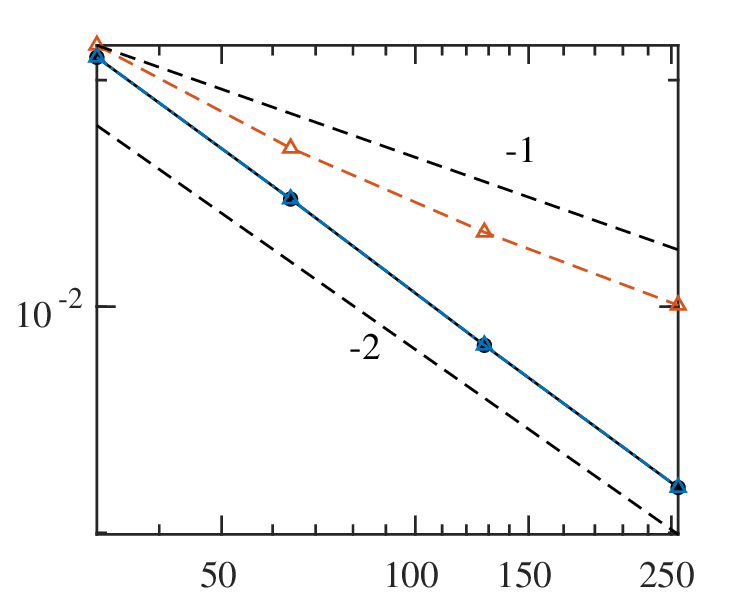}
    \begin{picture}(0,0)
        \put(-50,100){\small{(c)}}
        \put(-160,33){\small{\rotatebox{90}{Error in energy}}}
        \put(-109,-8){\small{Number of elements}}
    \end{picture}
\caption{\small{Convergence of $L_2$-norm errors with grid refinement for the compressible Euler equations: (a) density, (b) momentum, and (c) energy for the two-dimensional \textit{isentropic vortex} test case. Solid black lines denote the synchronous DG(1)-LSERK2 scheme, dashed orange lines denote the ADG(1)-LSERK2 scheme, and dash-dotted blue lines denote the ADG(1)-AT2-LSERK2 scheme. Black dashed reference lines indicate slopes of $-1$ and $-2$. The $x$-axis shows the number of elements in one direction ranging from 32 ($G = 3$) to 256 ($G = 6$). Simulation parameters: $t_{\text{final}} = 4.0$, $\sigma = 0.05$ for SA (DG) and CAA-AT (ADG-AT), and $0.03$ for CAA (ADG), $P = 256$, and $L = 4$.}}
\label{fig:accuracy-caa-dg1rk2}
\end{figure}

\begin{figure}[h]
    \centering
    \includegraphics[width=0.32\linewidth]{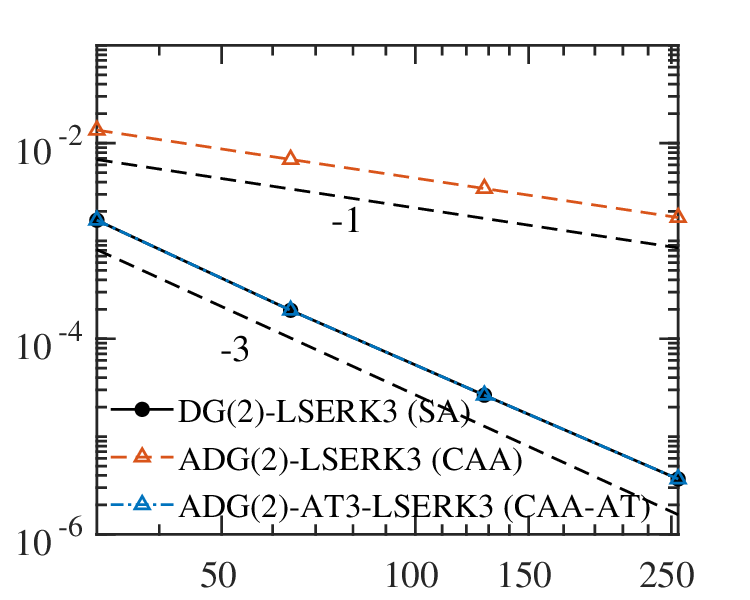}
    \begin{picture}(0,0)
        \put(-50,100){\small{(a)}}
        \put(-160,33){\small{\rotatebox{90}{Error in density}}}
        \put(-109,-8){\small{Number of elements}}
    \end{picture}
    \includegraphics[width=0.32\linewidth]{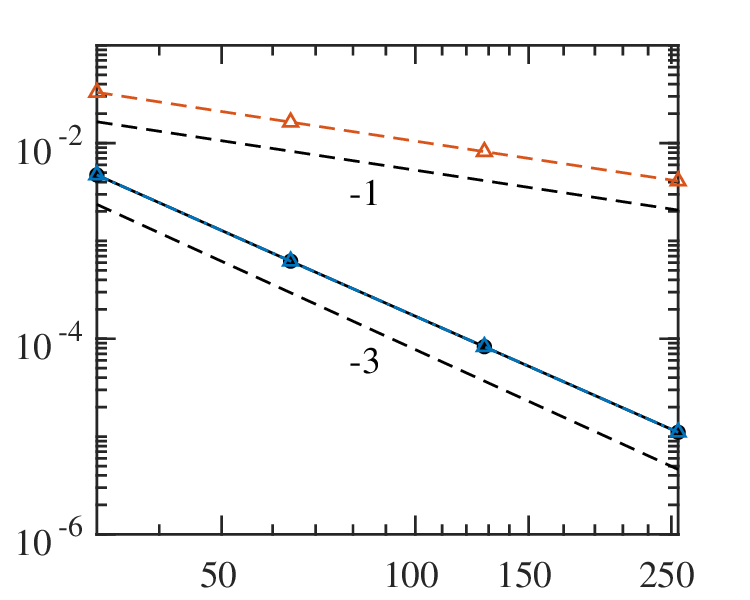}
    \begin{picture}(0,0)
        \put(-50,100){\small{(b)}}
        \put(-160,25){\small{\rotatebox{90}{Error in momentum}}}
        \put(-109,-8){\small{Number of elements}}
    \end{picture}
    \includegraphics[width=0.32\linewidth]{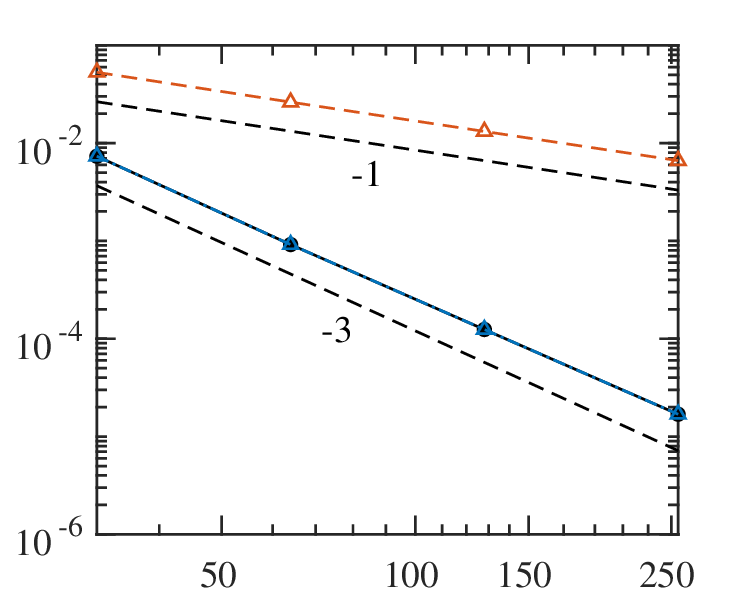}
    \begin{picture}(0,0)
        \put(-50,100){\small{(c)}}
        \put(-160,33){\small{\rotatebox{90}{Error in energy}}}
        \put(-109,-8){\small{Number of elements}}
    \end{picture}
\caption{\small{Convergence of $L_2$-norm errors with grid refinement for the compressible Euler equations: (a) density, (b) momentum, and (c) energy for the two-dimensional \textit{isentropic vortex} test case. Solid black lines denote the synchronous DG(2)-LSERK3 scheme, dashed orange lines denote the ADG(2)-LSERK3 scheme, and dash-dotted blue lines denote the ADG(2)-AT3-LSERK3 scheme. Black dashed reference lines indicate slopes of $-1$ and $-3$. The $x$-axis shows the number of elements in one direction ranging from 32 ($G = 3$) to 256 ($G = 6$). Simulation parameters: $t_{\text{final}} = 4.0$, $\sigma = 0.05$ for SA (DG) and CAA-AT (ADG-AT), and $0.03$ for CAA (ADG), $P = 256$, and $L = 4$.}}
\label{fig:accuracy-caa-dg2rk3}
\end{figure}

To quantify these observations, we compute the $L_2$-norm errors for density, momentum, and energy over a sequence of successively refined meshes. We first compare the ADG$(N_p)$-LSERK$q$ scheme using the standard flux with the corresponding synchronous DG$(N_p)$-LSERK$q$ scheme. The synchronous implementation is expected to achieve an accuracy of $\mathcal{O}(h^{N_p+1})$, whereas the ADG implementation with standard fluxes is theoretically limited to first-order accuracy due to the use of delayed boundary data~\cite{goswami2024-cmame-adg}. Figures~\ref{fig:accuracy-caa-dg1rk2} and~\ref{fig:accuracy-caa-dg2rk3} show the error convergence for $N_p = 1$ and $N_p = 2$, respectively, obtained using $P = 256$ MPI processes and $L = 4$. The slopes of the error curves for the synchronous DG schemes (solid black lines) correspond to second- and third-order accuracy for DG(1)-LSERK2 and DG(2)-LSERK3, respectively. In comparison, the ADG schemes with standard fluxes (dashed orange lines) exhibit a convergence rate of approximately first order in both cases, in agreement with theoretical predictions.

Figures~\ref{fig:accuracy-caa-dg1rk2} and~\ref{fig:accuracy-caa-dg2rk3} also show the corresponding results for the ADG schemes with second- and third-ordr AT fluxes (dash-dotted blue lines). Both approaches are expected to achieve an accuracy of $\mathcal{O}(h^{N_p+1})$ for basis functions of degree $N_p$~\cite{goswami2024-cmame-adg}.  In both cases, the error curves for the synchronous DG scheme and the ADG scheme with AT fluxes overlap closely and exhibit identical slopes, demonstrating second- and third-order convergence for the ADG(1)-AT2-LSERK2 and ADG(2)-AT3-LSERK3 schemes, respectively. These results confirm that the use of AT fluxes successfully restores the optimal order of accuracy of the asynchronous DG method while enabling communication avoidance.

\begin{figure}[h!]
    \centering
    \includegraphics[width=0.4\linewidth]{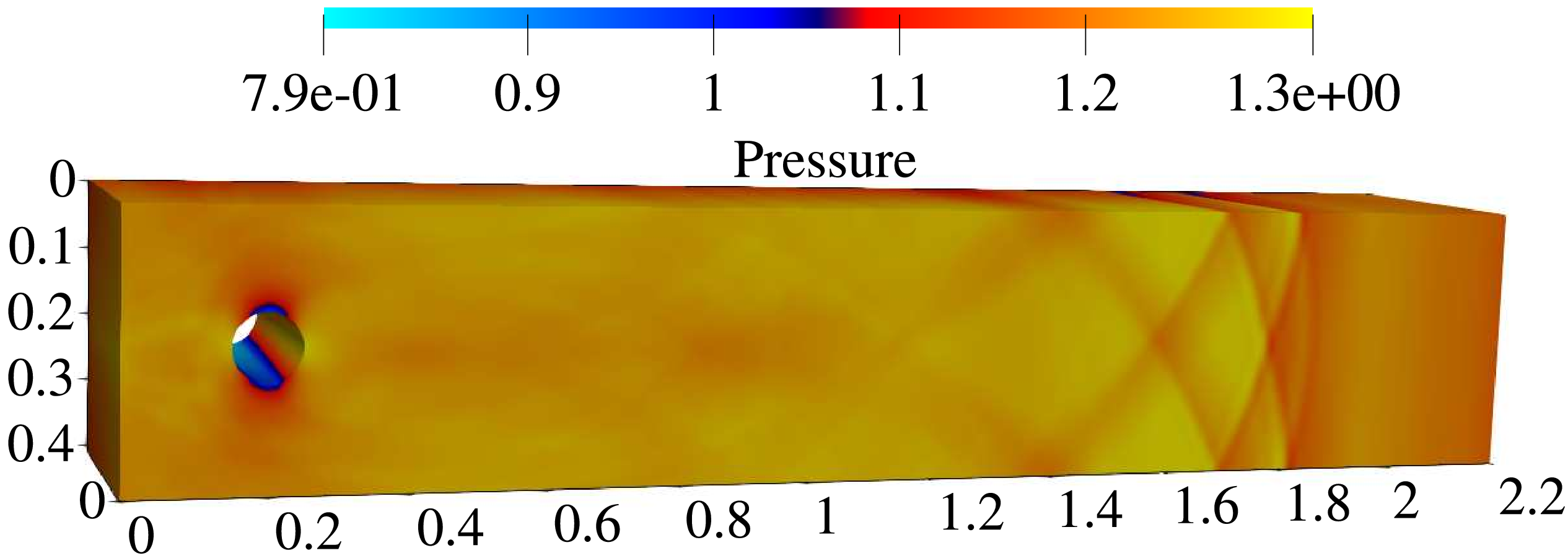}
    \begin{picture}(0,0)
        \put(-210,20){\small{(a)}}
        \put(-196,23){\small{{y}}}
        \put(-90,-4){\small{x}}
        \put(-185,6){\small{z}}
    \end{picture}\hspace{0.1cm}
    \begin{tikzpicture}
\node[anchor=south west, inner sep=0] (img) at (0,0)
{\includegraphics[width=0.4\linewidth]{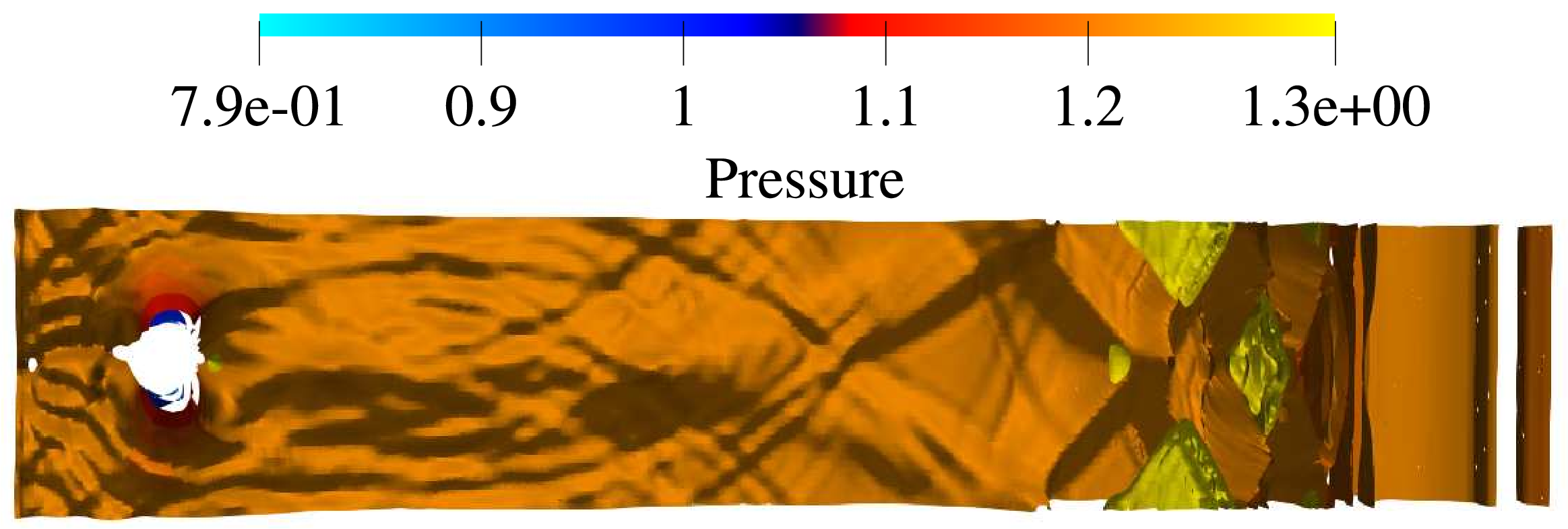}};
\begin{scope}[x={(img.south east)}, y={(img.north west)}]
    \node at (1.0,0.7) {\small (d)};
    
    \draw[red, thick] (0.8,0.3) ellipse (0.04 and 0.16);
    \draw[red, thick, rotate=13] (0.77,0.001) ellipse (0.05 and 0.09);
    \draw[red, thick] (0.135,0.32) ellipse (0.02 and 0.06);
\end{scope}
\end{tikzpicture}\\
    \vspace{0.2cm}
    \includegraphics[width=0.4\linewidth]{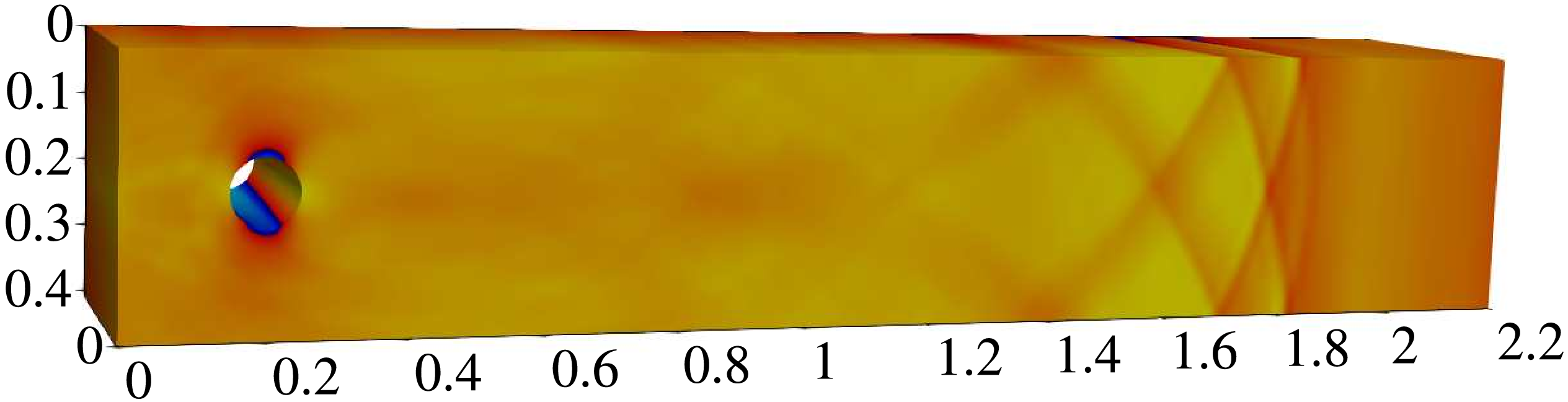}
    \begin{picture}(0,0)
        \put(-210,20){\small{(b)}}
        \put(-196,23){\small{{y}}}
        \put(-90,-4){\small{x}}
        \put(-185,6){\small{z}}
    \end{picture}\hspace{0.1cm}
    \begin{tikzpicture}
\node[anchor=south west, inner sep=0] (img) at (0,0)
{\includegraphics[width=0.4\linewidth]{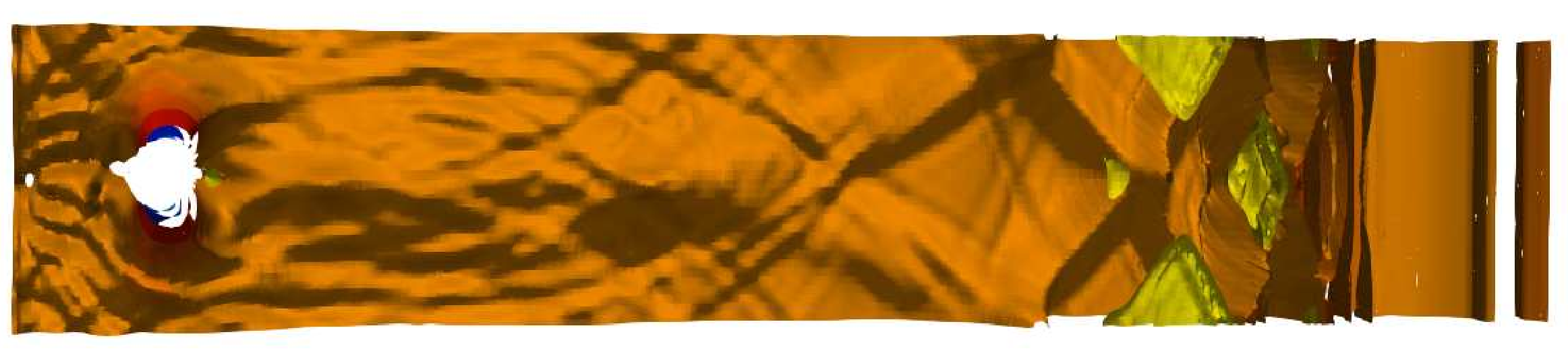}};
\begin{scope}[x={(img.south east)}, y={(img.north west)}]
    \node at (1.0,1.0) {\small (e)};
    
    \draw[red, thick] (0.798,0.45) ellipse (0.045 and 0.245);
    \draw[red, thick, rotate=13] (0.77,0.001) ellipse (0.06 and 0.13);
    \draw[red, thick] (0.135,0.5) ellipse (0.018 and 0.09);
\end{scope}
\end{tikzpicture}\\
    \vspace{0.2cm}
    \includegraphics[width=0.4\linewidth]{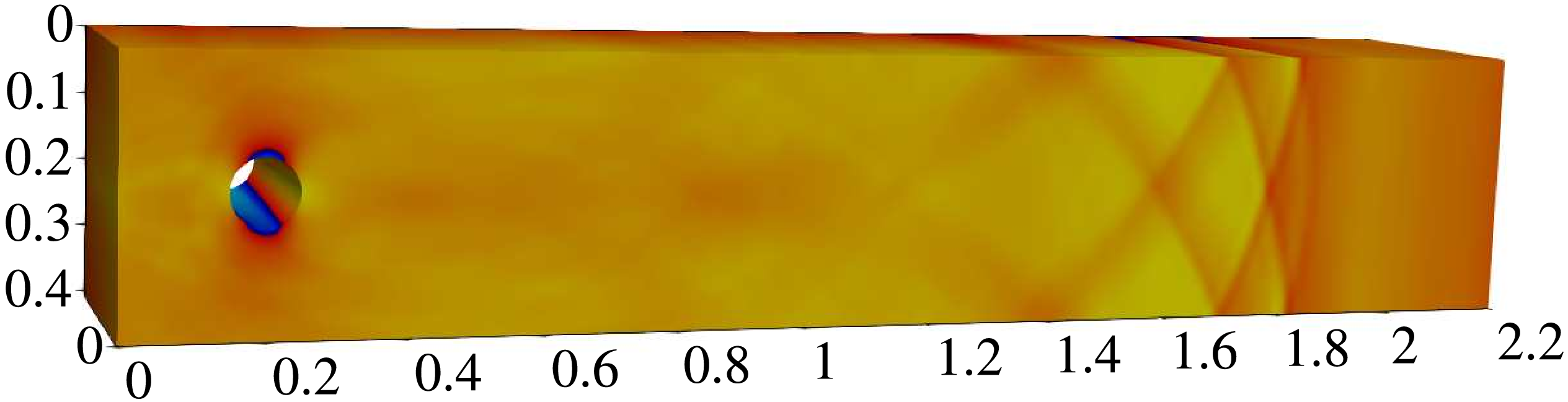}
    \begin{picture}(0,0)
        \put(-210,20){\small{(c)}}
        \put(-196,23){\small{{y}}}
        \put(-90,-4){\small{x}}
        \put(-185,6){\small{z}}
    \end{picture}\hspace{0.1cm}
    \begin{tikzpicture}
\node[anchor=south west, inner sep=0] (img) at (0,0)
{\includegraphics[width=0.4\linewidth]{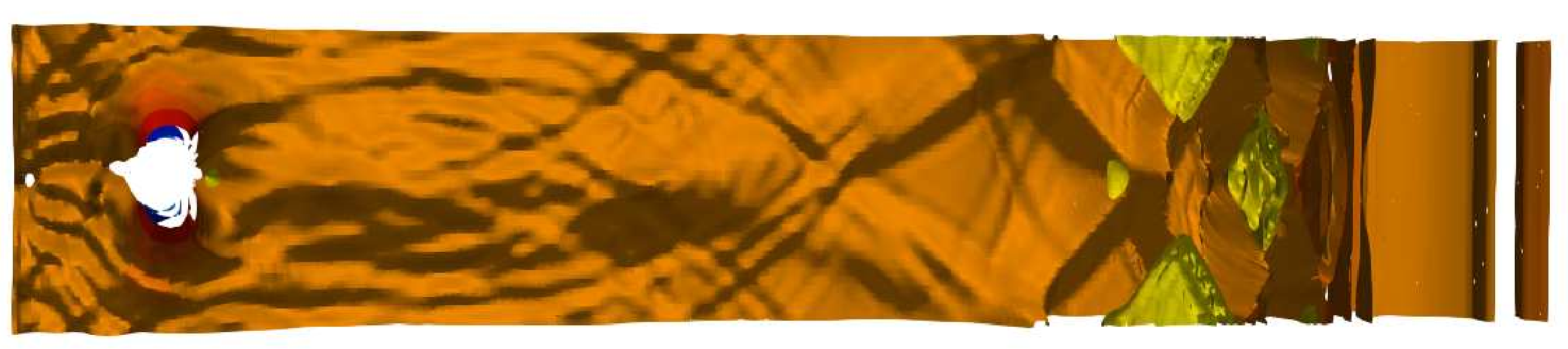}};
\begin{scope}[x={(img.south east)}, y={(img.north west)}]
    \node at (1.0,1.0) {\small (f)};
    
    \draw[red, thick] (0.798,0.45) ellipse (0.045 and 0.245);
    \draw[red, thick, rotate=13] (0.77,0.001) ellipse (0.06 and 0.13);
    \draw[red, thick] (0.135,0.5) ellipse (0.018 and 0.09);
\end{scope}
\end{tikzpicture}\\
\caption{\small{Pressure field (a–c) and the corresponding contour plots for a set of equidistant pressure levels (d–f) at $t_{\text{final}} = 1.0$ for the three-dimensional \textit{flow around a cylinder} test case obtained using synchronous DG(2)-RK3 (top row), ADG(2)-RK3 (middle row), and ADG(2)-AT3-RK3 (bottom row) schemes. Simulation parameters: $G = 3$ ($N_E = 204,800$, 27.6M DoFs), $P = 320$, $\sigma = 0.03$, and maximum allowable delay $L = 4$.}}
\label{fig:pressure-3D-dg2rk3}
\end{figure}

We next assess the accuracy of the asynchronous DG method for the three-dimensional \textit{flow around a cylinder} test case. Figure~\ref{fig:pressure-3D-dg2rk3} presents the pressure field at $t_{\text{final}} = 1.0$ obtained using the DG(2)-LSERK3, ADG(2)-LSERK3, and ADG(2)-AT3-LSERK3 schemes, along with the corresponding contour plots at a set of equidistant pressure levels. The simulations use a mesh refinement level of $G = 3$, resulting in 204,800 elements distributed across 320 MPI processes. For a fair comparison and to consistently visualize flow features in the three-dimensional domain, we use a common CFL value of $\sigma = 0.03$ for all three schemes.
While the pressure fields produced by all three schemes in parts (a), (b), and (c) appear visually indistinguishable, differences become apparent in the contour plots. In particular, the ADG(2)-LSERK3 scheme (Fig.~\ref{fig:pressure-3D-dg2rk3}(e)) exhibits localized distortions in the shock structures compared to the synchronous DG solution, indicating a loss of accuracy due to the use of delayed boundary data. These regions are highlighted by red ellipses. In contrast, the ADG(2)-AT3-LSERK3 scheme (Fig.~\ref{fig:pressure-3D-dg2rk3}(f)) closely matches the synchronous solution, with nearly identical shock patterns and contour structures.

\begin{figure}[h!]
    \centering
    \includegraphics[width=0.45\linewidth]{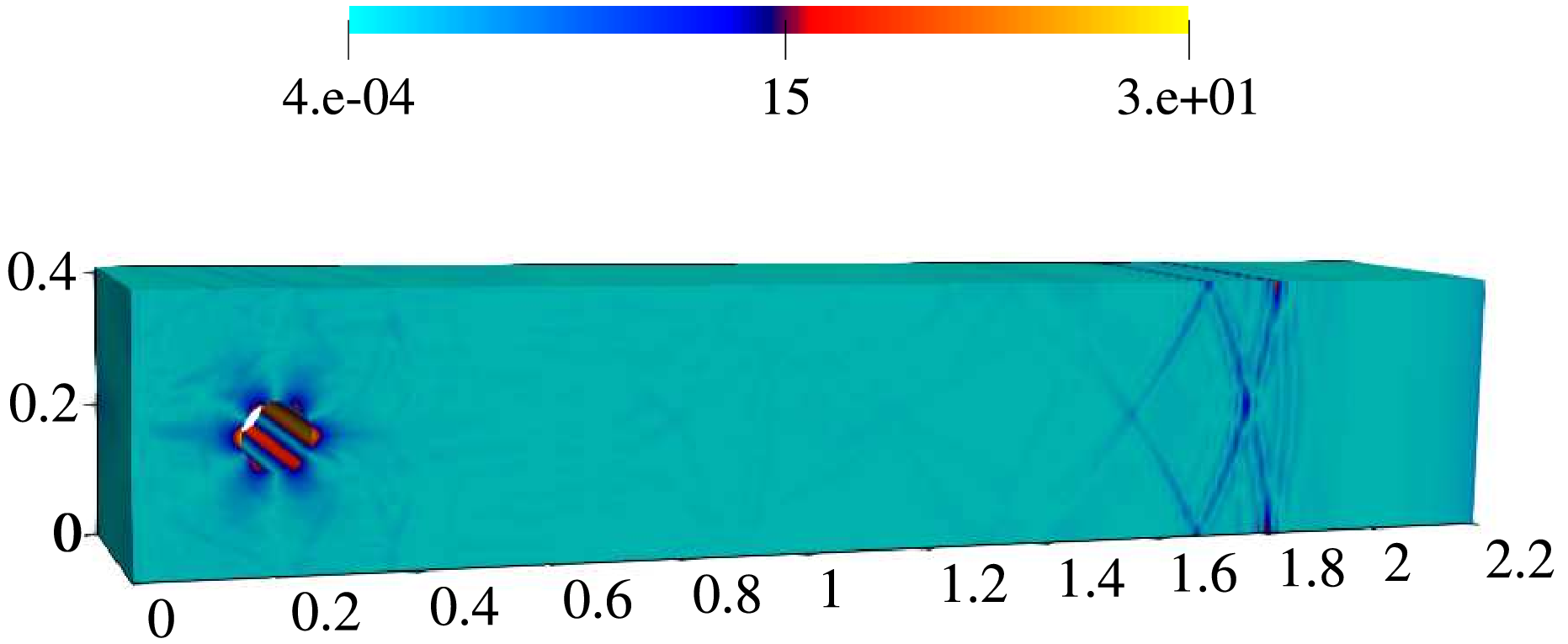}
    \begin{picture}(0,0)
        \put(-240,20){\small{(a)}}
        \put(-220,33){\small{{y}}}
        \put(-100,-4){\small{x}}
        \put(-205,6){\small{z}}
        \put(-42,3.2){\textcolor{red}{\mycirc{}}}
        \put(-35,25){\textcolor{red}{\Huge{$\longrightarrow$}}}
        \put(-118,63){\small{{$\|\nabla \rho \|$}}}
        \put(-41,13){\textcolor{red}{\rule[0.5ex]{0.4pt}{1.2cm}}}
        \put(-56,49){\rotatebox{-10}{\textcolor{red}{\rule[0.5ex]{1.5em}{0.4pt}}}}
    \end{picture}\hspace{0.1cm}
    \includegraphics[width=0.191\linewidth]{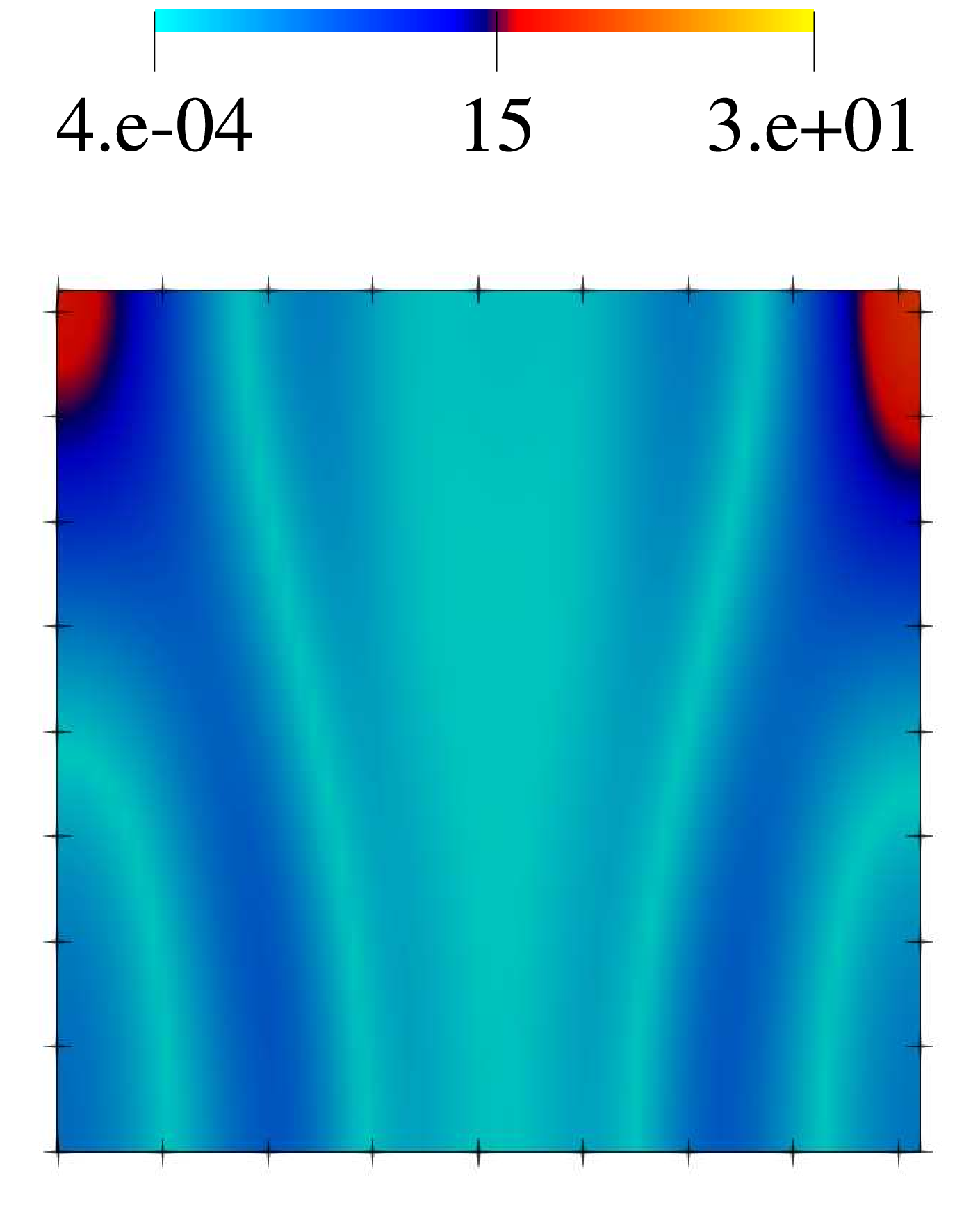}
    \begin{picture}(0,0)
        \put(5,20){\small{(d)}}
        \put(-54,90){\scriptsize{{$\|\nabla \rho \|$}}}
        \put(-87,-2){\scriptsize{0}}
        \put(-71,-2){\scriptsize{0.1}}
        \put(-52,-2){\scriptsize{0.2}}
        \put(-33,-2){\scriptsize{0.3}}
        \put(-14,-2){\scriptsize{0.4}}
        \put(-92,4){\scriptsize{0}}
        \put(-96,23){\scriptsize{0.1}}
        \put(-96,42){\scriptsize{0.2}}
        \put(-96,61){\scriptsize{0.3}}
        \put(-96,80){\scriptsize{0.4}}
        \put(-101,45){\scriptsize{z}}
    \end{picture}\\
    \vspace{0.1cm}
    \includegraphics[width=0.45\linewidth]{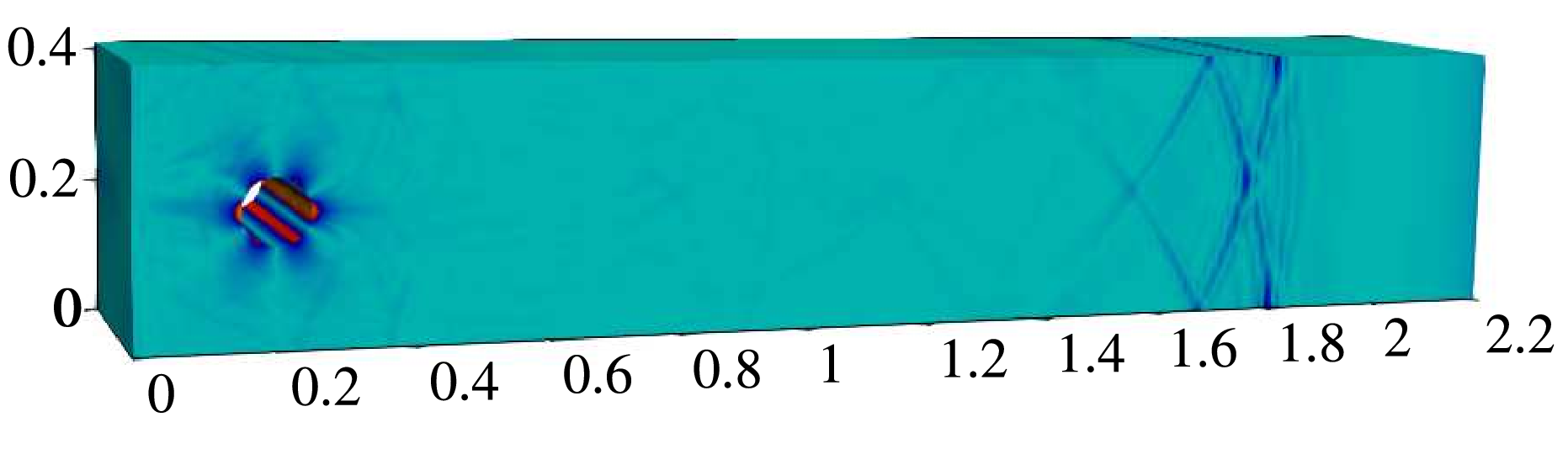}
    \begin{picture}(0,0)
        \put(-240,20){\small{(b)}}
        \put(-220,36){\small{{y}}}
        \put(-100,1){\small{x}}
        \put(-205,11){\small{z}}
        \put(-42,8.2){\textcolor{red}{\mycirc{}}}
        \put(-35,30){\textcolor{red}{\Huge{$\longrightarrow$}}}
    \end{picture}\hspace{0.1cm}
    \includegraphics[width=0.2\linewidth]{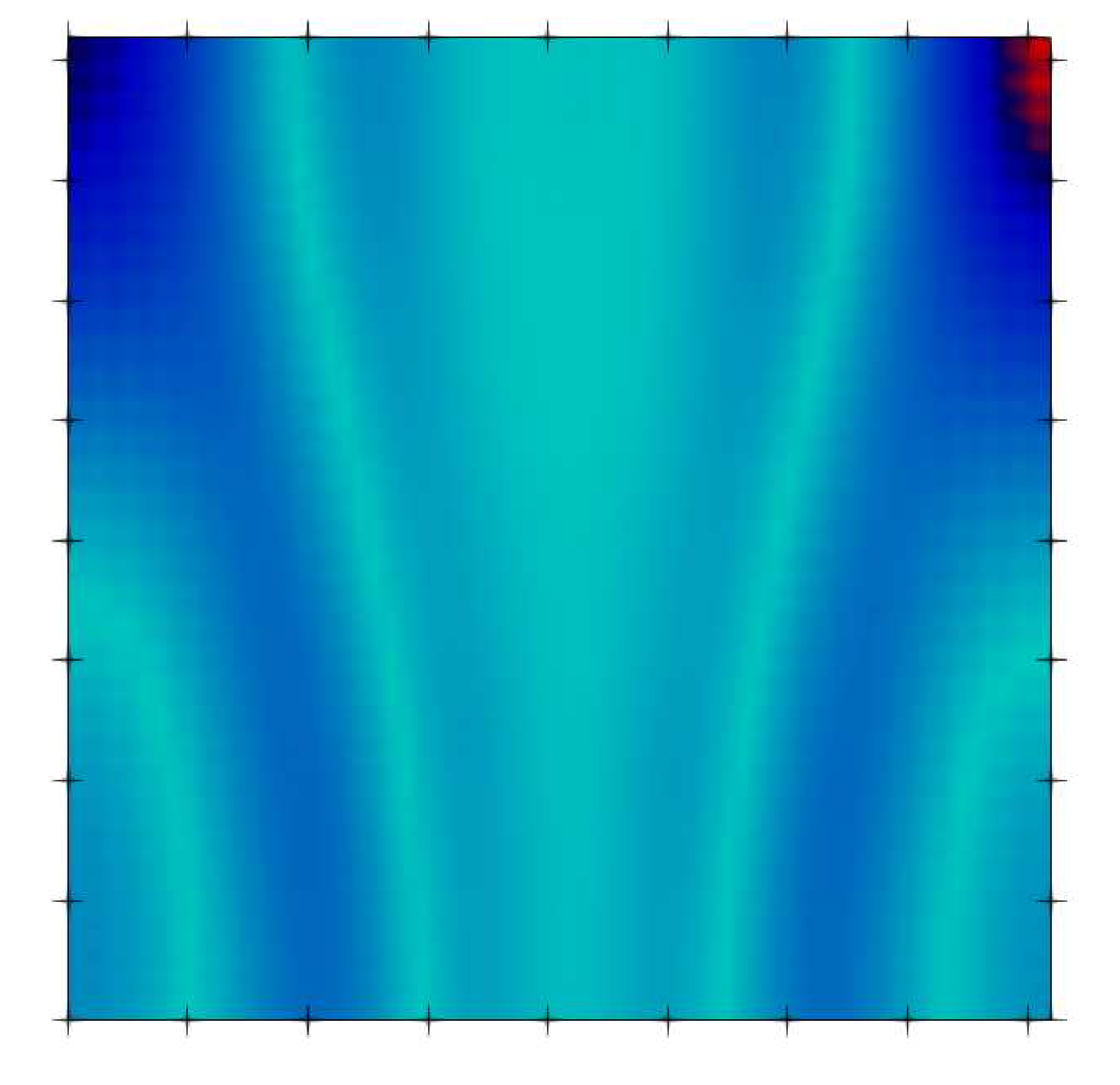}
    \begin{picture}(0,0)
        \put(5,20){\small{(e)}}
        \put(-90,-2){\scriptsize{0}}
        \put(-74,-2){\scriptsize{0.1}}
        \put(-55,-2){\scriptsize{0.2}}
        \put(-34,-2){\scriptsize{0.3}}
        \put(-14,-2){\scriptsize{0.4}}
        \put(-96,4){\scriptsize{0}}
        \put(-100,24){\scriptsize{0.1}}
        \put(-100,44){\scriptsize{0.2}}
        \put(-100,64){\scriptsize{0.3}}
        \put(-100,84){\scriptsize{0.4}}
        \put(-105,45){\scriptsize{z}}
    \end{picture}\\
    \vspace{0.1cm}
    \includegraphics[width=0.45\linewidth]{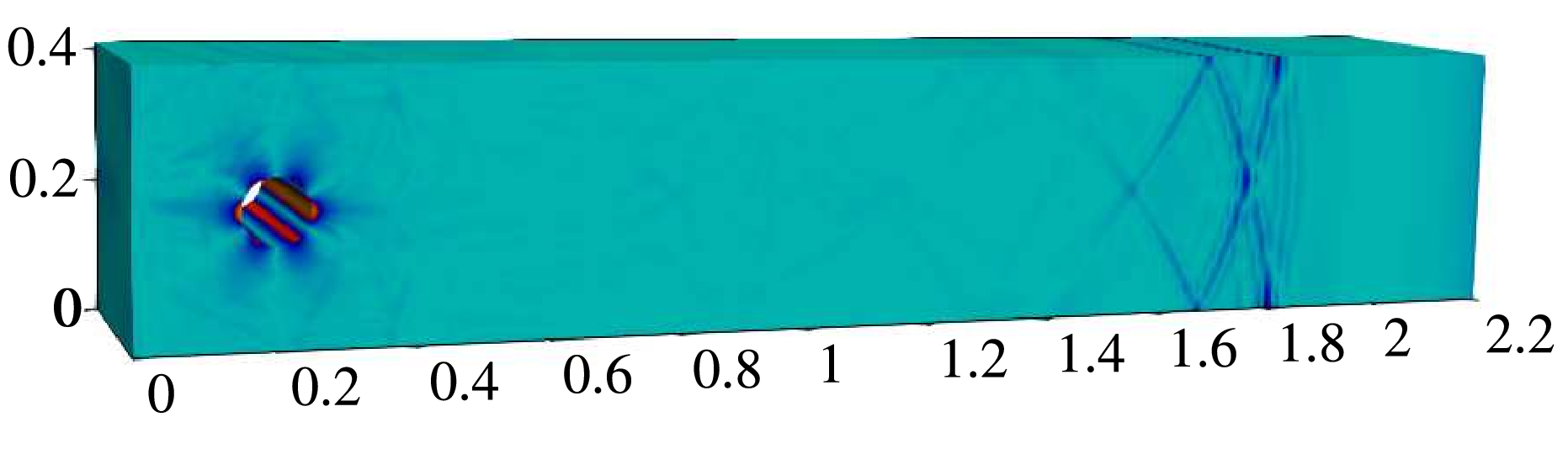}
    \begin{picture}(0,0)
        \put(-240,20){\small{(c)}}
        \put(-220,36){\small{{y}}}
        \put(-100,1){\small{x}}
        \put(-205,11){\small{z}}
        \put(-42,8.2){\textcolor{red}{\mycirc{}}}
        \put(-35,30){\textcolor{red}{\Huge{$\longrightarrow$}}}
    \end{picture}\hspace{0.1cm}
    \includegraphics[width=0.2\linewidth]{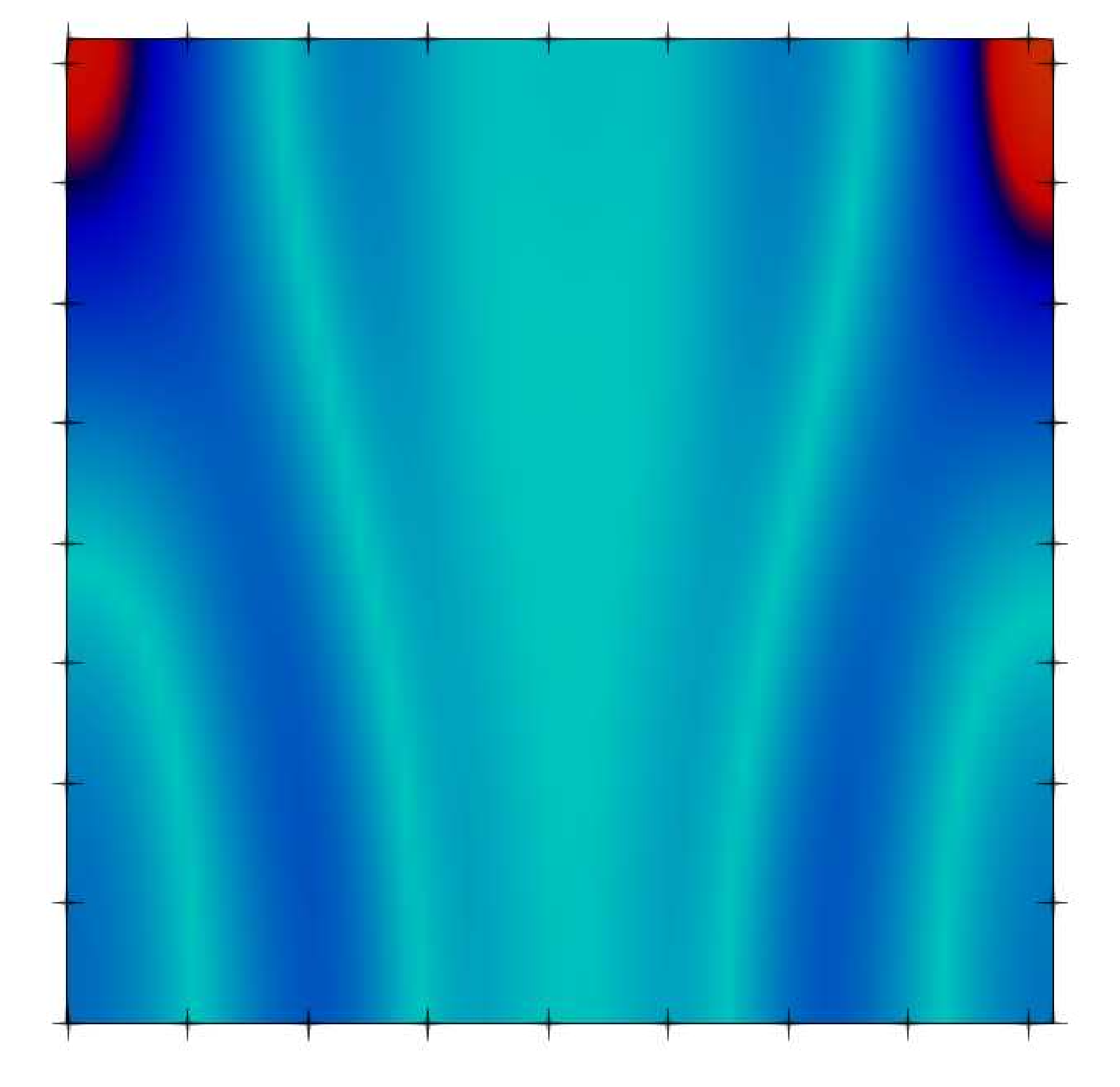}
    \begin{picture}(0,0)
        \put(5,20){\small{(f)}}
        \put(-90,-2){\scriptsize{0}}
        \put(-74,-2){\scriptsize{0.1}}
        \put(-55,-2){\scriptsize{0.2}}
        \put(-34,-2){\scriptsize{0.3}}
        \put(-14,-2){\scriptsize{0.4}}
        \put(-96,4){\scriptsize{0}}
        \put(-100,24){\scriptsize{0.1}}
        \put(-100,44){\scriptsize{0.2}}
        \put(-100,64){\scriptsize{0.3}}
        \put(-100,84){\scriptsize{0.4}}
        \put(-52,-8){\scriptsize{y}}
        \put(-105,45){\scriptsize{z}}
    \end{picture}\\
\caption{\small{Magnitude of the density gradient $\|\nabla \rho\|$ (a–c) and the corresponding slices at $x = 1.80$ (d–f) at $t_{\text{final}} = 1.0$ for the three-dimensional \textit{flow around a cylinder} test case obtained using synchronous DG(2)-RK3 (top row), ADG(2)-RK3 (middle row), and ADG(2)-AT3-RK3 (bottom row) schemes. Simulation parameters: $G = 3$ ($N_E = 204,800$, 27.6M DoFs), $P = 320$, $\sigma = 0.03$, and maximum allowable delay $L = 4$.}}
\label{fig:grad-density-3D-dg2rk3}
\end{figure}

To further highlight these differences, we next analyze the magnitude of the density gradient obtained using the three schemes, shown in Fig.~\ref{fig:grad-density-3D-dg2rk3}. The simulations use the same configuration as in Fig.~\ref{fig:pressure-3D-dg2rk3}. While the global fields in parts (a), (b), and (c) appear largely similar, noticeable differences emerge in the downstream region, particularly in the interval $x \in [1.6,\,1.8]$. In this region, the ADG(2)-LSERK3 solution exhibits slight distortions in the flow structures compared to the synchronous DG solution. To examine these differences more closely, parts (d), (e), and (f) present slices in the $yz$-plane at $x = 1.8$. The discrepancies are clearly visible in Fig.~\ref{fig:grad-density-3D-dg2rk3}(e), where the high-gradient regions near the top corners are significantly under-resolved compared to the corresponding structures in the synchronous solution in part (d) of the figure. Whereas, the ADG(2)-AT3-LSERK3 result in Fig.~\ref{fig:grad-density-3D-dg2rk3}(f) closely matches Fig.~\ref{fig:grad-density-3D-dg2rk3}(d), with nearly identical gradient distributions. This again demonstrates that the use of AT fluxes effectively mitigates the errors introduced by delayed PE-boundary data in the asynchronous DG method.

\subsection{Computational performance}
\label{sec:performance}

This subsection evaluates the computational performance and scalability of the asynchronous DG solver based on the communication-avoiding algorithm (CAA) with asynchrony-tolerant (AT) fluxes. We focus on strong-scaling behavior and quantify performance improvements relative to the synchronous algorithm (SA) implemented in \texttt{deal.II}. All results in this section use the CAA-AT formulation, which preserves high-order accuracy and therefore represents the practically relevant asynchronous approach.
Strong-scaling experiments are performed for both two- and three-dimensional test cases. For the two-dimensional \textit{isentropic vortex} test case, we consider mesh refinement levels ranging from $G = 4, N_p = 2$ (4096 elements and 147,456 DoFs) to $G = 7, N_p = 2$ (262,144 elements and 9.4M DoFs). For the three-dimensional \textit{flow around a cylinder} test case, we use configurations from $G = 1, N_p = 2$ (3200 elements and 432,000 DoFs) to $G = 4, N_p = 2$ (1.6M elements and 221.2M DoFs). The number of MPI processes ranges from 128 (4 compute nodes) to 16,416 (342 compute nodes) in two dimensions, and from 160 (4 compute nodes) to 20,400 (425 compute nodes) in three dimensions. The two-dimensional simulations are advanced for 5000 time steps, whereas the three-dimensional simulations use 1000 time steps.
Each configuration is executed five times, and for every run the solver records the average time per operation over the entire simulation. The reported timings are obtained by taking the mean of these average times across the five independent runs for each configuration and process count.
To assess performance, we report the \emph{speedup}
\[
\text{speedup} = \frac{T_{\text{SA}}}{T_{\text{CAA}}},
\]
and the \emph{parallel efficiency}
\[
\text{parallel efficiency} = \frac{T(P_0)}{T(P)} \frac{P_0}{P},
\]
where $T(P)$ denotes the total runtime on $P$ processes and $P_0$ is the reference process count.

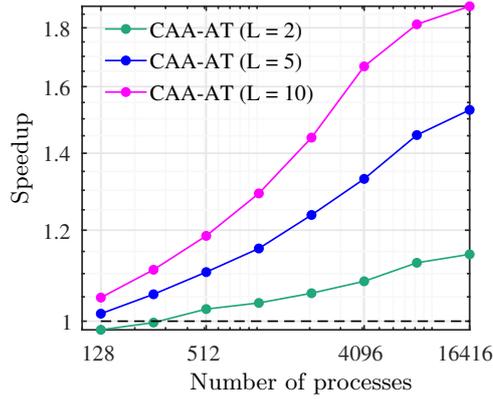
\begin{figure}[h!]
    \centering
    \includegraphics[width=0.4\linewidth]{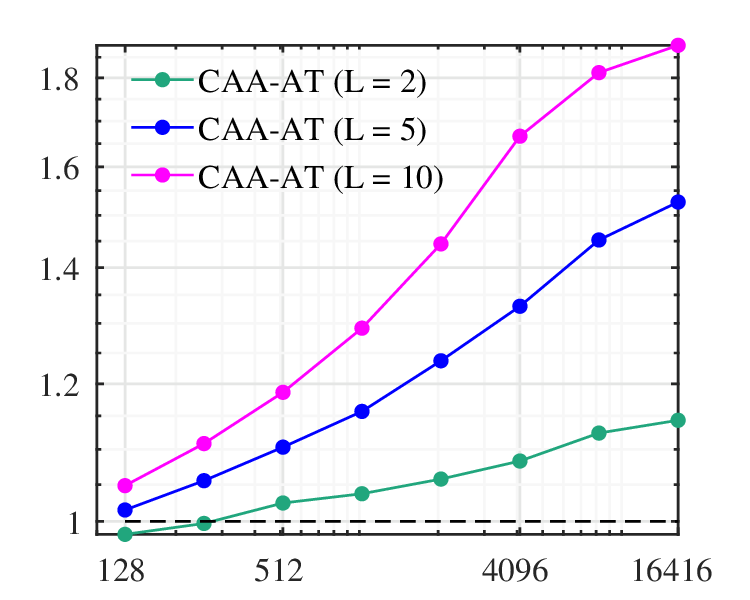}
    \begin{picture}(0,0)
        \put(-193,64){\small{\rotatebox{90}{Speedup}}}
        \put(-126,-6){\small{Number of processes}}
    \end{picture}
\caption{\small{Effect of communication delays on the speedup of the communication-avoiding algorithm with AT fluxes (CAA-AT: ADG(2)-AT3-LSERK3) relative to the synchronous algorithm (SA: DG(2)-LSERK3), obtained from strong-scaling experiments for the two-dimensional \textit{isentropic vortex} test case with configuration $G = 6$, $N_p = 2$ ($N_E = 65,536$, 2.4M DoFs).}}
\label{fig:delay-speedup-2d}
\end{figure}

Figure~\ref{fig:delay-speedup-2d} shows the speedup of the communication-avoiding algorithm with AT fluxes (CAA-AT) relative to the synchronous algorithm (SA) for the two-dimensional configuration $G = 6, N_p = 2$ (65,536 elements and 2.4M DoFs). In these experiments, the problem size is kept fixed while the number of MPI processes is increased from 128 to 16,416. A speedup greater than one indicates that the asynchronous solver outperforms the synchronous baseline.
The curves correspond to different values of the maximum allowable delay $L$, ranging from $L = 2$ to $L = 10$. For all values of $L$, the CAA-AT approach achieves a speedup greater than one across moderate to large process counts, demonstrating consistent performance gains over the synchronous solver. Moreover, the speedup increases with both the process count and the delay parameter $L$. As the process count increases, the fraction of elements located on process boundaries grows, and the runtime of the synchronous solver becomes increasingly dominated by communication and synchronization overhead due to frequent ghost-value exchanges and global synchronization. Consequently, larger values of $L$ lead to significantly higher speedups at extreme scales, as communication is performed less frequently and synchronization overhead is reduced.
These results clearly demonstrate that increasing the maximum allowable delay enhances performance by effectively mitigating communication costs. Based on this observation, the remainder of the performance study focuses on the case $L = 10$, which provides the highest speedup among the tested configurations. In this setting, communication is skipped for nine time steps and subsequently performed for $N_p + 1$ time steps to satisfy the AT-flux requirements.

\begin{figure}[h!]
    \centering
    \includegraphics[width=6.6cm]{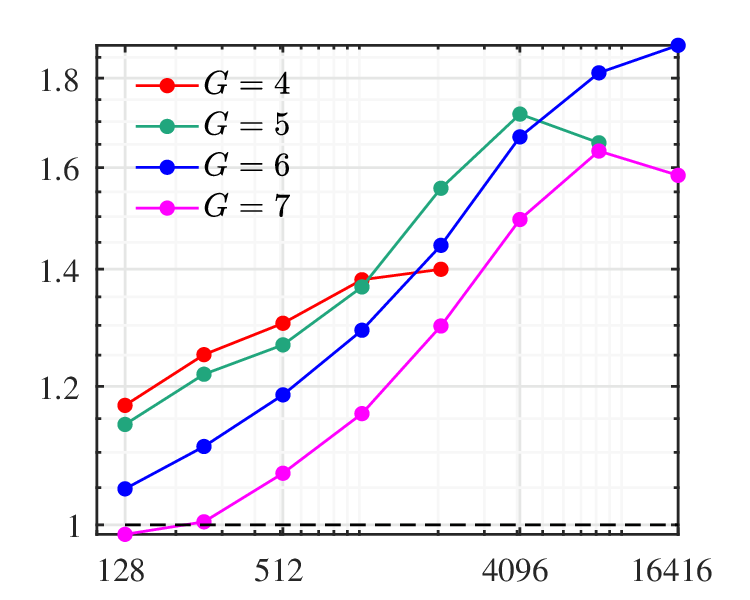}
    \put(-35, 95){(a)}
    \put(-193,64){\small{\rotatebox{90}{Speedup}}}
    \put(-126,-6){\small{Number of processes}}
    \hspace{1cm}
    \includegraphics[width=6.6cm]{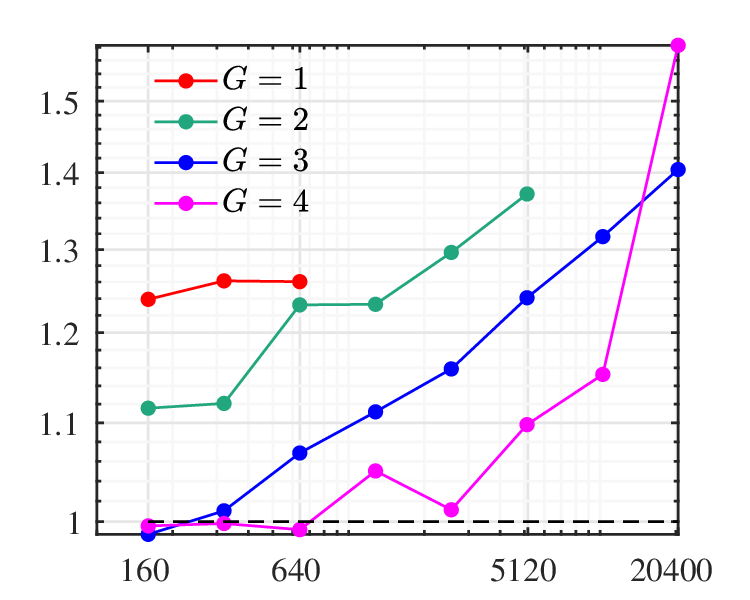}
    \put(-35, 95){(b)}
    \put(-193,64){\small{\rotatebox{90}{Speedup}}}
    \put(-126,-6){\small{Number of processes}}
\caption{\small{Speedup of the communication-avoiding algorithm with AT fluxes (CAA-AT: ADG(2)-AT3-LSERK3) relative to the synchronous algorithm (SA: DG(2)-LSERK3), obtained from strong-scaling experiments. (a) Two-dimensional \textit{isentropic vortex} test case: 147,456 DoFs (red), 589,824 DoFs (green), 2.4M DoFs (blue), and 9.4M DoFs (magenta). (b) Three-dimensional \textit{flow around a cylinder} test case: 432,000 DoFs (red), 3.5M DoFs (green), 27.6M DoFs (blue), and 221.2M DoFs (magenta). Results are shown for a maximum allowable delay $L = 10$. The dashed horizontal line denotes unit speedup for reference.}}
\label{fig:speedup-at}
\end{figure}
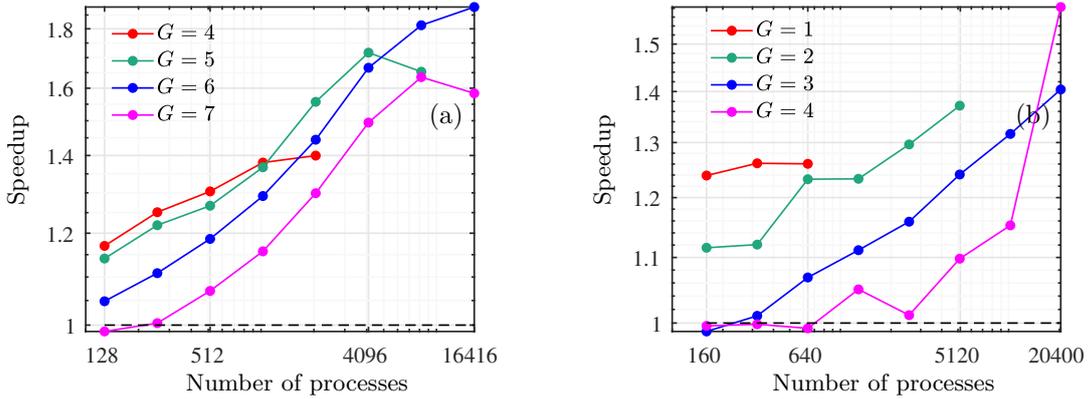

Figure~\ref{fig:speedup-at} compares the performance of the synchronous algorithm (SA) and the communication-avoiding algorithm with AT fluxes (CAA-AT) for multiple mesh resolutions in two and three dimensions using a fixed maximum allowable delay of $L = 10$. In all cases, strong-scaling experiments are performed with a fixed problem size while increasing the number of MPI processes. The black dashed horizontal lines denote unit speedup, corresponding to equal performance of SA and CAA-AT.
We first examine the two-dimensional results shown in Fig.~\ref{fig:speedup-at}(a), which include four mesh refinement levels: $G = 4$ (147,456 DoFs, red), $G = 5$ (589,824 DoFs, green), $G = 6$ (2.4M DoFs, blue), and $G = 7$ (9.4M DoFs, magenta). For the lowest resolution case ($G = 4$), the achievable speedup is limited due to the small amount of work per process. Nevertheless, CAA-AT consistently outperforms SA, achieving a speedup of approximately $1.4\times$ at the largest scale. As the mesh resolution increases, the performance benefit of CAA-AT becomes more pronounced. For the $G = 5$ and $G = 6$ cases, the speedup at extreme scale increases further, with the largest improvement observed for $G = 6$, where CAA-AT is approximately $1.9\times$ faster than SA at 16,416 processes.
Overall, CAA-AT delivers substantial performance gains across all two-dimensional configurations.

Figure~\ref{fig:speedup-at}(b) presents the corresponding three-dimensional results for the configurations $G = 1$ (432,000 DoFs, red), $G = 2$ (3.5M DoFs, green), $G = 3$ (27.6M DoFs, blue), and $G = 4$ (221.2M DoFs, magenta). Similar trends are observed in three dimensions, where CAA-AT consistently outperforms SA across all mesh resolutions and process counts. The advantage of the communication-avoiding approach becomes increasingly evident at large scales, with speedups growing from approximately $1.26\times$ for smaller problem sizes to about $1.6\times$ for the largest configuration ($G = 4$) at 20,400 processes. These results highlight the improved effectiveness of CAA-AT in communication-dominated regimes.

Table~\ref{tab:parallel-efficiency-3d} further quantifies these trends through parallel efficiency measurements for the three-dimensional test case. Across all problem sizes, the CAA-AT approach consistently achieves higher parallel efficiency than the synchronous solver, with the gap widening at larger process counts. In particular, for the largest configuration (221.2M DoFs), CAA-AT maintains a parallel efficiency of approximately $0.53$ at 20,400 processes, compared to only $0.13$ for the synchronous algorithm. These results confirm that the communication-avoiding algorithm significantly improves scalability in the communication-dominated regime.

\begin{table}[h!]
\centering
\caption{Parallel efficiencies for the three-dimensional \textit{flow around a clyinder} test case for the synchronous algorithm (SA: DG(2)-LSERK3) and the communication-avoiding algorithm with AT fluxes (CAA-AT: ADG(2)-AT3-LSERK3). Efficiencies are computed relative to the smallest process count ($P_{0}=160$).}
\setlength{\tabcolsep}{5pt}
\renewcommand{\arraystretch}{1.2}
\begin{tabular}{|c|cc|cc|cc|cc|}
\hline
& \multicolumn{8}{c|}{Parallel efficiency} \\
\cline{2-9}
$P$ 
& \multicolumn{2}{c|}{432,000 DoFs}
& \multicolumn{2}{c|}{3.5M DoFs}
& \multicolumn{2}{c|}{27.6M DoFs}
& \multicolumn{2}{c|}{221.2M DoFs} \\
& SA & CAA-AT & SA & CAA-AT & SA & CAA-AT & SA & CAA-AT \\
\hline
160   & 1.00 & 1.00 & 1.00 & 1.00 & 1.00 & 1.00 & 1.00 & 1.00 \\
320   & 0.61 & 0.62 & 0.89 & 0.89 & 0.98 & 1.01 & 0.98 & 0.98 \\
640   & 0.51 & 0.52 & 0.80 & 0.88 & 0.89 & 0.96 & 0.91 & 0.90 \\
1280  &  --  &  --  & 0.65 & 0.72 & 0.70 & 0.79 & 0.73 & 0.77 \\
2560  &  --  &  --  & 0.60 & 0.70 & 0.51 & 0.63 & 0.56 & 0.71 \\
5120  &  --  &  --  & 0.39 & 0.61 & 0.38 & 0.53 & 0.41 & 0.72 \\
10240 &  --  &  --  &  --  &  --  & 0.22 & 0.32 & 0.26 & 0.53 \\
20400 &  --  &  --  &  --  &  --  & 0.11 & 0.20 & 0.13 & 0.53 \\
\hline
\end{tabular}
\label{tab:parallel-efficiency-3d}
\end{table}

\begin{figure}[h!]
    \centering
    \includegraphics[width=6.6cm]{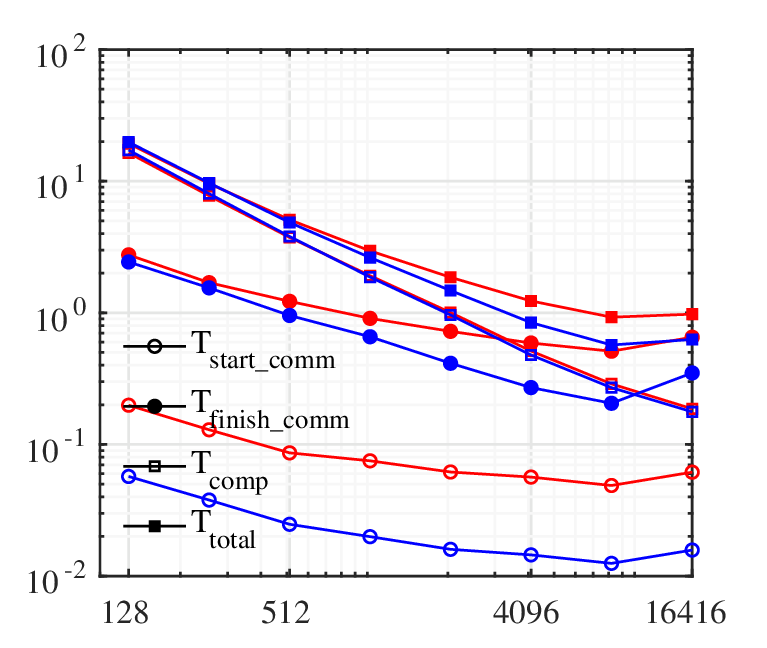}
    \put(-35, 95){(a)}
    \put(-193,64){\small{\rotatebox{90}{Time (s)}}}
    \put(-126,-6){\small{Number of processes}}
    \hspace{1cm}
    \includegraphics[width=6.6cm]{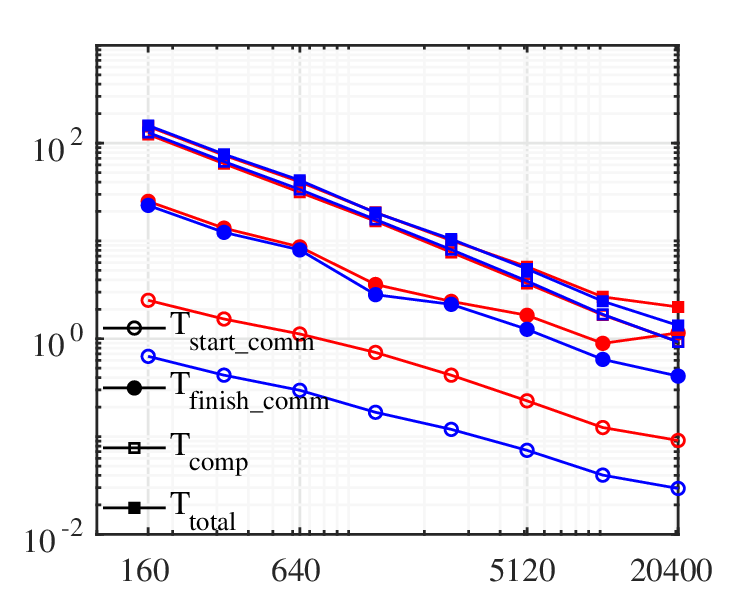}
    \put(-35, 95){(b)}
    \put(-193,64){\small{\rotatebox{90}{Time (s)}}}
    \put(-126,-6){\small{Number of processes}}
\caption{\small{Strong scaling of total communication and computation times for (a) the two-dimensional \textit{isentropic vortex} test case performed on up to $16,416$ processes across $342$ compute nodes, and (b) the three-dimensional \textit{flow around a cylinder} test case performed on up to $20,400$ processes across $425$ compute nodes, using the synchronous algorithm (SA: DG(2)-LSERK3, red) and the communication-avoiding algorithm with AT fluxes (CAA-AT: ADG(2)-AT3-LSERK3, blue). Simulation parameters: number of time steps $= 5000$ for 2D and $1000$ for 3D; refinement levels: (a) $G = 7$ (9.4M DoFs) and (b) $G = 4$ (221.2M DoFs).}}
\label{fig:communication}
\end{figure}
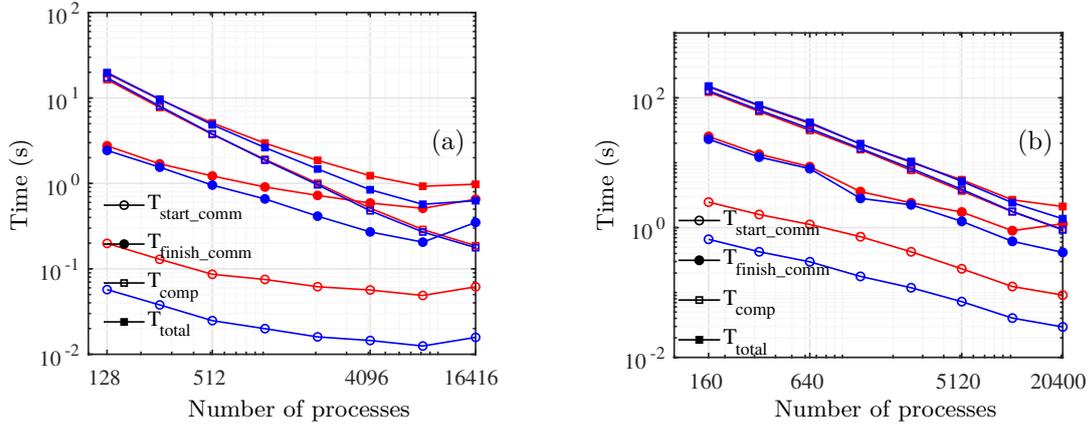

To better explain the observed performance improvements, we examine the relative contributions of computation and communication to the overall runtime. This analysis demonstrates that the improved scalability of CAA-AT directly results from its ability to reduce communication overheads at scale.
Figure~\ref{fig:communication} presents a breakdown of four key runtime components: the time required to initiate communication, $\text{T}_{\text{start\_comm}}$; the synchronization time associated with completing communication, $\text{T}_{\text{finish\_comm}}$; the total element-local computation time, $\text{T}_{\text{comp}}$, defined as the sum of \textit{part-0}, \textit{part-1}, and \textit{part-2} computations; and the overall execution time, $\text{T}_{\text{total}} = \text{T}_{\text{comp}} + \text{T}_{\text{comm}}$. In the figure, red curves correspond to the synchronous algorithm (SA), while blue curves represent the communication-avoiding algorithm with AT fluxes (CAA-AT).

Figure~\ref{fig:communication}(a) presents the breakdown for the two-dimensional test case for the configuration $G = 7, N_p = 2$ (9.4M DoFs). The total computation time exhibits near-ideal scaling across the entire range of MPI processes, decreasing almost linearly as the process count increases.
However, communication-related costs do not follow such linear reductions. For the synchronous solver, the synchronization time $\text{T}_{\text{finish\_comm}}$ becomes comparable to, and eventually exceeds the computation time at large process counts (beyond 2048 MPI processes), with the gap widening steadily at extreme scales. The CAA-AT approach, while exhibiting similar trends, consistently maintains substantially lower communication costs than SA. In particular, the growth of $\text{T}_{\text{finish\_comm}}$ is significantly delayed compared to the synchronous solver, allowing computation to remain the dominant cost over a larger range of process counts (up to 8192 MPI processes).
As a result, the total execution time $\text{T}_{\text{total}}$ for CAA-AT follows the ideal scaling trend more closely than that of SA.

Figure~\ref{fig:communication}(b) presents the corresponding results for the three-dimensional test case with configuration $G = 4, N_p = 2$ (221.2M DoFs). Similar behavior is observed, with computation initially scaling well and communication becoming dominant at larger process counts. For the synchronous solver, the synchronization time increases sharply at extreme scale (beyond 10,240 MPI processes), leading to a clear deviation of the total runtime from ideal scaling. Whereas, CAA-AT maintains significantly lower synchronization costs and continues to follow the expected scaling trend more closely. Consequently, CAA-AT achieves consistently lower total runtime than SA across the entire range of process counts.

\begin{figure}
    \centering
    \includegraphics[width=0.6\linewidth]{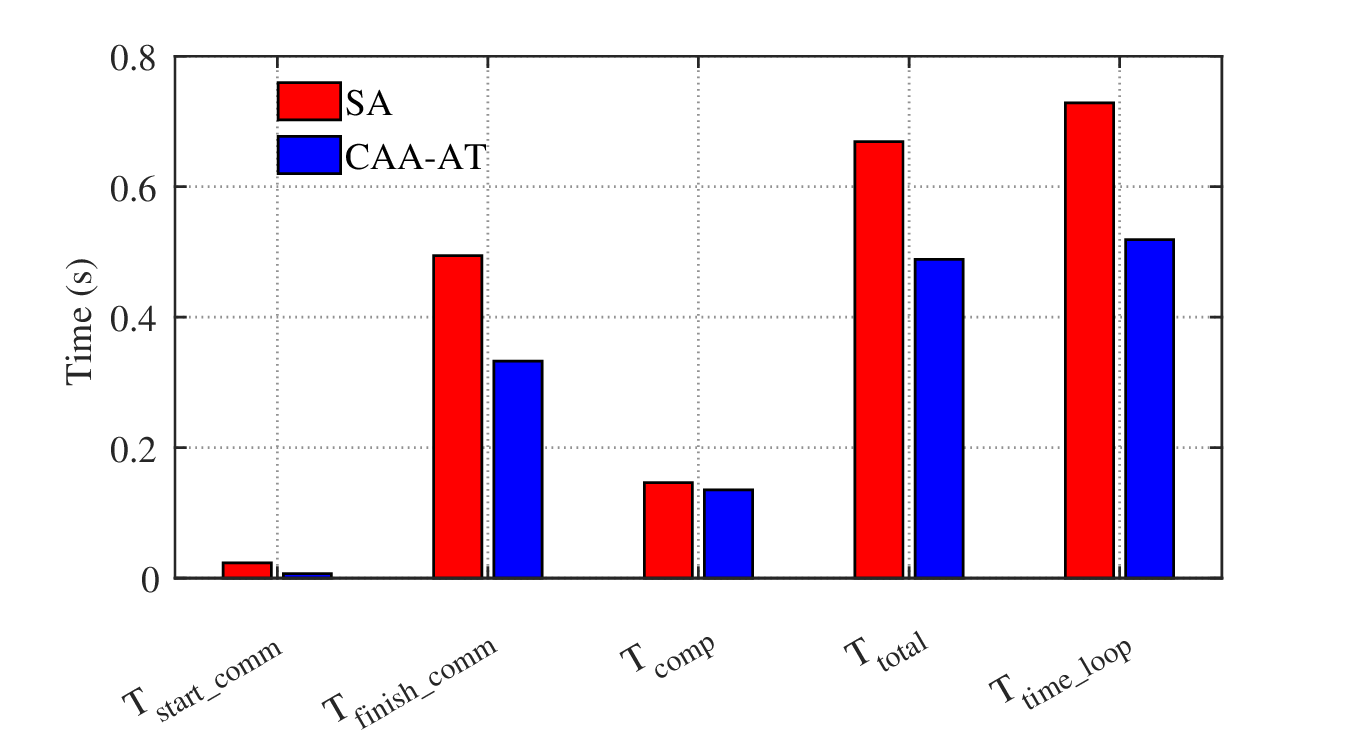}
\caption{\small{Breakdown of communication and computation times for the three-dimensional configuration $G = 4$ (221.2M DoFs) at $P = 20,400$.}}
\label{fig:communication-3d-G4}
\end{figure}

Figure~\ref{fig:communication-3d-G4} presents a further operation-level breakdown for the extreme three-dimensional configuration ($G = 4, N_p = 2$ with $P = 20,400$ processes). The bar plot compares the synchronous algorithm (SA) and the communication-avoiding algorithm with AT fluxes (CAA-AT) across individual runtime components, including communication costs, element-local computation, and the total time spent in the time-stepping loop. A clear reduction in communication and synchronization time is observed for CAA-AT, while the computation time remains comparable between the two methods.
Overall, these results confirm that the communication-avoiding asynchronous DG method effectively mitigates synchronization overheads and sustains favorable strong-scaling behavior at extreme process counts, thereby enabling improved performance of high-order DG solvers on large-scale parallel systems.

\section{Conclusions and discussion}
\label{sec:conclusions}
The scalability of high-order discretizations for time-dependent partial differential equations remains fundamentally constrained by communication and synchronization overheads on modern massively parallel computing systems. In particular, discontinuous Galerkin (DG) methods, while attractive for their accuracy, geometric flexibility, and local conservation properties, typically rely on frequent global or neighborhood-level communication at every stage of time integration. At extreme scales, such synchronization requirements significantly limit strong scaling and overall efficiency. Motivated by these challenges, this work has investigated the practical realization of mathematically asynchronous computing within a DG solver for compressible flow simulations.
Building on recent developments in asynchronous discontinuous Galerkin (ADG) methods, we implemented communication-avoiding algorithms (CAA) within the \texttt{deal.II} finite-element library, leveraging the matrix-free DG solver provided in \textit{step-76}. The key idea underlying this approach is to relax communication and synchronization requirements at a mathematical level by allowing controlled delays in the exchange of ghost values between processing elements (PEs). Within this framework, inter-process communication is performed only periodically, while the solver continues advancing in time using previously communicated data. This approach enables substantial overlap of computation and communication and reduces the frequency of global synchronization points that otherwise dominate execution time at scale.

A central challenge of asynchronous DG schemes is the degradation of accuracy caused by the use of delayed solution values in numerical flux evaluations at PE boundaries. Consistent with prior theoretical analysis, our results confirm that ADG schemes combined with standard numerical fluxes are restricted to first-order accuracy, irrespective of the polynomial degree of the basis functions. To overcome this limitation, we incorporated asynchrony-tolerant (AT) numerical fluxes, which reconstruct high-order accurate fluxes using a linear combination of previously computed fluxes from multiple time levels. The CAA with an appropriate AT flux preserves local conservation and restores the formal order of accuracy of the DG discretization in the presence of communication delays.
The accuracy of the proposed ADG schemes was validated using a two-dimensional isentropic vortex test case with an analytical solution. For polynomial degrees $N_p = 1$ and $2$, the ADG method with AT fluxes achieved second- and third-order convergence, respectively, matching the accuracy of the synchronous DG solver. In contrast, the ADG method with standard fluxes exhibited only first-order convergence, in agreement with theoretical predictions. These results demonstrate that AT fluxes are essential for maintaining accuracy in asynchronous DG formulations.

To assess scalability, we performed extensive strong-scaling experiments for both two- and three-dimensional compressible Euler test cases on a modern CPU-based supercomputing system. Detailed profiling of the synchronous solver revealed that ghost-value synchronization, particularly the completion of nonblocking MPI communication, becomes the dominant bottleneck beyond moderate process counts. While element-local DG operator evaluations scale favorably, the increasing ratio of PE-boundary elements and the loss of interior work severely limit further performance gains.
The communication-avoiding algorithm-based ADG solver with AT fluxes substantially alleviates these limitations. By increasing the maximum allowable communication delay, the CAA-AT approach reduces the frequency of synchronization and keeps communication costs significantly lower than computation over a wider range of process counts. For sufficiently large configurations considered, the ADG solver with AT fluxes achieved speedups of approximately $1.9\times$ in two dimensions and $1.6\times$ in three dimensions relative to the baseline synchronous DG solver. Operation-level breakdowns further confirmed that these gains arise primarily from suppressing synchronization overheads, while maintaining comparable computational costs per time step.

It is important to note that asynchronous DG schemes impose stricter stability constraints than their synchronous counterparts, with the allowable CFL number decreasing as the maximum communication delay $L$ increases. The use of asynchrony-tolerant (AT) fluxes partially alleviates this limitation by recovering stability of the ADG method. In this work, we restricted attention to second- and third-order accurate schemes and selected conservative CFL values to ensure stable execution across all configurations.
In general, increasing the delay parameter $L$ leads to more restrictive CFL limits. However, alternative implementations of the ADG method can mitigate this effect. In particular, the synchronization-avoiding algorithm discussed briefly before, in which communication is performed regularly while relaxing global synchronization, can effectively limit the impact of large delays. Previous studies have shown that communication delays in such settings follow a Poisson distribution with a mean close to unity, indicating that large delays occur infrequently in practice. This behavior allows the use of less restrictive CFL values compared to fully communication-avoiding implementations. While stricter stability limits may constrain the achievable time step in some applications, they are not restrictive for problems involving stiff source terms, such as chemically reacting flows, where the time step is already dictated by fast chemical time scales. In such regimes, the additional stability constraints of the ADG method become largely irrelevant, and the improved scalability of communication-avoiding schemes can be fully exploited.
Further improvements in the stability and robustness of asynchronous DG schemes, including adaptive control of the delay parameter $L$ and hybrid communication–synchronization strategies, remain active areas of research and will be explored in future work.

Several promising directions emerge from the present study. First, extending the current implementation to higher-order accurate schemes requires adopting more sophisticated coupling approaches between multi-level AT fluxes and multi-stage Runge-Kutta integrators. While this work employed a \emph{naive} coupling sufficient for second- and third-order accuracy, high-order formulations offer the potential for improved efficiency at scale and require further investigation. Second, the integration of ADG methods with fully coupled massively parallel reacting-flow solvers represents a particularly attractive application area, where the communication-avoiding approach can deliver significant performance benefits without compromising stability. Finally, extending the present approach to more complex physical models, including viscous and multi-physics systems on unstructured meshes as well as high-dimensional plasma equations such as the Vlasov equation, where the surface-to-volume ratio is significantly larger and communication overheads become particularly severe, will further broaden the applicability of asynchronous DG methods.

In summary, this work demonstrates that the asynchronous DG method with asynchrony-tolerant fluxes can be successfully integrated into large-scale, matrix-free DG solvers using different communication-avoiding approaches and deliver substantial scalability improvements for compressible flow simulations. By mitigating synchronization bottlenecks while preserving accuracy and conservation, the proposed approach represents an important step toward the development of efficient DG-based solvers for next-generation exascale computing platforms.

\section{Acknowledgments}
The authors benefited from discussions with Phani Motamarri, Martin Kronbichler, Katharina Kormann and Michał Wichrowski. Special acknowledgment is also due to the Council of Scientific and Industrial Research (CSIR), India, for awarding the doctoral fellowship to SKG.

%


\bibliographystyle{model1-num-names}
\bibliography{main.bib}




%



\end{document}